\documentclass[iop]{emulateapj}
\usepackage{epsfig}

\usepackage{amsfonts}
\usepackage{amsmath}
\usepackage{mathrsfs}
\usepackage{amssymb}
\usepackage{rotating}

\usepackage{color}
\usepackage{ulem}
\usepackage{xspace}
\usepackage{cancel}
\definecolor{lgray}{gray}{0.85}

\newcommand{\lya} {Ly$\alpha$\xspace}
\newcommand{\siiv} {\ion{Si}{4}\xspace}
\newcommand{\civ} {\ion{C}{4}\xspace}
\newcommand{\cv} {\ion{C}{5}\xspace}
\newcommand{\ciii} {\ion{C}{3}}
\newcommand{\oiii} {\ion{O}{3}}
\newcommand{\oii} {\ion{O}{2}}
\newcommand{\ovi} {\ion{O}{6}\xspace}
\newcommand{\heii}{\ion{He}{2}\xspace}
\newcommand{\heiii}{\ion{He}{3}\xspace}
\newcommand{\siiii}{[\ion{Si}{3}]\xspace}
\newcommand{\neiv}{[\ion{Ne}{4}]\xspace}
\newcommand{\nv} {\ion{N}{5}\xspace}
\newcommand{\niv} {[\ion{N}{4}]\xspace}

\newcommand{\unitcgssb}  {erg\,s$^{-1}$\,cm$^{-2}$\,arcsec$^{-2}$\xspace}

\def \cgssb {{\rm\,erg\,s^{-1}\,cm^{-2}\,arcsec^{-2}}}
\def\bea{\begin{eqnarray}}
\def\eea{\end{eqnarray}}
\begin{document}

\title{Deep HeII and CIV Spectroscopy of a Giant Ly$\alpha$ Nebula: Dense Compact Gas Clumps 
in the Circumgalactic Medium of a $\lowercase{z}\sim 2$ Quasar\footnotemark[\large $\star$]}\footnotetext[\large $\star$]{The data 
presented herein were obtained at the W.M. Keck Observatory, which is operated as a scientific partnership among 
the California Institute of Technology, the University of California and the National Aeronautics and Space Administration. 
The Observatory was made possible by the generous financial support of the W.M. Keck Foundation.}
 
\author{Fabrizio Arrigoni Battaia\altaffilmark{1,2}, %\& the Slug collaboration
        Joseph F. Hennawi\altaffilmark{1}, 
        J. Xavier Prochaska\altaffilmark{3,4}, 
        Sebastiano Cantalupo\altaffilmark{3,4,5}
	}
\altaffiltext{1}{Max-Planck-Institut f\"ur Astronomie, K\"onigstuhl 17, D-69117 Heidelberg, Germany; arrigoni@mpia.de}
\altaffiltext{2}{Member of the International Max Planck Research School for Astronomy \& Cosmic Physics at the University of Heidelberg (IMPRS-HD)}
\altaffiltext{3}{Department of Astronomy and Astrophysics, University of California, 1156 High Street, Santa Cruz, California 95064, USA}
\altaffiltext{4}{University of California Observatories, Lick Observatory, 1156 High Street, Santa Cruz, California 95064, USA}
\altaffiltext{5}{Institute for Astronomy, Department of Physics, ETH Zurich, CH-8093 Zurich, Switzerland}

\slugcomment{Submitted}
\shorttitle{Dense Compact Gas Clumps in the CGM of UM287}
\shortauthors{Arrigoni Battaia et al.}

\begin{abstract}
  The recent discovery by \citet{Cantalupo2014} of the largest
  ($\sim500$ kpc) and luminous (${L \simeq 1.43\times10^{45}}$~erg~s$^{-1}$) \lya nebula
    associated with the quasar UM287 ($z=2.279$) 
    poses a great challenge to our current understanding of the astrophysics
    of the halos hosting massive $z\sim2$ galaxies.
    Either an enormous reservoir of cool gas is required $M\simeq 10^{12}$
    M$_{\odot}$, exceeding the expected baryonic mass available, or
    one must invoke extreme gas clumping factors not present in high-resolution
    cosmological simulations. However, observations of \lya
    emission alone cannot distinguish between these two scenarios. We have
    obtained the deepest ever spectroscopic integrations in the
    \heii$\lambda1640$ and \civ$\lambda1549$ emission lines with the goal
    of detecting extended line emission, but detect neither
    line to a $3\sigma$ limiting SB$\simeq 10^{-18}$\unitcgssb. We
    construct simple models of the expected emission spectrum in the
    highly probable scenario that the nebula is powered by
    photoionization from the central hyper-luminous quasar. 
    The non-detection of HeII implies that the nebular
    emission arises from a mass $M_{\rm c} \lesssim
    6.4\times10^{10}$~M$_{\odot}$ of cool gas on $\sim 200\,{\rm kpc}$
    scales, distributed in a population of remarkably dense ($n_{\rm H}
    \gtrsim 3$~cm$^{-3}$) and compact ($R \lesssim 20$ pc) clouds, which would
    clearly be unresolved by current cosmological simulations.  
    Given the large gas motions suggested by the Ly$\alpha$ line ($v \simeq
    500\,{\rm km/s}$), it is unclear how these clouds survive without
    being disrupted by hydrodynamic instabilities.  
    Our work serves as
    a benchmark for future deep integrations with current and planned
    wide-field IFU spectrographs such as MUSE, KCWI, and KMOS.  Our
    observations and models suggest that a $\simeq 10\,{\rm hr}$
    exposure would likely detect $\sim 10$ rest-frame UV/optical
    emission lines, opening up the possibility of conducting detailed
    photoionization modeling to infer the physical state of gas in the
    circumgalactic medium.
\end{abstract}


%________________________________________________________________
\maketitle

\section{Introduction}

In the modern astrophysical lexicon, the intergalactic medium (IGM) is
the diffuse medium tracing the large-scale structure in the Universe,
while the so-called circumgalactic medium (CGM) is the material on
smaller scales within galactic halos ($r \lesssim 200\,{\rm kpc}$),
for which non-linear processes and the complex interplay
between all mechanisms that lead to galaxy formation take place.

Whether one is studying the IGM or the CGM, for decades the preferred
technique for characterizing such gas has been the analysis of
absorption features along background sightlines (e.g.,
\citealt{Croft2002, Bergeron2004,
  Hennawi2006,Hennawi2007,Prochaska2009, Rudie2012,
  Hennawi2013,Farina2013,Prochaska2013,Prochaska2013b,Lee2014}).
However, as the absorption studies are limited by the rarity of
suitably bright background sources near galaxies, and to the
one-dimensional information that they provide, they need to be
complemented by the direct observation of the medium in emission.

In particular, it has been shown that UV background
radiation could be reprocessed by these media and be detectable as
fluorescent \lya emission (\citealt{Hogan1987, Binette1993, Gould1996,
  Cantalupo2005}).  However, current facilities are still not capable
of revealing such low radiation levels, e.g. an expected surface
brightness (SB) of the order of SB$_{{\rm Ly}\alpha}\sim10^{-20}$
\unitcgssb (see e.g. \citealt{Rauch2008}).
Nonetheless, this signal can be boosted to observable
levels by the intense ionizing flux of a nearby quasar which, like a
flashlight, illuminates the gas in its surroundings
(\citealt{Rees1988, HaimanRees2001, Alam2002, Cantalupo2012}),
shedding light on its physical nature.

Detecting this fluorescence signal has been a subject of significant
interest, and several studies which specifically searched for emission
from the IGM in the proximity to a quasar (e.g., \citealt{Fynbo1999,
  Francis2004, Cantalupo2007, Rauch2008,Hennawi2013}) have so far a not straightforward interpretation.
But recently \citet{Cantalupo2012} identified a population
of compact Ly$\alpha$ emitters with rest-frame equivalent widths
exceeding the maximum value expected from star-formation, $EW_0^{{\rm
    Ly}\alpha}>240$\AA\ (e.g., \citealt{Charlot1993}), which are the
best candidates to date for fluorescent emission powered by a proximate quasar. 

Besides illuminating nearby clouds in the IGM, a quasar may irradiate
gas in its own host galaxy or CGM. A number of studies have reported
the detection of extended \lya emission in the vicinity of $z \sim
2-4$ quasars (e.g., \citealt{HuCowie1987, Heckman1991spec,
  Heckman1991, Christensen2006, Hennawi2009, North2012}), but detailed
comparison is hampered by the different methodologies of these
studies. Although extended \lya nebulae on scales of $\sim100$ kpc (up
to $250$ kpc) have been observed around high-redshift radio galaxies
(HzRGs; e.g.  \citealt{McCarthy1993, vanOjik1997, Reuland2003,
  VillarM2003, VillarM2003b, Reuland2007, Miley2008}), these objects
have the additional complication of the interaction between the
powerful radio jets and the ambient medium, complicating the
interpretation of the observations.  Nebulae of comparable size and
luminosity have similarly been observed in a distinct population
of objects known as `Ly$\alpha$ blobs' (LABs) (e.g.,
\citealt{Steidel2000, Matsuda2004, Dey2005, Smith2007, Geach2007,
  Prescott2009, Yang2011, Yang2012, Yang2014, FAB2014}) which do not show
direct evidence for an AGN. Despite increasing evidence
that the LABs are also frequently associated with obscured AGN 
(\citealt{Geach2009, Overzier2013, Prescott2015})
(although lacking powerful radio jets), the
mechanism powering their emission remains controversial with at least
four proposed which may even act together: (i) photoionization by a
central obscured AGN (\citealt{Geach2009, Overzier2013}), (ii) shock heated gas
by galactic superwinds (\citealt{Taniguchi2001}), (iii) cooling
radiation from cold-mode accretion (e.g., \citealt{Fardal2001,
  Yang2006,Faucher2010, Rosdahl12}), and (iv) resonant scattering of \lya from
star-forming galaxies (\citealt{Dijkstra2008, Hayes2011, Cen2013}).

The largest and most luminous Ly$\alpha$ nebula known is that around
the quasar UM287 ($i$-mag=17.28) at $z=2.279$, recently discovered by
\citet{Cantalupo2014} in a narrow-band imaging survey of
hyper-luminous quasars (\citealt{FABProceeding}). Its size of 460 kpc and average \lya surface
brightness of $SB_{{\rm Ly}\alpha}=6.0\times10^{-18}$ \unitcgssb (from
the $2\sigma$ isophote), which corresponds to a total luminosity of
$L_{Ly\alpha}=(2.2\pm0.2)\times10^{44}$ erg s$^{-1}$, make it the largest reservoir ($M_c\sim10^{12}$
M$_{\odot}$) of cool ($T\sim10^4$ K) gas ever observed around a
QSO. The emission has been explained as recombination and/or
scattering emission from the central quasar and has been regarded as
the first direct detection of a cosmic web filament
(\citealt{Cantalupo2014}).

However, as discussed in \citet{Cantalupo2014}, \lya emission alone
does not allow to break the degeneracy
between the clumpiness or density of the gas, and the total gas
mass. Indeed, in the scenario where the nebula is ionized by the
quasar radiation, the total cool gas mass scales as $M_{\rm c}\sim
10^{12} C^{-1/2}$~M$_{\odot}$, where ${C=\langle n_{\rm H}^2 \rangle/\langle n_{\rm H}\rangle^2}$ is a clumping factor
introduced by \citet{Cantalupo2014} to account for the possibility of
higher density gas unresolved by the cosmological simulation used to
model the emission. Thus, if one assume $C=1$, the 
implied cool gas mass in the extended nebula is exceptionally high for the 
expected dark matter halo inhabited by a $z\sim2-3$ quasar, i.e. 
$M_{DM}=10^{12.5}$ M$_{\odot}$ (\citealt{White2012}). This is further 
aggravated by the fact that current
cosmological simulations show that only a small fraction ($\sim 15\%$,
\citealt{Fumagalli2014, Faucher-Giguere2014, Cantalupo2014}) of the total baryons
reside in a phase ($T<5\times10^4$ K) sufficiently cool to emit in the
\lya line. A possible solution to this discrepancy would be to assume
a very high clumping factor up to $C\simeq 1000$, which would then imply a
large population of cool, dense clouds in the CGM and extending into
the IGM, which are unresolved by current cosmological simulations
(\citealt{Cantalupo2014}).

Both of these scenarios, whether it be too much cool gas, or a large
population of dense clumps, are reminiscent of a similar problem that
has emerged from absorption line studies of the quasar CGM
(\citealt{Hennawi2006,Hennawi2007,Prochaska2009,Prochaska2013,Prochaska2013b,Prochaska2014}). This work reveals substantial
reservoirs of cool gas $\gtrsim 10^{10}$ M$_{\odot}$, manifest as a high
covering factor $\simeq 50\%$ of optically thick absorption, several
times larger than predicted by hydrodynamical
simulations (\citealt{Fumagalli2014,Faucher-Giguere2014}). This conflict most likely
indicates that current simulations fail to capture essential aspects
of the hydrodynamics in massive halos at $z\sim 2$
(\citealt{Prochaska2013b,Fumagalli2014}), perhaps failing to resolve the formation of
clumpy structure in cool gas, which in the most extreme cases give
rise to giant nebulae like UM287.

In an effort to better understand the mechanism powering the
emission in UM287, and further constrain the physical properties of
the emitting gas, this paper presents the result of a sensitive search for emission
in two additional diagnostics, namely
\heii$\lambda$1640\AA\
\footnote{The \heii$\lambda$1640\AA\ is the first line of the Balmer
  series emitted by the Hydrogen-like atom He$^+$, i.e. corresponding
  to the H$\alpha$ line.}, and \civ$\lambda$1549\AA.
The detection of either of these high-ionization emission lines in the extended nebula,
would indicate that the nebula is `illuminated' by an intense source of hard ionizing
photons $E \gtrsim 4 {\rm Ryd}$, 
and would thus establish that photoionization by the quasar 
is the primary mechanism powering the giant \lya
nebula.
As we will show in this work, in a photoionization scenario where \heii emission
results from recombinations, the strength of this line is sensitive to the density of the
gas in the nebula, which can thus break the degeneracy between gas density and gas mass
described above. 
In addition, because \heii is not a resonant line, a comparison of its morphology and kinematics to
the \lya line can be used to test whether Ly$\alpha$ photons are resonantly scattered.
On the other hand, a detection of extended emission in the \civ line
can provide us information on the metallicity of
the gas in the CGM, and simultaneously constrain the size at which the halo is metal-enriched.
To interpret our observational results, we exploit the models
presented by \citet{Hennawi2013} and already used in the context of
extended \lya nebulae (i.e. LABs) in \citet{FAB2014}, and show how a sensitive
search for diffuse emission in Ly$\alpha$ and additional diagnostic lines
can be used to constrain the physical properties of the quasar CGM.

This paper is organized as follows.
In \S\ref{sec:data}, we describe our deep spectroscopic observations,
the data reduction procedures, and the surface brightness limits of
our data.  In \S\ref{sec:results}, we present our constraints on
emission in the \civ and \heii lines, and our analysis of the
kinematics of the \lya line.  In \S\ref{sec:model} we present the
photoionization models for UM287 and  in \S\ref{sec:comparison} we compare them with our observational
results and with absorption spectroscopic studies (\S\ref{conAbsL} and \S\ref{absStudies}). In \S\ref{sec:sens} we discuss which other lines might
be observable with current facilities.  In
\S\ref{sec:caveats} we further discuss some of the assumptions made in
our modeling.  Finally, \S\ref{sec:Conclusion} summarizes our
conclusions.

Throughout this paper, we adopt the cosmological parameters $H_0 = 70$ km s$^{-1}$ Mpc$^{-1}$, $\Omega_M=0.3$ and $\Omega_{\Lambda}=0.7$. 
In this cosmology, 1\arcsec\ corresponds 
to 8.2 physical kpc at $z=2.279$. All magnitudes are in the AB system (\citealt{Oke1974}).

\section{Observations and Data Reduction}
\label{sec:data}

Two moderate resolution (FWHM$\sim 300$ km s$^{-1}$) spectra of the UM287 nebula
were obtained using the Low Resolution Imaging Spectrograph (LRIS; \citealt{Oke1995}) 
on the Keck I telescope on UT 2013 Aug 4, 
in multi-slit mode with custom-designed slitmasks. We used the 600 lines mm$^{-1}$ grism blazed at 4000 \AA\ 
on the blue side, resulting in wavelength coverage of $\approx 3300 - 5880$ \AA, which allows us to 
cover the location of the \civ and \heii lines. The dispersion of this grism is $\sim$ 4 \AA\ per pixel and 
our 1\arcsec slit give a resolution of FWHM $\simeq$ 300 km s$^{-1}$. 
We observed each mask for a total of $\sim 2$ hours in a series of 4 exposures.

\begin{figure}
\centering
\epsfig{file=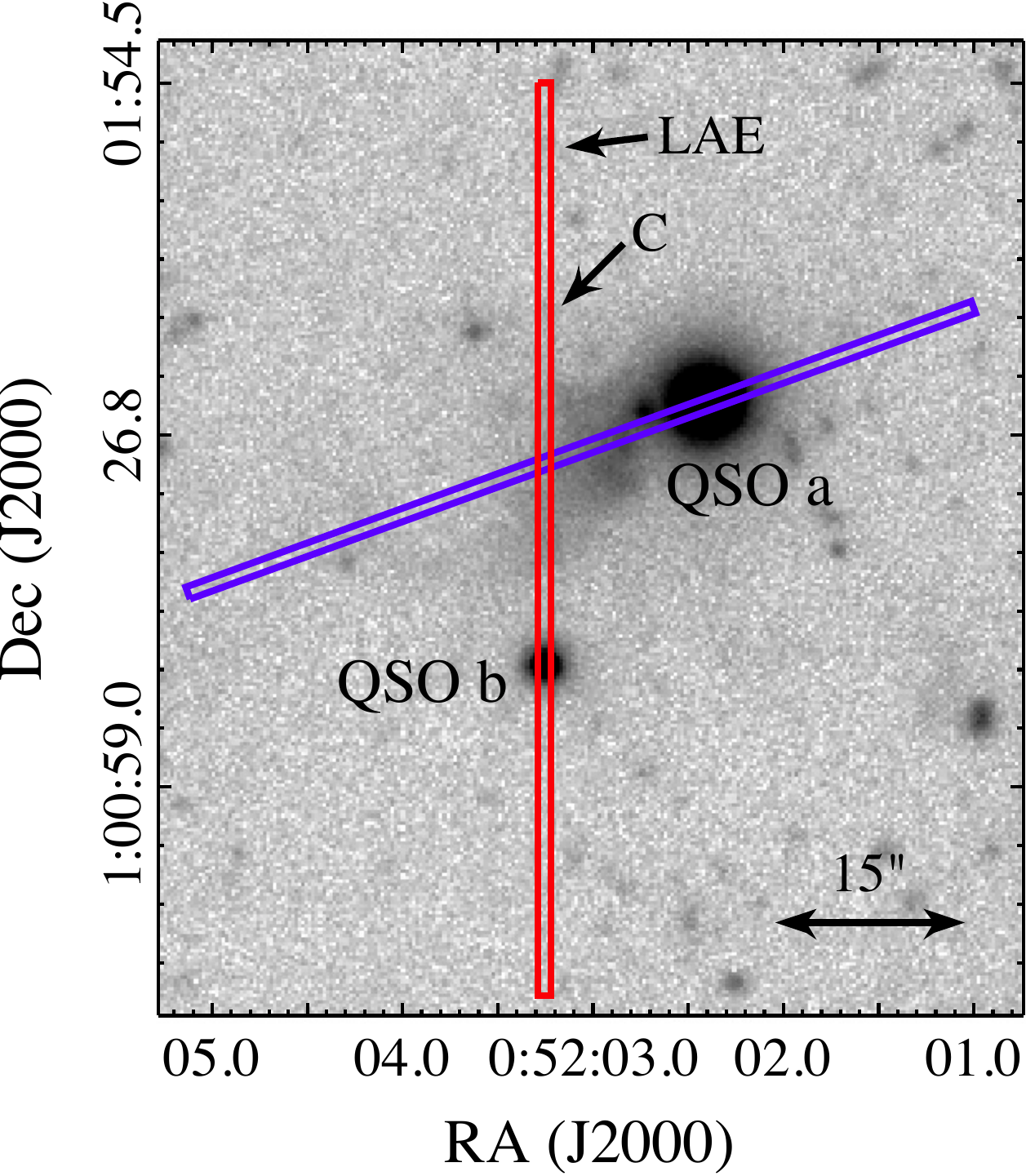, width=0.85\columnwidth, clip} 
\caption{10-hours narrow-band image matching the \lya line at the redshift of UM287 (adapted from Figure 1 of \citealt{Cantalupo2014}). 
`QSO a' is the quasar UM287, while `QSO b' is the faint companion quasar . The red and blue lines highlight the position of the 1\arcsec slits 
chosen to study the extended emission in this work. Note that a Lyman Alpha Emitter (`LAE') and a continuum source 
(`C') fall within the `red' slit (see Figure \ref{Fig2}).}
\label{Fig1}
\end{figure}

Figure \ref{Fig1} shows the position of the two 1\arcsec-slits (red
and blue) on top of the narrow-band image 
(matching the \lya line at the redshift of UM287) presented by \citet{Cantalupo2014}.
We remind the reader that \citet{Cantalupo2014} found
a optically faint ($V=21.54\pm0.06$) radio-loud quasar (`QSO b')
at the same redshift, and at a projected distant of 24.3 arcsec
($\sim200$ kpc) from the bright UM287 quasar (`QSO a').  The
first slit orientations was chosen to simultaneously cover the
extended \lya emission and the UM287 quasar (blue slit), whereas the
second (red slit) was chosen to cover the companion quasar `b'
together with the diffuse nebula.  By covering one of the quasars with each
slit orientation we are thus able to cleanly subtract
the PSF of the quasars from our data (see Section
\S\ref{sec:results}).  

The 2-d spectroscopic data reduction is performed exactly as described
in \citet{Hennawi2013} and we refer the reader to that work for additional
details. In what follows, we briefly summarize the key elements of the data
reduction  procedure.  All data were reduced using the LowRedux
pipeline\footnote{http://www.ucolick.org/$\sim$xavier/LowRedux}, which
is a publicly available collection of custom codes written in the
Interactive Data Language (IDL) for reducing slit
spectroscopy. Individual exposures are processed using standard
techniques, namely they are overscan and bias subtracted and flat
fielded. 
Cosmic rays and bad
pixels are identified and masked in multiple steps. 
Wavelength solutions are determined from low order
polynomial fits to arc lamp spectra, and then a wavelength map is
obtained by tracing the spatial trajectory of arc lines across each
slit. 

We then perform the sky and PSF subtraction as a coupled problem,
using a novel custom algorithm that we briefly summarize here (see
\citealt{Hennawi2013} for additional details). We adopt an iterative
procedure, which allows us to obtain the sky background, the 2-d spectrum of
each object, and the noise, as follows.  First, we identify objects in
an initial sky-subtracted image\footnote{By
    construction, the sky-background has a flat spatial profile
    because our slits are flattened by the slit illumination
    function.}, and trace their trajectory across the detector. We
then extract a 1-d spectrum, normalize these sky-subtracted images by
the total extracted flux, and fit a B-spline profile to the normalized
spatial light profile of each object relative to the position of its
trace.  Given this set of 2-d basis functions, i.e. the flat sky and
the object model profiles, we then minimize chi-squared for the best
set of spectral B-spline coefficients which are the spectral
amplitudes of each basis component of the 2-d model.  The result of
this procedure are then full 2-d models of the sky-background, all
object spectra, and the noise ($\sigma^2$).  We then use this model
sky to update the sky-subtraction, the individual object profiles are
re-fit and the basis functions updated, and chi-square fitting is
repeated. We iterate this procedure of object profile fitting and
subsequent chi-squared modeling four times until we arrived at our
final models.

For each slit, each exposure is modeled according to
the above procedure, allowing us to subtract both the sky and the PSF
of the quasars. These images are registered to a common frame by
applying integer pixel shifts (to avoid correlating errors), and are
then combined to form final 2-d stacked sky-subtracted and
sky-and-PSF-subtracted images. The individual 2-d frames are optimally
weighted by the ${\rm (S\slash N)}^2$ of their extracted 1-d
spectra. The final result of our data analysis are three images: 1) an
optimally weighted average sky-subtracted image, 2) an optimally
weighted average sky-and-PSF-subtracted image, and 3) the noise model
for these images $\sigma^2$.  The final noise map is propagated from
the individual noise model images taking into account weighting and
pixel masking entirely self-consistently.

Finally, we flux calibrate our data following the procedure in \citet{Hennawi2013}.
As standard star spectra were not typically taken immediately before/after our 
observations, we apply an archived sensitivity function for the LRIS B600/4000 grism
to the 1-d extracted quasar spectrum for each slit, and then integrate
the flux-calibrated spectrum against the SDSS $g$-band filter
curve. The sensitivity function is then rescaled to yield the correct
SDSS $g$-band photometry. Since the faint quasar is not clearly detected in SDSS, 
we only used the $g$-band magnitude of the UM287 quasar to calculate this correction. 
Note that this procedure is effective for
point source flux-calibration because it allows us to account for the
typical slit-losses that affect a point source.  However, this
procedure will tend to underestimate our sensitivity to extended
emission, which is not affected by these slit-losses. Hence, our
procedure is to apply the rescaled sensitivity functions (based on
point source photometry) to our 2-d images, but reduce them by a
geometric slit-loss factor so that we properly treat extended
emission. To compute the slit-losses we use the measured spatial FWHM
to determine the fraction of light going through our $1.0\arcsec$
slits, but we do not model centering errors (see Section
\S\ref{sec:results} for a test of our calibration, and see
\citealt{Hennawi2013} for more details).

Given this flux calibration, the 1$\sigma$ SB limit of our
observations are $SB_{1\sigma} = 1.3\times10^{-18}$ \unitcgssb, and $SB_{1\sigma} = 1.5\times10^{-18}$ \unitcgssb for
\civ and \heii, respectively.  This
limits are obtained by averaging over a 3000 km s$^{-1}$ velocity
interval, i.e. $\pm 1500$ km s$^{-1}$ on either side of the systemic
redshift of the UM287 quasar, i.e. $z=2.279\pm0.001$ (\citealt{McIntosh1999}), at the \civ and \heii locations,
and a 1\arcsec\,$\times$\,1\arcsec\ aperture\footnote{Obviously, if we use a smaller
velocity aperture we get a more sensitive limit, i.e. $SB_{\rm limit}=SB_{1\sigma} \sqrt{\frac{\Delta v_{\rm new}}{3000 \ {\rm km
      \ s^{-1}}}}$, e.g. we obtain $SB_{1\sigma} = 7.3\times10^{-19}$ \unitcgssb for a 700 km s$^{-1}$ velocity interval.}.
This limits (approximately independent of wavelength) are about $3\times$ the 
  $1\sigma$ limit in 1 arcsec$^2$ quoted by \citet{Cantalupo2014} for their $\sim10$ hours narrow-band 
  exposure targeting the \lya line, i.e. $5\times10^{-19}$ \unitcgssb\,
\footnote{Note that spatial averaging allow us to achieve more sensitive limits. 
If we consider an aperture of 1\arcsec$\times$20\arcsec, we reach $SB_{1\sigma}^{A=20} = 3.7\times10^{-19}$ \unitcgssb at the location 
of the \lya line.}.
Note that we choose this velocity range to enclose all the extended \lya emission, 
even after smoothing (see next Section \S\ref{sec:results}), and because the narrow-band image of 
\citet{Cantalupo2014} covers approximately this width, i.e. $\Delta v\sim2400$ km s$^{-1}$.

Further, it is important to stress here that the line ratios 
we use in this work are only from the spectroscopic data 
(we do \textit{not} use the NB data for the \lya line), and hence they are independent of 
any errors in the absolute calibration. Although we do not use the 
NB data in our analysis, we show in the next section that our results are
consistent with the NB imaging, and thus robustly calibrated.

\section{Observational Results}
\label{sec:results}

Following \citet{Hennawi2013}, we search for extended \lya, \civ, and \heii emission 
by constructing a $\chi$ image  
\begin{equation} 
\chi^2 = \sum_{i}^{N_{\rm pix}} \frac{\left({\rm DATA}_i - {\rm
    MODEL}_i\right)^2}{\sigma_i^2}\label{chi} 
\end{equation} 
where the sum is taken over all $N_{\rm pix}$ pixels in the image, `DATA' is the
image, `MODEL' is a linear combination of 2-d basis functions multiplied by
B-spline spectral amplitudes, and $\sigma$ is a model of the noise in
the spectrum, i.e. $\sigma^2 = {\rm SKY} + {\rm OBJECTS} + {\rm READ NOISE}$. 
The `MODEL' and the $\sigma^2$ are obtained during our data reduction procedure (see Section \S\ref{sec:data}, 
and \citealt{Hennawi2013} for details).
  
\begin{figure*}
\centering
\epsfig{file=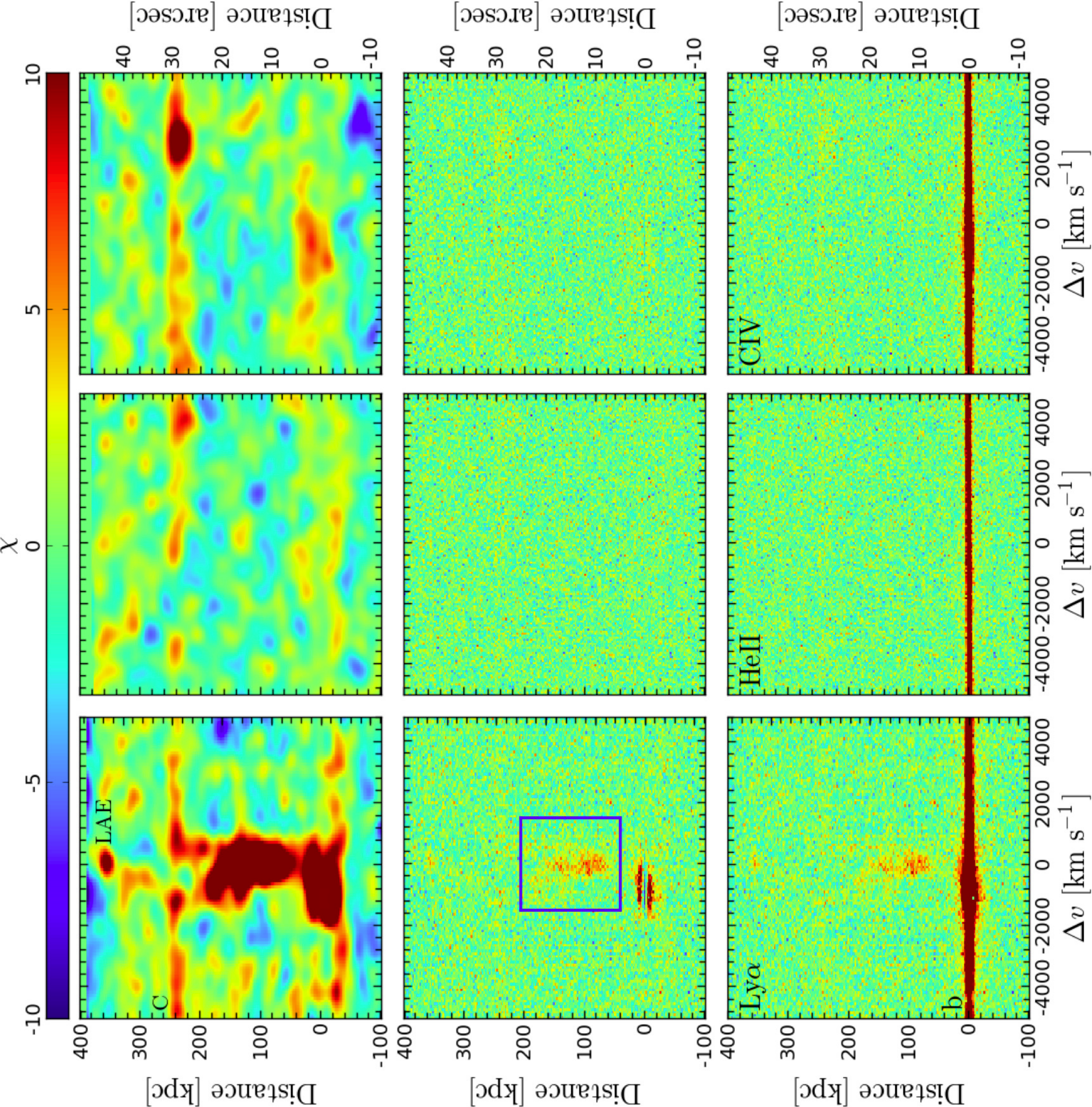, width=0.9\textwidth, clip,angle=270} 
\caption{Two-dimensional spectra for the red slit shown in Figure \ref{Fig1}, plotted as $\chi$-maps following \citet{Hennawi2013}. 
In all panels, $v=0$ km~s$^{-1}$ indicates the systemic redshift of the UM287 quasar, while the distance is computed from the companion 
quasar, i.e. `QSO b'.
{\bf Bottom row:} $\chi_{\rm sky}$ (sky-subtracted only) at the 
location of \lya, \heii, and \civ. {\bf Middle row:} $\chi_{\rm sky+PSF}$ (sky and PSF subtracted) at the 
location of \lya, \heii, and \civ. {\bf Upper row:} smoothed maps $\chi_{\rm smth}$ after the PSF subtraction of the companion 
QSO (`QSO b' in Figure \ref{Fig1}). As expected, the extended \lya emission is well visible in these panels up to 200 kpc from the companion QSO. Note also 
that within this slit we have a continuum source (source `C' in Figure \ref{Fig1}) at $\sim230$ kpc, and a Lyman Alpha emitter 
(`LAE', also highlighted in Figure \ref{Fig1}) at $\sim350$ kpc (see Section \S\ref{sec:results} for details).
The blue box indicates the aperture used to compute the SB$_{\rm Ly\alpha}$, and the limits on \heii/\lya and \civ/\lya 
line ratios, i.e. 1\arcsec$\times$20\arcsec and $\Delta v=3000$ km~s$^{-1}$.} 
\label{Fig2}
\end{figure*}

\begin{figure*}
\centering
\epsfig{file=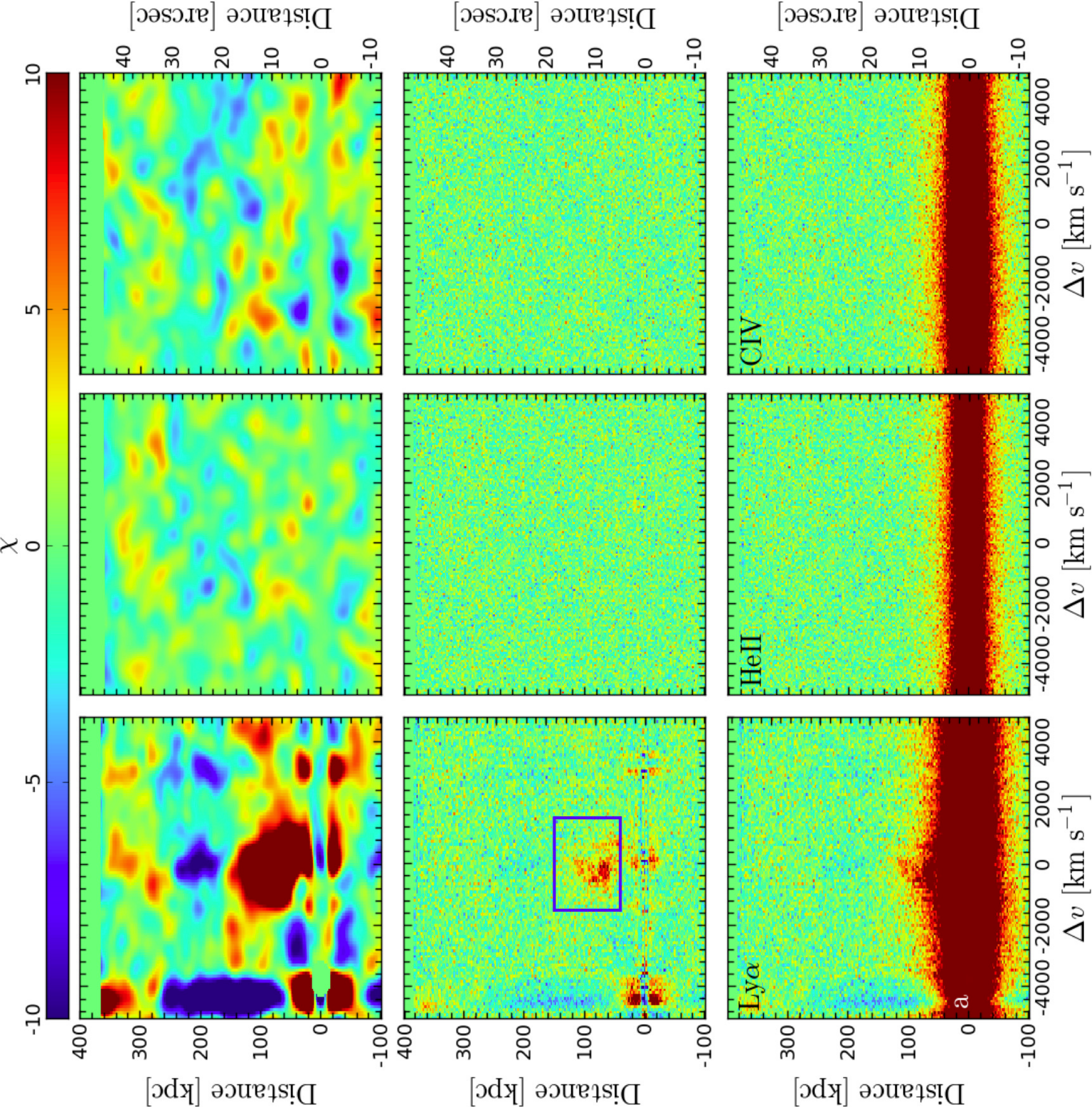, width=0.9\textwidth, clip,angle=270} 
\caption{Two-dimensional spectra for the blue slit shown in Figure \ref{Fig1}, plotted as $\chi$-maps following \citet{Hennawi2013}. 
In all panels, $v=0$ km~s$^{-1}$ indicates the systemic redshift of the UM287 quasar. The distance is also computed from the UM287 
quasar, i.e. `QSO a'.
{\bf Bottom row:} $\chi_{\rm sky}$ (sky-subtracted only) at the 
location of \lya, \heii, and \civ. {\bf Middle row:} $\chi_{\rm sky+PSF}$ (sky and PSF subtracted) at the 
location of \lya, \heii, and \civ. {\bf Upper row:} smoothed maps $\chi_{\rm smth}$ after the PSF subtraction of the UM287 
QSO (`QSO a' in Figure \ref{Fig1}). As expected, also along this slit we detect extended \lya emission. Given our sensitivity limits, the \lya line is detected 
up to $\sim$150 kpc from the UM287 QSO. Note that for such a bright QSO, it is difficult to cleanly subtract its PSF. 
The blue box indicates the aperture used to compute the SB$_{\rm Ly\alpha}$ as outlined in section \S\ref{sec:results}.} 
\label{Fig3}
\end{figure*}

Figures \ref{Fig2} and \ref{Fig3} show the two-dimensional spectra for the slits in Figure \ref{Fig1} plotted as $\chi$-maps.
Note that if our noise model is an accurate description of the data, 
the distribution of pixel values in the $\chi$-maps should be a Gaussian with unit variance. In these images, emission 
will be manifest as residual flux, inconsistent with being Gaussian
distributed noise.  
The bottom row of each figure shows the $\chi_{\rm sky}$ map (only sky subtracted) 
at the location of the \lya, \civ, and \heii, respectively.
Even in these unsmoothed data the extended \lya emission is clearly visible  
up to $\sim200$ kpc ($\sim24$\arcsec) from `QSO b', along the `red' slit (Figure \ref{Fig2}). 
This emission has SB$_{\rm Ly\alpha}= (6.3 \pm 0.4)\times10^{-18}$ \unitcgssb, 
calculated in a 1\arcsec$\times$20\arcsec aperture\footnote{Note 
that one spatial dimension is set by the width of the slit, i.e. 1\arcsec.} 
and over a 3000 km s$^{-1}$ velocity interval (blue box in Figure 2). 
This value is in agreement with the emission detected in the continuum-subtracted image presented in
\citet{Cantalupo2014} within a 1\arcsec$\times$20\arcsec aperture at the same position within the slit, 
i.e. SB$_{\rm Ly\alpha}=(7.0 \pm 0.1)\times10^{-18}$ \unitcgssb.
Along the 'blue' slit, the extended emission is inevitably mixed with the PSF of the hyper-luminous 
UM287 QSO, making PSF subtraction much more challenging.  Nevertheless, we compute the emission in the extended \lya line in an aperture 
of about 1\arcsec$\times$13\arcsec aperture (from 40 to 150 kpc) and within 3000 km~s$^{-1}$, 
after subtracting the PSF of the quasar (see Figure \ref{Fig3}).
Again, we find that surface brightness measured from spectroscopy
(SB$_{\rm Ly\alpha}=1.4\times10^{-17}$ \unitcgssb), and from
narrow-band imaging (SB$_{\rm Ly\alpha}=1.7\times10^{-17}$ \unitcgssb)
agree within the uncertainties\footnote{We do not
    quote errors for these second set of measurements because there
    are significant systematics associated with the PSF subtraction in
    both imaging and spectroscopic data.}.  The agreement between
the Ly$\alpha$ spectroscopic and narrow band imaging surface brightnesses for
both slit orientations confirm that our spectroscopic calibration procedure is
robust.

We do {\it not} detect any
extended emission in either the \civ or in the \heii line, for either
of the slit orientations.  To better visualize the presence of
extended emission, we first subtract the PSF of the QSOs for each
position angle (see middle rows in Figures \ref{Fig2}, and
\ref{Fig3}), and finally, we show in the upper rows the smoothed
$\chi_{\rm smth}$ maps. These smoothed maps are of great assistance
in identifying faint extended emission (see \citealt{Hennawi2013} for
more details on the PSF subtraction and the calculation of the
smoothed $\chi$-maps). The lack of compelling emission features in the
PSF-subtracted smoothed maps confirm the absence of extended \civ and \heii at
our sensitivity limits in both slit orientations.

As our goal is to measure line ratios between the \lya emission and the \civ and \heii lines, 
we compute the surface brightness limits within the same aperture in which we calculated the \lya emission along the 'red' slit, 
i.e. 1\arcsec$\times$20\arcsec and $\Delta v=3000$ km~s$^{-1}$.
Because the companion quasar is much fainter than the UM287 quasar, the 
PSF subtraction along the `red' slit does not suffer from systematics, whereas
the large residuals in the left panel of Figure \ref{Fig3} indicate that there are significant PSF subtraction
systematics for the Ly$\alpha$ emission in the `blue' slit covering the UM287 quasar.
We have thus decided to focus on the line ratios  
obtained from the `red' slit,
although the constraints we obtain from the `blue' slit are comparable. 
We find SB$_{1\sigma, CIV}^{A=20}= 3.3\times10^{-19}$ \unitcgssb and SB$_{1\sigma, HeII}^{A=20}= 3.7\times10^{-19}$ \unitcgssb, respectively at the CIV and HeII locations.

\begin{figure}
\centering
\epsfig{file=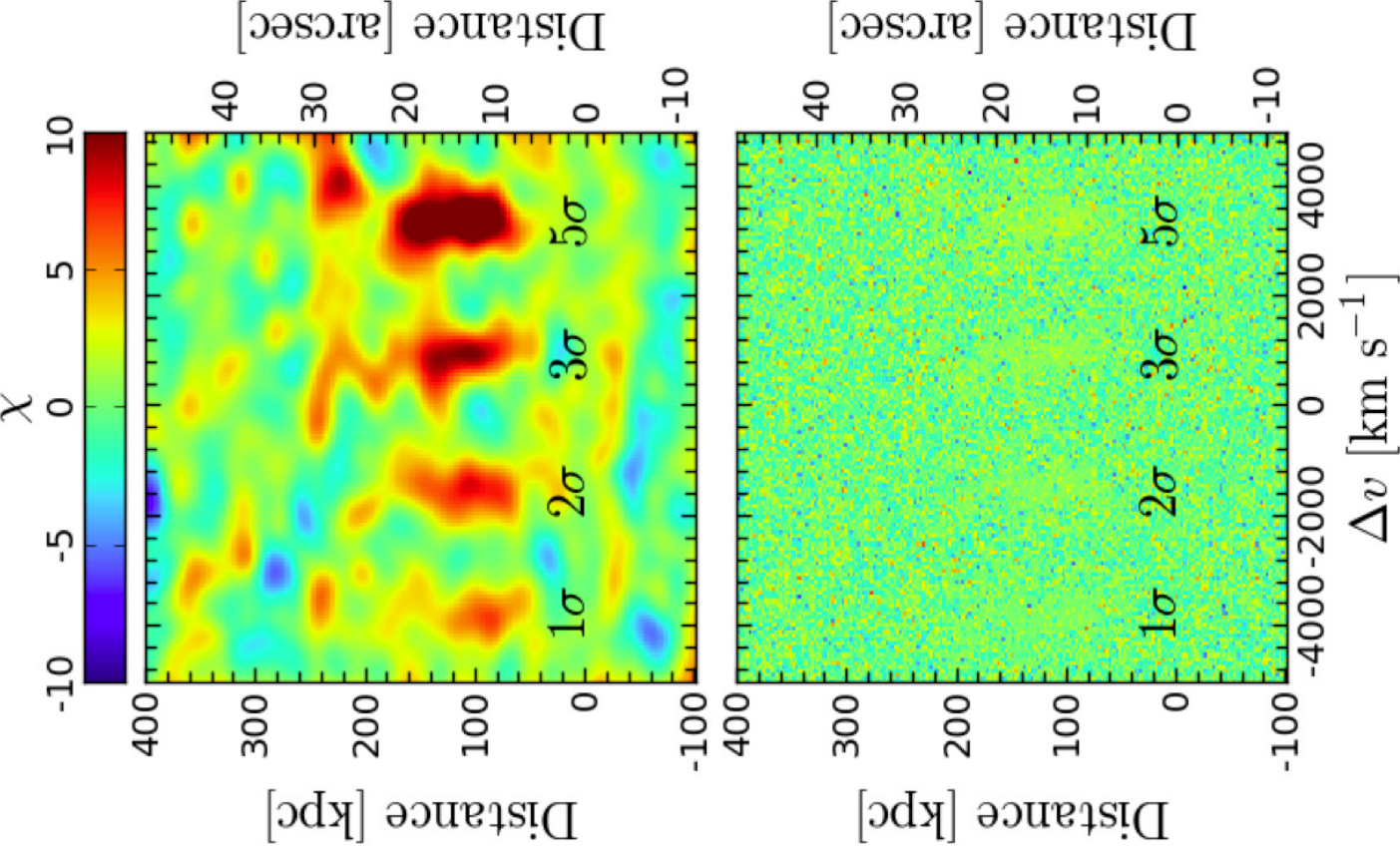, width=1.5\columnwidth, angle=270, clip} 
\caption{Illustration of detection significance of scaled models of the \lya emission at the location of the HeII line along 
the `red' slit (Figure \ref{Fig2}).
The synthetic sources corresponds to 1, 2, 3, and 5$\times$ SB$_{1\sigma, HeII}^{A=20}$. The bottom panel shows the 
$\chi_{\rm sky}$ (sky-subtracted only) map, while the upper panel shows the smoothed map. This figure suggests that we should be able to
clearly detect extended emission $\gtrsim 3\sigma$ on the scale of the \lya line. } 
\label{Fig4}
\end{figure}

To better understand how well we can recover emission in the \heii and
\civ lines in comparison to the \lya, we visually estimate the
detectability of extended emission in these lines by inserting fake
sources as follows.  First, we select the \lya emission above its
local $1\sigma$ limit along the `red' slit, we smooth it and scale it
to be 1, 2, 3, and 5$\times$ SB$_{1\sigma}^{A=20}$ at the location of
the HeII and CIV line. Finally, we add Poisson realizations of these
scaled models into our 2-d PSF and sky-subtracted images.  In Figure
\ref{Fig4} we show the $\chi$-maps for this test at the location of
HeII.  This test suggests that we should be able to
clearly detect extended emission on the same scale as the \lya line if
the source is $\gtrsim 3 \times$ SB$_{1\sigma}^{A=20}$. Thus, in the
remainder of the paper we use 3$\sigma$
($\sigma\equiv$SB$_{1\sigma}^{A=20}$) upper limits on the \heii/\lya
and \civ/\lya ratios. Given the values for the SB$_{\rm Ly\alpha}$ and
the surface brightness limits at the location of the CIV and HeII
lines (within the 1\arcsec$\times$20\arcsec aperture and 3000
km~s$^{-1}$ velocity window for the red slit) we get
(\heii/\lya)$_{3\sigma}\lesssim0.18$ and
(\civ/\lya)$_{3\sigma}\lesssim0.16$. Note that given the brighter 
\lya emission at the location of the `blue' slit, the limits implied 
are about 2$\times$ lower than these quoted limits for the `red' slit.

It is important to note that we detect extended \civ emission around
the faint companion quasar `b' (see the smoothed maps in Figure
\ref{Fig2}). As this line is physically distinct from the UM287 nebula
and essentially follows the extended \lya emission around the faint
quasar (compare the smoothed maps for \lya and \civ), this suggests
that we have detected the extended narrow emission line region (EELR) of this
source. This kind of emission, produced by the gas excited by an AGN on scales of tens of kpc, 
is usually observed around low redshift $z<0.5$ type-I (e.g. \citealt{Stockton2006, Husemann2013}) and type-II quasars 
(e.g. \citealt{Greene2011}), traced by [\oiii] and Balmer lines.
We do not quote a value for the emission
because, given the much smaller scales in play here, its accuracy depends
on the PSF-subtraction. However note that this detection, near the
limit of our sensitivity, clearly demonstrates that we
could have detected faint extended emission in the
\civ and \heii lines within the \lya nebula itself if this emission were
characterized by higher line ratios.

Finally, we briefly comment on the nature of two other sources which fall within 
the `red' slit, i.e. a Lyman Alpha emitter (LAE) (i.e. $EW_{{\rm Ly}\alpha}^{rest}>20$\AA)
and a continuum source (see Figure \ref{Fig2}).
Indeed, this slit orientation was also chosen to confirm the presence of 
a LAE at about 350 kpc northward of `QSO b', clearly visible in the narrow-band image in
Figure 2 of \citet{Cantalupo2014} and in our Figure~\ref{Fig1}.
Our LRIS data confirm the presence of a line emission from a LAE at a redshift $z=2.280\pm0.002$,
which is consistent with the redshift of the UM287 quasar, within our 
uncertainties.
We ascribe this emission to the \lya line, and we compute a flux of 
$F_{Ly\alpha}=(9.2\pm0.9)\times10^{-18}$ erg~s$^{-1}$~cm$^{-2}$ (in an aperture of $\Delta v=1400$ km~s$^{-1}$, and 4 arcsec$^2$), 
in agreement with $F_{Ly\alpha}=(8.4\pm0.4)\times10^{-18}$ erg~s$^{-1}$~cm$^{-2}$ 
computed in an aperture of 4 arcsec$^2$ in the map of \citet{Cantalupo2014}.
We also serendipitously obtained a spectrum of a source at $\sim230$ kpc from quasar `b',
which coincides with a continuum sources in our deep V-band image (\citealt{Cantalupo2014}).
In our 2-d spectrum, we detect a faint continuum associated with this source and an emission
line at a wavelength of 5123\AA, which appears at a velocity $\sim2750$ km~s$^{-1}$
from the \civ line in the right panel of Figure~\ref{Fig2}. However, given the low signal-to-noise ratio
of the continuum, and the detection of a single emission line, we are unable to determine
the redshift of this source.

\subsection{Kinematics of the Nebula}

With these slit spectra for two orientations, we can begin to study the
kinematics of the Ly$\alpha$ emission of the UM287 giant nebula. 
We first focus on the `red' slit (see Figure \ref{Fig2}), which covers the companion 
quasar (`QSO b') and the extended \lya emission
at a projected distance of $100$$-160$ kpc ($\sim13$\arcsec$-19$\arcsec) from UM287 (`QSO a').
We tested the kinematics of the detected emission by measuring 
the flux-weighted line centroid and the flux-weighted velocity dispersion ($\sigma$) around
the centroid velocity in 2-pixels wide bins ($\sim0.54$\arcsec) across the 
spatial slit direction. We then converted the velocity dispersion to
a gaussian-equivalent FWHM$_{\rm gauss}$ assuming FWHM$_{\rm gauss}\sim2.35\sigma$.
Note that, because of the resonant nature of the \lya emission, the line width
may be broadened by radiative transfer effects (e.g., \citealt{Cantalupo2005}) and 
representing, thus, only an upper limit for the thermal or kinematical broadening.
The extended emission has an average FWHM$_{\rm gauss}\sim500$~km~s$^{-1}$ at a redshift of $z=2.279$, 
which is centered on the systemic redshift of the UM287 quasar. Although the emission appears 
coherent on this large scales, the gaussian FWHM calculated at each location ranges between $\sim370$~km~s$^{-1}$ and $\sim600$~km~s$^{-1}$, 
suggesting the need of higher resolution data to better characterize its width and shape. 
The line emission is red-shifted by $\sim750$ km/s from quasar `b'.  However note that our estimate
for the redshift of quasar `b' $z=2.275$ has a large 800 km~s$^{-1}$ error, 
because it is estimated from broad rest-frame UV emission lines which are poor tracers of the systemic frame (\citealt{Cantalupo2014}).

As for the `blue' slit, statements about kinematics are limited by the				         
challenge of accurately subtracting the PSF of the bright UM287 				         
quasar.  Given our SB limit, we detect the \lya emission out to 				         
$\sim150$ kpc. As expected from the narrow-band imaging, the \lya is				         
stronger at this location in comparison with the other slit					         
orientation.  In particular, the emission shows a peak at $\sim$$63$				         
kpc ($\sim7.7$\arcsec) in agreement with the narrow-band data (see				         
Figure \ref{Fig1} or Figure 2 in \citealt{Cantalupo2014}).  At this				         
second location, the \lya line appears broader ${\rm FWHM}_{\rm gauss}\sim 920$~km~s$^{-1}$ and appears  
to vary more with distance along the slit. This larger width may arise from the 		         
fact we are probing smaller distances from the UM287 quasar than in the `red' slit.

Note that, at our spectral resolution (${{\rm FWHM}\sim320}$ km~s$^{-1}$), there is no evidence for ``double-peaked'' kinematics characteristic of resonantly-trapped \lya 
(e.g. \citealt{Cantalupo2005}) along either slit. This may indicate 
that resonant scattering of Ly$\alpha$ photons does not play an important role in the Ly$\alpha$
kinematics,  however, data at a higher resolution are needed to confirm this conclusion.

These estimates for the widths of \lya\ emission are
comparable to the velocity widths observed in absorption 
in the CGM surrounding $z \sim 2$ quasars 
(${\Delta v \approx 500}$~km~s$^{-1}$; \citealt{Prochaska2009, QPQ8}),
pheraps suggesting that the kinematics traced in emission are
dominated by the motions of the gas as opposed to the
effects of radiative transfer. 
Both the emission and absorption kinematics are comparable to 
the virial velocity $\sim300$~km~s$^{-1}$ of the massive dark matter halos 
hosting quasars (M$_{\rm DM}\sim10^{12.5}$~M$_{\odot}$, \citealt{White2012}), and thus appear consistent with 
gravitational motions.

\section{Modeling the Ly$\alpha$, CIV and HeII Emission around UM287}
\label{sec:model}

As shown by \citet{Cantalupo2014}, the extended \lya emission nebula 
around UM287 can be explained by photoionization from 
the central quasar, implying a large amount of
cool ($T\sim10^4$ K) 
gas, i.e. ${M_{c} \simeq 10^{12}\,C^{-1/2}}$~M$_{\odot}$. 
To further constrain the properties of the gas in this huge nebula,
in this section we exploit the simple model for cool clouds in a quasar halo
introduced by \citet{Hennawi2013} and the consequent photoionization
modeling procedure introduced by \citet{FAB2014}.  Our main goal is to
show how our line ratio constraints on \civ/\lya and \heii/\lya can be
used to constrain the physical properties of the gas in the UM287
nebula, such as the volume density ($n_{\rm H}$), column density
($N_{\rm H}$), and gas metallicity ($Z$).

We reiterate that as in our previous work (\citealt{Cantalupo2014}), model the
\lya emission alone cannot break the degeneracy between the 
clumpiness or density of the gas, and the total gas mass.
In the next sections we show how information on additional lines (in particular \heii) can
constrain the density of the emitting gas and thus break this degeneracy.

\subsection{Photoionization Modeling}
\label{theory}

In the following, we briefly outline the simple model for cool halo gas
introduced by \citet{Hennawi2013} for the case of UM287.
We assume a simple picture where UM287
has a spherical halo populated with spherical clouds of cool
gas ($T \sim 10^4$\,K) at a single uniform hydrogen volume density
$n_{\rm H}$, and uniformly distributed throughout the halo. 
We model a scale length of $R=160$ kpc from the central quasar, which 
approximately corresponds to the distance probed by the `red' slit, and represents
the expected virial radius for a dark matter halo hosting 
a quasar at this redshift.
In this configuration, the spatial distribution of the gas is completely specified
by $n_{\rm H}$, $R$, the hydrogen column density $N_{\rm H}$, and the
cloud covering factor $f_C$.

Note that the total mass of cool gas in our simple model can be written as
(\citealt{Hennawi2013}):

\bea
\label{mass}
M_{\rm c} &=& \pi R^2 f_{\rm C} N_{\rm H} \frac{m_p}{X}\\
&=& 2.7\times10^{10}\!\left(\frac{R}{160\,{\rm kpc}}\right)^2 
\!\!\left(\frac{N_{\rm H}}{10^{19.5}\,{\rm cm}^{-2}}\right)\!\!
\left(\frac{f_{\rm C}}{1.0}\right)\,M_{\odot} \nonumber
\eea
where $m_p$ is the mass of the proton and $X$ is the hydrogen mass fraction.

In this simple model, the \lya SB is determined by simple relations
which depend only on $n_{\rm H}$, $N_{\rm H}$, $f_C$, and the
luminosity of the QSO at the Lyman limit ($L_{\nu_{\rm LL}}$)(see
\citealt{Hennawi2013} for details). To build intuition, it is useful
to consider two limiting regimes for the recombination emission, for
which the clouds are optically thin ($N_{\rm HI} \ll 10^{17.2}$
cm$^{-2}$) and optically thick ($N_{\rm HI} \gg 10^{17.2}$ cm$^{-2}$)
to the Lyman continuum photons, where $N_{\rm HI}$ is the neutral
column density of a single spherical cloud.  We argue below, that
given the luminosity of the UM287 quasar, the optically thick case is
however unrealistic.

\begin{itemize}

\item[-] Optically thin to the ionizing radiation:
\begin{equation}
\label{SBthin}
{\rm SB}_{{\rm Ly}\alpha}^{\rm thin} = \frac{\eta_{\rm thin}h\nu_{{\rm Ly}\alpha}}{4\pi(1+z)^4}~\alpha_{\rm A}\left(1 + \frac{Y}{2X} \right) n_{\rm H}f_{\rm C} N_{\rm H}, 
\end{equation}
where $\eta_{\rm thin}=0.42$ is the fraction of recombinations 
which result in a Ly$\alpha$ photon 
in the optically thin limit (\citealt{OF2006}), $h$ is the Planck constant, 
$\nu_{{\rm Ly}\alpha}$ 
is the frequency of the \lya line, $\alpha_{\rm A}=4.18 \times 10^{-13}\,{\rm cm^{-3}\,s^{-1}}$ 
is the case A recombination coefficient 
at $T=10,000$\,K (\citealt{OF2006})\footnote{Note that this equation hides a 
dependence on temperature through the recombination coefficient $\alpha_{\rm A}$, which 
usually is neglected, but that can be important, i.e $\alpha_{\rm A}$ is decreasing by a factor 
of $\sim$6 from $T=10^4$\,K to $T=10^5$\,K (CHIANTI database, \citealt{Dere1997,Landi2013}). 
}, and $X=0.76$ and $Y=0.24$ are the respective hydrogen and helium 
mass fractions implied by Big Bang Nucleosynthesis (\citealt{Boesgaard1985,Izotov1999,Iocco2009,PlanckColl2014}).

\item[-] Optically thick to the ionizing radiation 
\begin{equation}
\label{SBthick}
{\rm SB}_{{\rm Ly}\alpha}^{\rm thick} = \frac{\eta_{\rm thick}h\nu_{{\rm Ly}\alpha}}{4\pi(1+z)^4}f_{\rm C}\Phi_{\rm LL}\left(R/\sqrt{3}\right),
\end{equation}
where $\eta_{\rm thick}=0.66$ is the fraction of ionizing photons 
converted into Ly$\alpha$ photons, and where $\Phi_{\rm LL}$ $({\rm [phot \ s^{-1} \ cm^{-2}}])$ is 
the ionizing photon number flux, 
\begin{equation}
\label{phi}
\Phi_{\rm LL}= \int_{\nu_{\rm LL}}^{\infty}\frac{F_{\nu}}{h\nu} d\nu = \frac{1}{4\pi r^2}\int_{\nu_{\rm LL}}^{\infty} \frac{L_{\nu}}{h\nu} d\nu.
\end{equation}
\end{itemize}
Thus, in the optically thick case the Ly$\alpha$ surface brightness scales with the 
luminosity of the central source, $SB_{{\rm Ly}\alpha}^{\rm thick} \propto f_C L_{\nu_{\rm LL}}$, while in the optically 
thin regime the SB does not depend on $L_{\nu_{\rm LL}}$,  $SB_{{\rm Ly}\alpha}^{\rm thin} \propto f_C n_{\rm H} N_{\rm H}$, 
provided the AGN is bright enough to keep the gas in the halo ionized enough
to be optically thin.

We now argue that the \lya emitting gas is unlikely to be optically
thick $N_{\rm HI} \gtrsim 10^{17.2}$~cm$^{-2}$.  Equations \ref{SBthick} and \ref{phi} can be combined to express the
SB in terms of $L_{\nu_{\rm LL}}$, the luminosity at the Lyman edge.
To compute this luminosity, we assume that the quasar spectral energy
distribution obeys the power-law form $L_{\nu} = L_{\nu_{\rm LL}}
(\nu\slash \nu_{\rm LL})^{\alpha_{\rm UV}}$, blueward of $\nu_{\rm
  LL}$ and adopt a slope of $\alpha_{\rm UV}= -1.7$ consistent with
the measurements of \citet{Lusso2015}. The quasar ionizing luminosity
is then parameterized by $L_{\nu_{\rm LL}}$, the specific luminosity
at the Lyman edge\footnote{We describe in detail the assumed
    quasar spectral-energy distribution (SED) in Section
    \S\ref{SED}.}. We determine the normalization $L_{\nu_{\rm LL}}$
by integrating the \cite{Lusso2015} composite spectrum against the
SDSS filter curve, and choosing the amplitude to give the correct
$i$-band magnitude of the UM287 quasar ($i$-mag$=17.28$), which gives
a value of $L_{\nu_{\rm LL}} = 5.4\times10^{31}$ erg s$^{-1}$
Hz$^{-1}$.

Substituting this value of $L_{\nu_{\rm LL}}$ for UM287 into
equation \ref{SBthick}, 
we thus obtain
\bea
\label{SBthickValues}
{\rm SB}_{{\rm Ly}\alpha}^{\rm thick}&=& 8.8\times 10^{-16}\left(\frac{1+z}{3.279}\right)^{-4}\!\!
\left(\frac{f_{\rm C}}{1.0}\right)\!\!
\left(\frac{R}{160\,{\rm kpc}}\right)^{-2}\\ 
&\times&\left(\frac{L_{\nu_{\rm LL}}}{10^{31.73}\,{\rm erg\,s^{-1}\,Hz^{-1}}}\right)\cgssb\nonumber.
\eea
This value is over two order of magnitude larger than 
the observed SB value of the \lya emission at 160 kpc from UM287.
Even if we consider a larger radius, $R=250$ kpc, in order to get the 
observed $SB_{{\rm Ly}\alpha}$ we would need a very low covering factor,
i.e. $f_{\rm C}\sim0.02$. 
Such a small covering factor would be strictly at odds with the observed
smooth morphology of the diffuse nebula as seen in Figure~\ref{Fig1}.
We directly test this assumption as follows. We randomly populate an area comparable to the extent
of the \lya nebula
with point sources such that $f_C=0.1-1.0$, and we
convolve the images with a Gaussian kernel with a FWHM equal to our
seeing value, in order to mimic the effect of seeing in the
observations. We find that the smooth morphology observed cannot be
reproduced by images with $f_C<0.5$, as they appear too clumpy.
Thus, the smooth morphology of the emission in the Ly$\alpha$ nebula
implies a covering factor of $f_{C}\gtrsim 0.5$.

In the following sections we construct photoionization models for a grid of
parameters governing the physical properties of the gas
to estimate the expected \heii and \civ emission. 
Following the discussion here, we shall see that the models which reproduce the observed
Ly$\alpha$ SB will be optically thin, because given the high covering factor
optically thick models would be too bright.

\subsection{The Impact of Resonant Scattering}

It is important to stress at this point that the \lya photons should
be subject to substantial resonant scattering under most of the
astrophysical conditions, given the large optical depth at line center
(see e.g. \citealt{Gould1996}). Thus, typically, a \lya photon
experiences a large numbers of scattering before escaping the system
in which it is produced. This process thus leads to double-peaked
emission line profiles as \lya photons must diffuse in velocity space
far from the line center to be able to escape the system
(e.g. \citealt{Neufeld1990,Gould1996,Cantalupo2005,Dijkstra2006,Verhamme2006}).  Although our
models are optically thin at the Lyman limit, i.e. to ionizing
photons, for the model parameters required to reproduce the SB of the
emission, they will almost always be optically thick to the \lya
transition (i.e. $N_{\rm HI} \gtrsim 10^{14}$ cm$^{-2}$). Hence one
should be concerned about the resonant scattering of \lya photons
produced by the central quasar itself.  However, radiative transfer
simulations of radiation from the UM287 quasar through a simulated gas
distribution have shown that the scattered \lya line photons from the
quasar do not contribute significantly to the \lya surface brightness
of the nebula on large scales, i.e. $\gtrsim100$ kpc
(\citealt{Cantalupo2014}).  This is because the resonant scattering
process results in very efficient diffusion in velocity space, such that
the vast majority of resonantly scattered photons produced by the quasar
itself escape the system at very small scales $\lesssim 10$ kpc,
and hence do not propagate at larger
distances (e.g. \citealt{Dijkstra2006,Verhamme2006,Cantalupo2005}).  For this reason,
based on the results of the radiative transfer simulations of \citet{Cantalupo2014}, we 
do not model the contribution of resonant scattering of the quasar photons to the \lya
emission. Similar considerations also apply to the resonant \civ line, however we note
that resonant scattering of \civ is expected to be much less efficient, because the much
lower abundance of metals imply the gas in the nebula is much less likely to be optically thick
to \civ.

To avoid a contribution to the 
\lya and \civ emission from scattering of photons from the QSO we have thus masked both lines 
in our assumed input quasar spectrum. 
Note that with this approach we do not neglect the `scattered' Ly$\alpha$ photons
arising from the diffuse continuum produced by the gas itself,
which however turn out to be insignificant
\footnote{Note that this value depends on the broadening of the line due to 
turbulent motions of the clouds. Given current estimates of typical equivalent widths of optically thick absorbers in quasar spectra, 
i.e. $\sim1$\AA\ (\citealt{Prochaska2013b}), in our calculation we consider turbulent motions of 30 km~s$^{-1}$. 
However, note that our results are not sensitive to this parameter.}.

\subsection{Modeling the UM287 Quasar SED}
\label{SED}

We assume that the spectral energy distribution (SED) of UM287 has the
form shown in Figure \ref{Fig5}.  As we do not have complete coverage
of the spectrum of this quasar, we adopt the following assumptions to model the
full SED. Given the ionization energies for the
species of interest to us in this work, i.e.  1~Ryd=13.6 eV for
Hydrogen, 4~Ryd=54.4 eV for \heii, and 64.5 eV for \civ, we have
decided to stick to power-law approximations above 1~Ryd.
However, note that the UV range of the SED is so far not well
constrained (see \citealt{Lusso2015} and reference therein).
In particular, we model the quasar SED using a
composite quasar spectrum which has been corrected for IGM
absorption (\citealt{Lusso2015}). 
This IGM corrected composite is important because it
allows us to relate the $i$-band magnitude of the UM287 quasar to
the specific luminosity at the Lyman limit $L_{\nu_{\rm LL}}$. For
energies greater than one Rydberg, we assume a power law form
$L_{\nu} = L_{\nu_{\rm LL}} (\nu\slash \nu_{\rm LL})^{\alpha_{\rm
    UV}}$ and adopt a slope of $\alpha_{\rm UV}= -1.7$, consistent
with the measurements of \cite{Lusso2015}, while in the Appendix we
test also the cases for $\alpha_{\rm EUV}=-1.1$, and $-2.3$.  We
determine the normalization $L_{\nu_{\rm LL}}$ by integrating the
\citet{Lusso2015} composite spectrum against the SDSS filter curve,
and choosing the amplitude to give the correct $i$-band magnitude of
the UM287 quasar (i.e. $i$=17.28), which gives a value of
$\L_{\nu_{LL}}=5.4\times10^{31}$ erg~s$^{-1}$~Hz$^{-1}$.  We extend
this UV power law to an energy of 30 Rydberg, at which point a
slightly different power law is chosen $\alpha = -1.65$, such that
we obtain the correct value for the specific luminosity at 2 keV
$L_{\nu}({\rm 2\,keV})$ implied by measurements of $\alpha_{\rm
  OX}$, defined to be $L_\nu(2\,{\rm keV})\slash L_\nu(2500\,{\rm
  \AA}) \equiv (\nu_{\rm 2\,keV}\slash \nu_{2500\,{\rm
    \AA}})^{\alpha_{\rm OX}}$. We adopt the value $\alpha_{\rm OX} =
-1.5$ measured by \cite{Strateva2005} for SDSS quasars. An X-ray
slope of $\alpha_{\rm X}=-1$, which is flat in $\nu f_{\nu}$ is
adopted in the interval of 2-100\,keV, and above 100\,keV, we adopt
a hard X-ray slope of $\alpha_{\rm HX} = -2$. For the rest-frame
optical to mid-IR part of the SED, we splice together the composite
spectra of \cite{Lusso2015}, \cite{VandenBerk2001}, and
\cite{Richards2006}.  These assumptions about the SED are
essentially the standard ones used in photoionization modeling of
AGN (e.g. \citealt{Baskin2014}).  Summarizing, given the lack of
information, for energies greater than one Rydberg we parametrized
the SED of the UM287 quasar with a series of power-laws as
\begin{equation}
f_\nu \propto \begin{cases} \nu^{\alpha_{\rm EUV}}, & \text{if } h\nu\geq 1 \,{\rm Ryd} \\ 
\nu^{\alpha}, & \text{if } 30 \, {\rm Ryd}\leq h\nu<2 \, {\rm keV} \\
\nu^{\alpha_{\rm X}}, & \text{if } 2 \, {\rm keV}\leq h\nu<100 \, {\rm keV} \\
\nu^{\alpha_{\rm HX}}, & \text{if } h\nu\geq100 \, {\rm keV}.
\end{cases}
\end{equation}

\subsection{Input Parameters to Cloudy}

Having established our assumptions on the UM287 SED, and on the resonant scattering, 
we now explain how we choose the range of our model parameter grid.
We perform our calculations with the Cloudy
photoionization code (v10.01), last described by \citet{Ferland2013}.
Because the emitting clouds are expected to be much smaller than their distance 
$r\sim R_{\rm vir} = 160\,{\rm kpc}$ from the central ionizing source, we assume 
a standard plane-parallel geometry for the emitting clouds illuminated
by the distant quasar. 
In order to keep the models as simple as possible, and because we are
primarily interested in understanding how photoionization 
together with the observed line ratios can
constrain the physical properties of the gas (i.e. $n_{\rm H}$ and $N_{\rm H}$), 
without resorting to extreme parameter combinations, we proceed as follows. 
We focus on reproducing the SB$_{\rm Ly\alpha}\sim7\times10^{-18}$ \unitcgssb at 160 kpc distance from the
UM287 quasar, which is basically the distance probed by the `red' slit\footnote{Note that we have decided to model a single distance from 
the UM287 quasar. The sensitivity of our results to this simple assumption is discussed in Section \ref{sec:caveats}.}.
In particular, eqn.~(\ref{SBthin}) implies that a certain combination of $N_{\rm H}$ and 
$n_{\rm H}$ are thus required. Further, given the dependence on metallicity ($Z$) of the \civ and
\heii lines, and of the gas temperature which determine the amount of collisional excitation in the \lya line, 
we also consider variations in $Z$. 
Thus, we run a uniform grid of models with this wide range of parameters:

\begin{figure*}
\centering
\epsfig{file=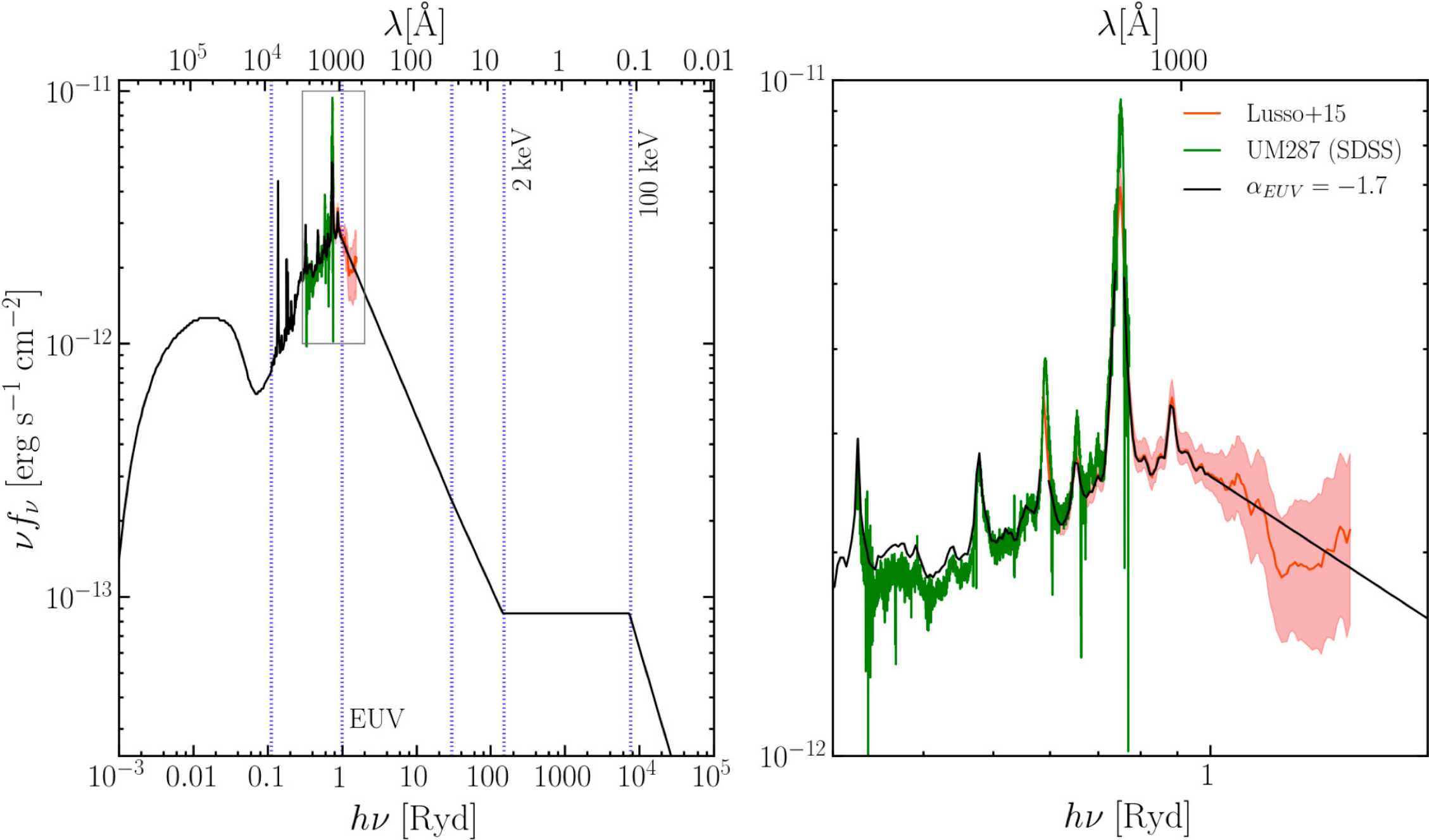, width=0.5\textwidth, angle=270} 
\caption{Spectral energy distribution (SED) of UM287 used as incident radiation field in our modeling. The black solid line 
indicate our fiducial input spectrum characterized by a slope in the EUV of $\alpha_{\rm EUV}=-1.7$ (\citealt{Lusso2015}).
The right panel is a zoomed version of the box highlighted in the left panel.
Note the agreement between the composite spectrum used as input in our calculation and the SDSS spectrum of UM287 (green solid line). 
To prevent a contribution from resonantly scattered photons, we mask the emission from the line center of \lya and \civ.} 
\label{Fig5}
\end{figure*}

\begin{itemize}

\item[--] $n_{\rm H} = 10^{-2}$ to 10$^2$ cm$^{-3}$ (steps of 0.2 dex);
\item[--] $N_{\rm H} = 10^{18}$ to $10^{22}$ (steps of 0.2 dex); 
\item[--] $Z=10^{-3}\, Z_{\odot}$ to $Z_{\odot}$ (steps 0.2 dex).

\end{itemize}

\noindent Note that by exploring this large
parameter range, some of the models that we consider result in clouds optically thick at
the Lyman limit, but as explained in the previous Section \S\ref{theory}, these
parameter combinations result in nebulae which are too bright and thus inconsistent with the
observed \lya surface brightness.
In what follows, we only consider the models which closely reproduce the observed \lya surface brightness, 
i.e. $5.5\times10^{-18}$~\unitcgssb$<$~SB$_{\rm Ly\alpha}<8.5\times10^{-18}$~\unitcgssb.

Photoionization models are self-similar 
in the ionization parameter $U \equiv \frac{\Phi_{LL}}{c n_{\rm H}}$, 
which is the ratio of the number density of ionizing photons to hydrogen atoms.
As the luminosity of the central QSO is known, the variation
in  the ionization parameter $U$ results from the variation of the volume number density
$n_{\rm H}$ for the models in our grid. The range of ionization parameters that we
cover is comparable to those in previous analysis of photoionization
around AGNs, e.g. in the case of the narrow line regions (NLR; e.g. \citealt{Groves2004}) and in 
the case of extended emission line regions (EELR; e.g. \citealt{Humphrey2008}).
Finally, we emphasize that once we fix the 
source luminosity and define the ionizing spectrum, the line ratios
we consider are described 
by two model parameters, namely the density $n_{\rm H}$ of the gas and its metallicity $Z$. We 
will see this explicitly in the next section.

\section{Models vs Observations}
\label{sec:comparison}

\begin{figure}
\epsfig{file=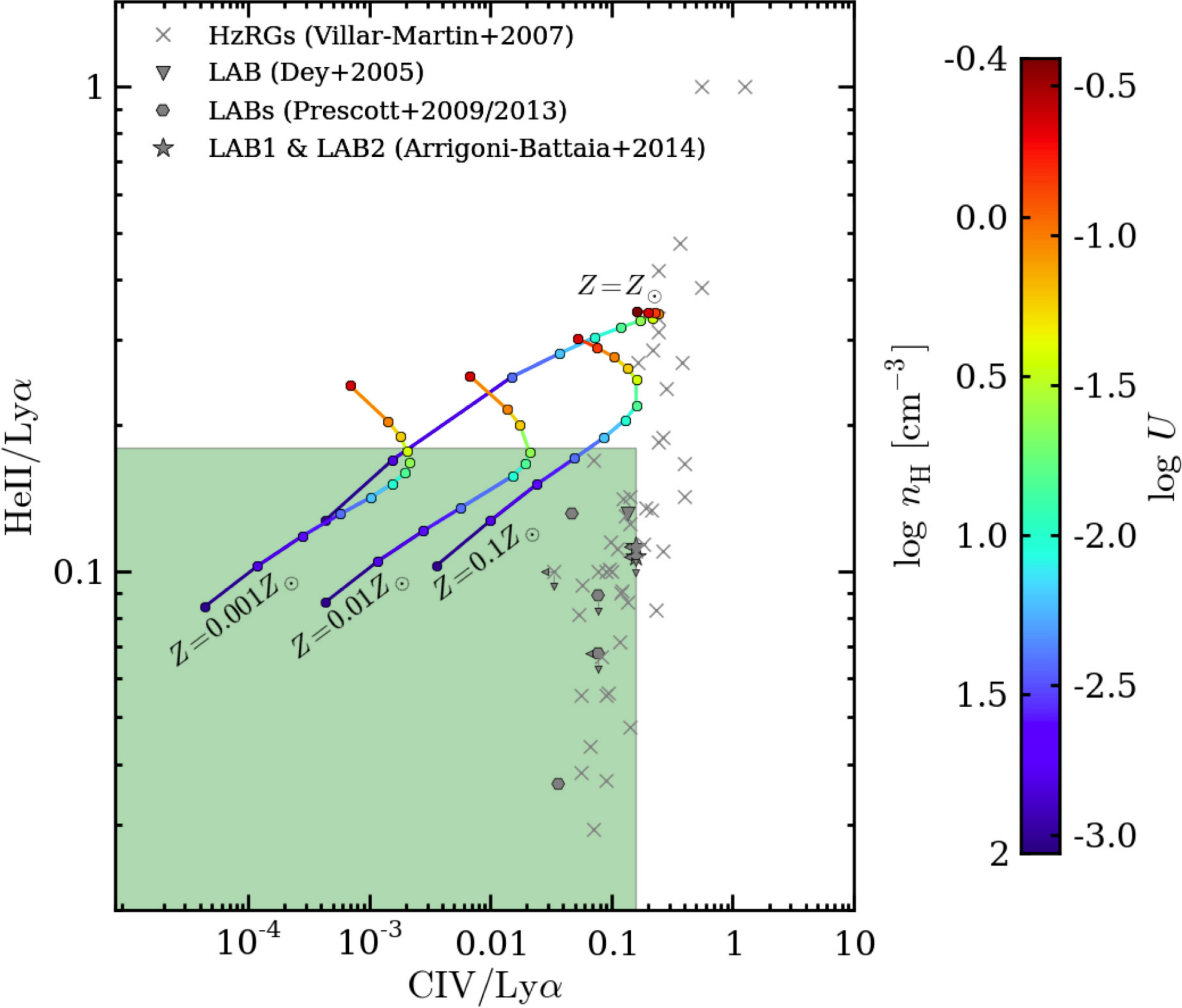, width=0.9\columnwidth, angle=270, clip} 
\caption{HeII/Ly$\alpha$ versus CIV/Ly$\alpha$ log-log plot. Our upper limits on the HeII/Ly$\alpha$ and CIV/Ly$\alpha$ ratios are compared with 
the Cloudy photoionization models that reproduce the observed SB$_{\rm Ly\alpha}\sim7\times10^{-18}$ \unitcgssb. 
For clarity, we plot only the models with $Z=0.001,0.01,0.1,$ and $1 Z_{\odot}$.
The models are color coded following the ionization parameter $U$, and thus the volume density $n_{\rm H}$ (see color bar on the right).
The green shaded area represents the region defined by the upper limits of the UM287 nebula. Note that these upper limits 
favor models with $n_{\rm H} \gtrsim 3$~cm$^{-3}$, $N_{\rm H} \lesssim 10^{20}$~cm$^{-2}$, and $\log U \lesssim -1.5$. 
This is even more clear in Figure \ref{Fig7}.}
\label{Fig6}
\end{figure}

As we discuss in Section \S\ref{sec:results}, our LRIS observations
provide upper limits on the \civ/\lya and \heii/\lya ratios,
i.e. (\civ/\lya)$_{3\sigma}\lesssim0.16$ and
(\heii/\lya)$_{3\sigma}\lesssim0.18$.  On the other hand, each
photoionization model in our grid predicts these line ratios, and
Figure \ref{Fig6} shows the trajectory of these models in the \heii/\lya vs
\civ/\lya plane. The region allowed given our observational constraints on the line
ratios is indicated by the green shaded area.
We remind the reader that we select only the models
which produce the observed \lya emission of SB$_{\rm
  Ly\alpha}\sim7\times10^{-18}$ \unitcgssb, which to lowest order
requires a combination of $N_{\rm H}$ and $n_{\rm H}$ as shown by
eqn.~(\ref{SBthin}).  Since the luminosity of the central source is known, these models can be
thought to be parametrized by either $n_{\rm H}$ or the ionization
parameter $U$, as shown by the color coding on the color-bar. In the
same plot we show trajectories for different metallicities $Z = 1,$~$0.1,$~$0.01,$~$10^{-3}\,Z_{\odot}$.

We now reconsider the covering factor. 
We argued in \S\ref{theory} that based on
the morphology of the nebula, the covering factor need to be $f_{\rm C} \gtrsim
0.5$, and that optically thick gas clouds would tend to overproduce
the Ly$\alpha$ SB for such high covering factors. Our models provide a
confirmation of this behavior.  For a covering factor of $f_{\rm
  C}=1.0$ a large number of models are available, whereas if we lower
the covering factor to $f_{\rm C} = 0.3$, we find that only two models
in our extensive model grid can satisfy the Ly$\alpha$ SB of the
nebula. This results because as we decrease $f_{\rm C}$, assuming the
gas is optically thin, eqn.~(\ref{SBthin}) indicates we must
correspondingly increase the product $N_{\rm H}n_{\rm H}$ by $1\slash
f_{\rm C}$ in order to match the observed Ly$\alpha$ SB. However, note
that the neutral fraction also scales with this product $x_{\rm
  HI}\propto N_{\rm H}n_{\rm H}$ such that for low enough values of
$f_{\rm C}$ increasing $N_{\rm H}n_{\rm H}$ would result in
self-shielding clouds that are optically thick. We already argued in
\S\ref{theory} that if the clouds are optically thick the covering factor must
be much lower $f_{\rm C} \simeq 0.02$, which is ruled out by the
diffuse morphology of the nebula. Hence our constraint on the covering factor
$f_{\rm C}\gtrsim 0.5$ can also be motivated by the simple fact that
gas distributions with lower covering factors would over-produce
the Ly$\alpha$ SB. Henceforth, for simplicity, we assume a covering factor of
$f_{\rm C} = 1.0$ throughout this work, but in \S\ref{sec:caveats} we
test the sensitivity of our results to this assumption.

The gray symbols in Figure \ref{Fig6} also show a compilation of
measurements of the \heii/\lya and \civ/\lya line ratios from the literature for other giant
\lya nebulae from the compilation in \citet{FAB2014}.  Specifically,
we show measurements or upper limits for the two line ratios for seven
\lya blobs (\citealt{Dey2005}; \citealt{Prescott2009, Prescott2013,
  FAB2014})\footnote{From the sample of \citet{FAB2014}.  We decide to
  plot only the upper limits of LAB1 and LAB2, which set the tighter
  constraints for that sample.}, and \lya nebulae associated with 53
high redshift radio galaxies (\citealt{Humphrey2006};
\citealt{VillarM2007}).
Note that we show measurements from the literature in
Figure~\ref{Fig6} for reference, but these measurements cannot be directly
compared to our observations or our models for several reasons.
First, the emission arising from the narrow line region of the
central obscured AGN is typically included for the HzRGs,
contaminating the line ratios for the nebulae.  In addition, the
central source UV luminosities are unknown for both LABs and
HzRGs, and thus they cannot be directly compared to our models, which
assume a central source luminosity.  See \citealt{FAB2014} and
references therein for a discussion on this dataset and its
caveats.

The trajectory of our optically thin models through the \heii/\lya  and \civ/\lya 
diagram can be understood as follows. We first focus on the curve for $Z=Z_{\odot}$
and follow it from low to high $U$ (i.e. from high to low volume density $n_{\rm H}$).
First consider the trend of the \heii/\lya ratio. \heii is a recombination line and thus, 
once the density is fixed, its emission depends basically on the fraction of Helium that is doubly ionized.
For this reason, the \heii/\lya ratio is increasing from log$U\sim-3$ and `saturates', reaching a peak at a value
of $\sim0.34$ which is set by atomic physics and in particular by the ratio of the recombination
coefficients of \lya and \heii.
Indeed, if we neglect the contribution of collisional excitation to the \lya line emission, which is
a reasonable assumption near solar metallicity,  then both the \heii and \lya are produced primarily
by recombination and the recombination emissivity can be written as
\begin{equation}
j_{\rm line} = f_{\rm V}^{\rm elem}\frac{h\nu_{\rm line}}{4\pi}n_{\rm e}n_{\rm ion}\alpha_{\rm line}^{\rm eff}(T) \label{eqn:jthin}, 
\end{equation}
where $n_{\rm ion}$ is the volume density of He$^{++}$ and H$^+$ for the case of HeII and 
\lya, respectively. Here $\alpha_{\rm line}^{\rm eff}(T)$ is the temperature dependent recombination coefficient
for HeII or \lya, and the factor $f_{\rm V}^{\rm elem}=3f_{\rm C}N_{\rm elem}/(4Rn_{\rm elem})$ takes into account that the emitting clouds 
fill only a fraction of the volume (see \citealt{Hennawi2013}). 
Thus, once the Helium is completely doubly ionized, i.e. $n_{\rm p}\sim n_{\rm H}$ and $n_{\rm He^{++}}\sim(Y/2X)n_{\rm H}$, the ratio between the two lines is given 
by the relation
\bea
\frac{j_{\rm HeII}}{j_{\rm Ly\alpha}} &=& 0.34 \left(\frac{\alpha_{\rm HeII}^{\rm eff}(20,000 K)}{1.15\times10^{-12}\, {\rm cm^3\,s^{-1}}}\right) \label{ratioHeIILyalpha} \\
&\times& \left(\frac{\alpha_{\rm Ly\alpha}^{\rm eff}(20,000 K)}{2.51\times10^{-13}\, {\rm cm^3\,s^{-1}}}\right)^{-1}, \nonumber
\eea 
Note that eqn.~(\ref{ratioHeIILyalpha}) 
depends slightly on temperature, with a decrease of the ratio at higher temperatures.
Before reaching this maximum line ratio, \heii/\lya is lower because Helium is
not completely ionized, and is roughly given by 
${{\rm HeII}/{\rm Ly}\alpha\sim x_{\rm He^{++}}\times (j_{\rm HeII}/j_{\rm Ly\alpha})_{\rm max}}$, 
where $x_{\rm He^{++}}$ is the fraction of doubly ionized Helium.
As stated above, this simple argument does not take into account collisional excitation
of Ly$\alpha$. 
In particular, at lower metallicities when metal line coolants are lacking,
the temperature of the nebula is increased, and collisionally excited
\lya, which is extremely sensitive to temperature, becomes
an important coolant, boosting the \lya emission over the pure recombination
value. Thus metallicity variations result in a change of the level of the asymptotic HeII/Lya ratio as seen in
Figure~\ref{Fig6}.

Our photoionization models indicate that the \civ emission line is 
an important coolant and is powered primarily by collisional excitation.
The efficiency of \civ as a coolant depends on the amount of Carbon in the C$^{+3}$ ionic state.
For this reason, the \civ/\lya ratio is increasing from log$U\sim-3$, reaches a peak due to 
a maximum in the $C^{+3}$ fraction,
and lowers again at higher $U$ where Carbon is excited to yet higher
ionization states, e.g. \cv.  For example, for the $Z=0.1 \ Z_{\odot}$
models, the \civ/\lya ratio peaks at log~$U=-1.4$ and then decreases at higher $U$.  
Given that \civ is a
coolant, the strength of its emission depends on the metallicity of
the gas. Indeed, for metallicities lower than solar, \civ becomes a
sub-dominant coolant with respect to collisionally excited \lya (and for very low 
metallicity, e.g. $Z=10^{-3}Z_{\odot}$, also to He
\lya), and its emission becomes metallicity dependent as can be seen in Figure \ref{Fig6}.

At lower metallicities the \lya line becomes an important coolant. For
the $Z=0.001Z_{\odot}$ grid, the collisional contribution to
Ly$\alpha$ has an average value of $\sim40$\%, while it decreases to
$\sim37$\%, $\sim25$\%, $\sim1$\% for the $Z=0.01,0.1, 1 \,Z_{\odot}$
cases, respectively.  Given that the strength of the
collisionally excited Ly$\alpha$ emission increases with density along
each model trajectory, this slightly dilutes the aforementioned trends in
the \heii and \civ line emission.  Specifically, the density
dependence of collisionally excited \lya emission moves the line
ratios to lower values for log$\,U\gtrsim-1.5$, which would otherwise
asymptote at the expected \heii/\lya
ratio in eqn.~(\ref{ratioHeIILyalpha}).  Thus the effect of collisionally
excited \lya emission tend to mask the
`saturation' of the \heii/\lya ratio due to
recombination effects alone, and results in a continuous increase of \heii/\lya
with $U$.

\begin{figure*}
\centering
\epsfig{file=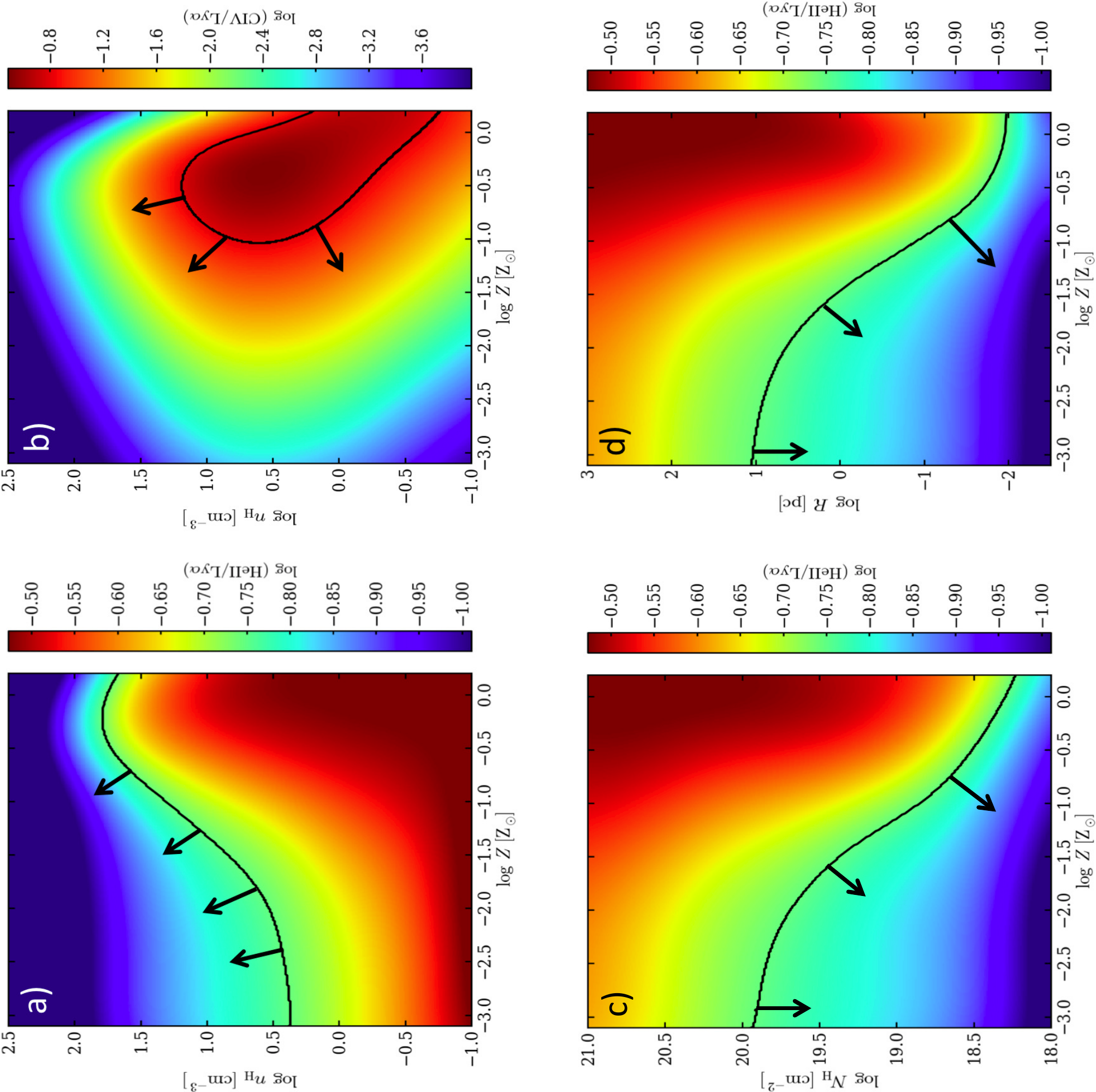, width=0.97\textwidth, angle=270, clip} 
\caption{Constraints on the physical parameters of the gas clouds from our photoionization models 
that reproduce the observed SB$_{\rm Ly\alpha}\sim7\times10^{-18}$ \unitcgssb in the case of an input spectrum with $\alpha_{\rm EUV}=-1.7$. 
Given the known luminosity of the central source, the assumed SED, and the fixed SB$_{\rm Ly\alpha}$, 
our models can be thought to be parametrized by only two parameters, namely $n_{\rm H}$ and $Z$.
{\bf Panel `a':} map of the \heii/\lya ratio in the $n_{\rm H}$-$Z$ plane.
The black solid line indicate our 3$\sigma$ upper limit \heii/\lya$<0.18$ (i.e. log(\heii/\lya)$<-0.74$). 
{\bf Panel `b':} map of the \civ/\lya ratio in the $n_{\rm H}$-$Z$ plane. The black solid line indicate our 3$\sigma$ upper limit \civ/\lya$<0.16$ (i.e. log(\heii/\lya)$<-0.79$).
Note that the constraints from the \heii/\lya ratio are stronger.
{\bf Panel `c':} map of the \heii/\lya ratio in the $N_{\rm H}$-$Z$ plane. 
The black solid line indicate our 3$\sigma$ upper limit.
Models with $N_{\rm H} \lesssim 10^{20}$ cm$^{-2}$ are selected.
{\bf Panel `d':} map of the \heii/\lya ratio in the $R$-$Z$ plane. The black solid line indicate our 3$\sigma$ upper limit.
Note that really small cloud sizes are favored, i.e. $R \lesssim 20$ pc.
}
\label{Fig7}
\end{figure*}

Overall, Figure~\ref{Fig6} illustrates that our simple photoionization
models can accommodate the constraints implied by our observed upper
limits on the \heii/\lya and \civ/\lya ratios of UM287. In particular,
our non-detections are satisfied (green shaded region) for models with
high volume densities $n_{\rm H}$ and low metallicities $Z$. These
constraints can be more easily visualized in Figure \ref{Fig7}, where
we show the allowed regions in the $n_{\rm H}$-$Z$ plane implied by
our limits on the \heii/\lya (panel `a') and \civ/\lya ratios (panel
`b').  Specifically, in these panels the solid black line indicate the
upper limits in the case of the UM287 nebula, i.e. \heii/\lya$<0.18$
(or log(\heii/\lya)$<-0.74$), and \civ/\lya= 0.16 (or
log(\civ/\lya)$<-0.79$), while the arrows indicate the region of the
parameter space that is allowed.  It is evident that our limits on the
extended emission in the \heii/\lya ratio give us stronger constraints
than those from the \civ/\lya ratio. The \heii/\lya ratio provides a
constraint on the volume density which is metallicity dependent,
however even if we assume a $\log_{10} Z\simeq -2-3$, which are the
lowest possible values comparable to the background metallicity of the
IGM (e.g. \citealt{Schaye2003}), we obtain a conservative lower limit
on the volume density of $n_{\rm H} \gtrsim 3$ cm$^{-3}$.

Given this constraint on $n_{\rm H}$, and the fact that we know the
\lya emission level, which in turns approximately scales as $n_{\rm H}N_{\rm H}$
(see eqn.~(\ref{SBthin})), we can use our lower limit on $n_{\rm H}$ to
place an upper limit on $N_{\rm H}$ or equivalently on the total cool
gas mass because it scales as $f_{\rm C}\, N_{\rm H}$ once the radius is fixed (see eqn.~(\ref{mass})).
Panel `c' of Figure \ref{Fig7} shows that our limit on the
\heii/\lya ratio combined with the total SB$_{{\rm Ly}\alpha}$ implies
the emitting clouds have column densities $N_{\rm H} \lesssim 10^{20}$ cm$^{-2}$.  Thus, if
we assume that the physical properties of the slab modeled at 160 kpc
are representative of the whole nebula, we can compute a rough
estimate for the total cool gas mass. With this strong assumption,
that $n_{\rm H} \gtrsim 3$ cm$^{-3}$ is valid over the entire area of
the nebula, i.e. $911$ arcsec$^2$ (from the $2\sigma$ isophote of the \lya
map; \citealt{Cantalupo2014}), we then deduce that $N_{\rm H} \lesssim 10^{20}$
cm$^{-2}$ over this same area, and hence the total cool gas mass is
${M_{\rm c}\lesssim 6.4\times10^{10}}$~M$_{\odot}$.

Further, by combining the lower limit on volume density $n_{\rm H}$ and upper
limit on column density $N_{\rm H}$, we can also obtain an upper limit
on the sizes of the emitting clouds defined as ${R\equiv N_{\rm H}/n_{\rm H}}$.  
Panel `d' in Figure \ref{Fig7} shows that this upper limit
is constrained to be $R \lesssim 20$ pc. Assuming a unit
covering factor $f_{\rm C}=1.0$, this constraint on cloud sizes
implies $\gtrsim 53,500$ clouds per square arcsec on the sky, 
and each cloud should have a cool gas mass $M_{\rm c} \lesssim
1.3\times10^{3}$ M$_{\odot}$. Assuming these clouds have the
same properties throughout the whole nebula, 
we find that $\gtrsim 4.9\times10^7$ clouds are needed to cover 
the extent of the \lya emission ($\sim911$ arcsec$^2$)\footnote{We quote a lower limit on the number of clouds per arcsec$^{2}$ 
because we calculate this value without taking into
account the possible overlap of clouds along the line of sight, and also because we use the maximum radius allowed by our constraints. 
In other words, we simply estimate the number of clouds with radius $R=20$ pc needed to cover the area of
a square arcsec on the sky at the systemic redshift of the UM287 quasar.}.

The foregoing discussion indicates that we are able to break the
degeneracy between the volume density of the gas $n_{\rm H}$ and the
total cool gas mass presented in \citet{Cantalupo2014}. As a reminder, this degeneracy arises because the
\lya surface brightness scales as ${\rm SB_{{\rm Ly}\alpha}} \propto n_{\rm H}N_{\rm H}$, whereas the total
cool gas mass is given by $M_{\rm c}\propto N_{\rm H} $. Thus observations of
the \lya alone cannot independently determine the cool gas mass.
\citet{Cantalupo2014} modeled the \lya 
emission in the UM287 nebula in a way that differs from our simple model of cool clouds in the
quasar CGM. Specifically, they used the gas distribution in a massive
dark matter halo $M= 10^{12.5}$~M$_\odot$ meant to represent the quasar host, and 
carried out ionizing and \lya radiative transfer simulations under the
assumption the gas is highly ionized by a quasar with the same 
luminosity as UM287, and the extended \lya emission is
dominated by recombinations, similarly to our simpler Cloudy
models\footnote{Although note that our Cloudy models treat
  collisionally exited \lya emission properly, whereas this effect
  cannot be properly modeled via the method in \citet{Cantalupo2014}.}.
Under these assumptions, they are
\textit{not able to reproduce the
  observed \lya surface brightness of the nebula.} This arises
because only $\sim15\%$ of the total gas in the simulated halo
is cool enough to emit \lya recombination radiation ($T<5\times10^4$ K),
because the vast majority of the baryons in the halo have been shock-heated
to the virial temperature of the halo, i.e. $T\sim 10^7$ K. 
Even if they assume all of the gas in the simulated halo can 
produce the \lya line ($M_{\rm gas}\approx 10^{11.3}$~M$_{\odot}$ 
for the dark matter halo; \citealt{Cantalupo2014}), the
surface brightness of the resulting nebula is still too faint.
As a result, \citet{Cantalupo2014} postulated that the emission
in the simulated halo must be boosted by a clumping factor $C=\langle n_{\rm H}^2\rangle/\langle n_{\rm H}\rangle^2$,
which represents the impact of clumps of cool gas which are not resolved
by the simulation.  They then determined the scaling relation between the simulated \lya emission and 
the column density of the simulated gas distribution, i.e. 
$N_{\rm H}\propto {\rm SB_{{\rm Ly}\alpha}}^{1/2} \, C^{-1/2}$ \footnote{In \citealt{Cantalupo2014} 
this relation is quoted as $N_{\rm HII}\propto {\rm SB_{{\rm Ly}\alpha}}^{1/2} \, C^{-1/2}$, 
but $N_{\rm H}\sim N_{\rm HII}$ in this simulated case where the gas is highly ionized.}
(\citealt{Cantalupo2014}), as expected for recombination radiation. Note
that accordingly, ${\rm SB_{{\rm Ly}\alpha}}^{1/2}  \propto N_{\rm H}\,C^{1/2} \propto M_{\rm c}\,C^{1/2}$
and one sees that this is identical to the scaling implied by eqn.~(\ref{SBthin}),
${\rm SB_{{\rm Ly}\alpha}}\propto N_{\rm H} n_{\rm H}$, if one identifies $n_{\rm H}$ with
$C^{1/2} \langle n_{\rm H}\rangle$. Our simple cloud model adopts a single density for all the clouds $n_{\rm H}$,
whereas in the clumping picture, there could be a range of densities present, but the emission
is dominated by gas with $n_{\rm H} \simeq C^{1/2}\,\langle n_{\rm H}\rangle$.
In this context, \citet{Cantalupo2014} inferred that if $C = 1$, the high observed SB$_{{\rm Ly}\alpha}$,
implies very high column densities up to $N_{\rm H}\approx 10^{22}$~cm$^{-2}$ 
corresponding to cool gas masses $M_{\rm c}=10^{12}$~M$_\odot$,
in excess of the baryon budget of the simulation. More generally, in the presence
of clumping this constraint becomes $M_{\rm c}=10^{12}\, C^{-1/2}$~M$_\odot$.

By introducing the constraint on the volume density $n_{\rm
  H}\gtrsim3$~cm$^{-3}$ using the \heii/\lya ratio, our work (i)
breaks the degeneracy between density $n_{\rm H}$ (or equivalently
$C$) and total column density $N_{\rm H}$ (or equivalently $M_{\rm
  c}$), (ii) allows us to then constrain the total cool gas mass
${M_{\rm c}\lesssim 6.4\times10^{10}}$~M$_{\odot}$ without making any
assumptions about the quasar host halo mass, and (iii) demands the
existence of a population of extremely compact ($R \lesssim 20$~pc) dense
clouds in the CGM/IGM. The ISM-like densities and extremely small
sizes of these clouds clearly indicate that they would be unresolved
by current cosmological hydrodynamical simulations, given their
resolution on galactic scales 
(\citealt{Fumagalli2014,Faucher-Giguere2014,Crighton2015,Nelson2015}). Indeed, our
measurements would imply a clumping factor $C\gtrsim200$ for the
simulation of \citet{Cantalupo2014}, in agreement with the value they
required in order to reproduce the observed \lya from their simulated
halo.

\begin{figure}
\centering
\epsfig{file=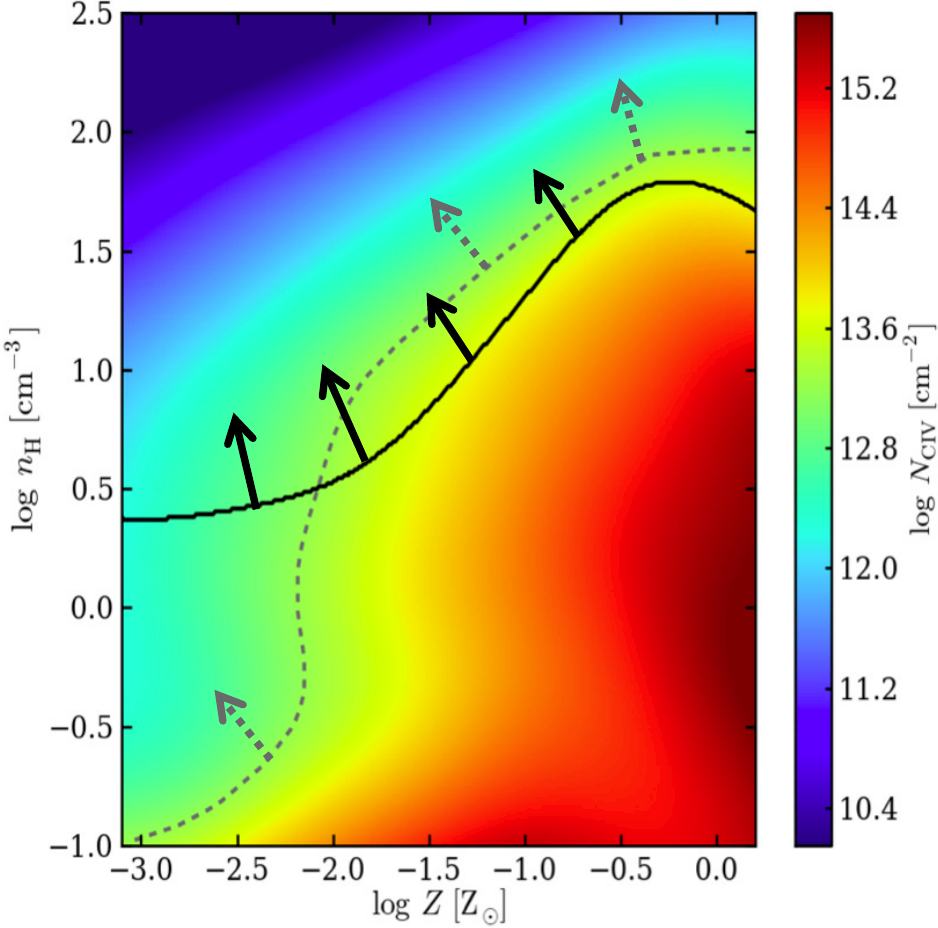, width=0.97\columnwidth, clip}\\ 
\epsfig{file=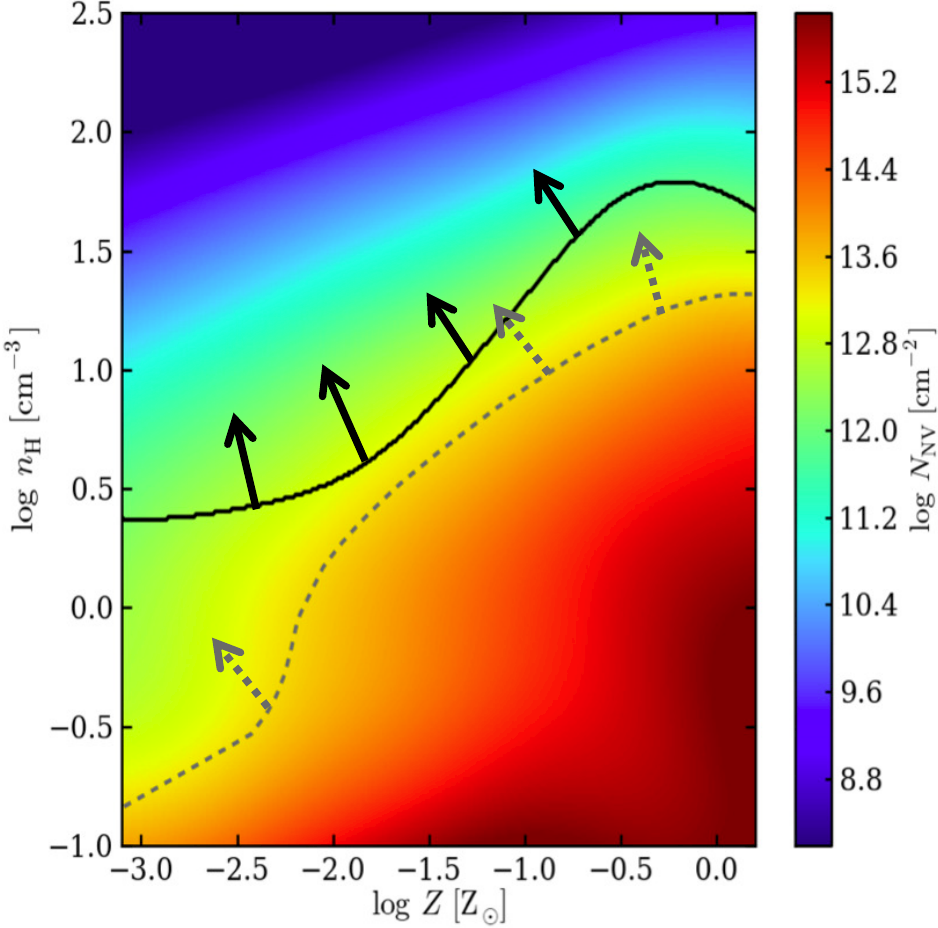, width=0.97\columnwidth, clip} 
\caption{{\bf Top Panel:} Map of the \civ column density $N_{\rm CIV}$ in the $n_{\rm H}$-$Z$ plane built from our photoionization models 
that reproduce the observed SB$_{\rm Ly\alpha}\sim7\times10^{-18}$ \unitcgssb in the case of an input spectrum with $\alpha_{\rm EUV}=-1.7$.
The black solid line indicate our 3$\sigma$ upper limit in the \heii/\lya ratio, while the 
gray dashed line indicate our limit of $N_{\rm CIV}<10^{13.2}$ cm$^{-2}$ implied by the absence 
of absorption at the resolution of the SDSS spectrum of UM287.
{\bf Bottom Panel:} Map of the \nv column density $N_{\rm NV}$ in the $n_{\rm H}$-$Z$ plane built from our photoionization models 
that reproduce the observed SB$_{\rm Ly\alpha}\sim7\times10^{-18}$ \unitcgssb in the case of an input spectrum with $\alpha_{\rm EUV}=-1.7$.
The black solid line indicate our 3$\sigma$ upper limit in the \heii/\lya ratio, while the 
gray dashed line indicate our limit of $N_{\rm NV}<10^{13.4}$ cm$^{-2}$ implied by the absence 
of absorption at the resolution of the SDSS spectrum of UM287.
The spectroscopic constraints for both species imply that the gas along the sightline, if present, 
is in a similar state as the observed nebula, being illuminated by the bright quasar as well. 
}
\label{Fig8}
\end{figure}

\subsection{Constraints from Absorption Lines}
\label{conAbsL}

A source lying in the background of the UM287 nebula that pierces the
gas at an impact parameter of $\simeq 160$ kpc may also exhibit
absorption from high-ion UV transitions like \civ and \nv, which can
be constrained from absorption spectroscopy.  In Figure \ref{Fig8} we
show a map for the column density of the \civ and \nv ionic states
($N_{\rm CIV}$, $N_{\rm NV}$) for our model grid that reproduces the
observed SB$_{\rm Ly\alpha}\sim7\times10^{-18}$ \unitcgssb.  Given our
non-detection of \heii emission, our upper limits on the \heii/\lya
ratios (indicated by the black solid line in both panels), imply
$N_{\rm CIV}\lesssim 10^{13.8}$~cm$^{-2}$ and $N_{\rm NV} \lesssim
10^{13.0}$~cm$^{-2}$, respectively.  The quasar UM287 resides at the
center of the nebula, and our narrow band image indicates it is
surrounded by \lya emitting gas. It is thus natural to assume that the
UM287 quasar pierces the nebular gas over a range of radial distances\footnote{This would 
not be the case if the emitting gas is all behind the 
quasar. Given that the quasar shines towards us and contemporary on the gas, 
this configuration seems unlikely.}. 
Thus a non-detection of absorption in these transitions places
further constraints on the physical state of the absorbing gas in the
nebula. 

To this end, we examined the high signal-to-noise 
${\rm S\slash N}\simeq$~70~pix$^{-1}$ 
SDSS spectrum of the UM287 quasar, which has a resolution of
$R \simeq 2000$. We find no evidence
for any metal-line absorption within a $\sim2000$~km~s$^{-1}$ window
of the quasar systemic redshift coincident with the velocity of the
\lya emitting nebula (see Figure~\ref{Fig2}-\ref{Fig3}), implying
$N_{\rm CIV}<10^{13.2}$ cm$^{-2}$ (EW$_{\rm CIV}<15$ m\AA), and
$N_{\rm NV}<10^{13.4}$ cm$^{-2}$. These limits constrain the amount of
gas in these ionic states intercepted by the quasar at all distances,
but in particular at $\simeq 160$ kpc, where we conducted our detailed
modeling of the emission.  As such, directly analogous to our
constraints from the emission line ratios,  we can similarly determine the
constraints in the $n_{\rm H}$-$Z$ plane from the non-detections of \civ and
\nv absorption, which are shown as the gray dashed lines in
Figure~\ref{Fig8}.
As expected these metal absorption constraints depend sensitively
on the enrichment of the gas, but the region of the $n_{\rm H}$-$Z$ plane
required by our non-detections are consistent with that required
by our \heii/\lya emission
constraint. Specifically, for log $Z>-2.3$, the absence of absorption provides a comparable
lower-limit on the density as the non-detection of emission, whereas at lower
metallicities the absorption constraint allows lower volume densities $n_{\rm
  H}>0.1$~cm$^{-3}$ (Figure \ref{Fig8}),  which are already ruled out by \heii/\lya.
To conclude, in the context of our simple model,
both high-ion metal-line absorption and  \heii and \civ emission paint
a consistent picture of the physical state of the gas.

For completeness, we also searched for metal-line absorption along the
companion quasar `QSO b' sightline in our Keck/LRIS spectrum
(resolution $R\simeq 1000$ and ${\rm S\slash N}\simeq 60$~pix$^{-1}$). 
We do detect strong, saturated \civ absorption with $N_{\rm CIV}>10^{14.4}$
cm$^{-2}$ and $z=2.2601$.  This implies, however, 
a velocity offset of $\approx -1700$ km~s$^{-1}$ with
respect to the systemic redshift of the UM287 quasar, and thus 
from the extended \lya emission detected in the slit
spectrum of Figure~2.  Given this large kinematic displacement from
the nebular \lya emission, we argue that this absorption is probably
not associated with the UM287 nebulae, and is likely to be a
narrow-associated absorption line system associated with
the companion quasar. 
This is further supported by the strong detection of the rarely
observed \nv doublet.
The large negative velocity offset $-1370$~km~s$^{-1}$ 
between the absorption and our best estimate for the
redshift of QSOb $z =2.275$ (from the \siiv emission line)
suggests that this is outflowing gas, but
given the large error $\sim 800 {\rm km \, s^{-1}}$ on the latter, and the unknown
distance of this absorbing gas along the line-of-sight, we do not
speculate further on its nature.

Finally, note that at the time of writing, there is no existing
echelle spectrum of UM287 available, although given that this quasar
is hyper-luminous $r \simeq 17$, a high signal-to-noise ratio high
resolution spectrum could be obtained in a modest integration.  Such a
spectrum would allow us to obtain much more sensitive constraints on
the high-ion states \civ and \nv, corresponding to $N_{\rm CIV} <
10^{12}$~cm$^{-2}$ and $N_{\rm NV} < 10^{12.5}$~cm$^{-2}$, respectively, and additionally
search for \ovi absorption down to $N_{\rm OVI} < 10^{13}$~cm$^{-2}$. If for
example \civ were still not detected at these low column densities,
this would raise our current constraint on $n_{\rm H}$ by 0.5 dex to
$n_{\rm H}\gtrsim 10\,{\rm cm^{-3}}$ as shown in Figure~\ref{Fig8}.
Furthermore, the detection of metal-line absorption (at a velocity consistent with the
nebular \lya emission) would determine the metallicity of the 
gas in the nebula, and Figure~\ref{Fig8} suggests we would be sensitive down to metallicities
as low as $Z\simeq -3$, i.e. as low as the background
metallicity of the IGM (e.g. \citealt{Schaye2003}).

\subsection{Comparison to Absorption Line Studies}
\label{absStudies}

It is interesting to compare the high volume densities
($n_{\rm H}>3$~cm$^{-3}$) implied by our analysis to independent 
absorption line measurements of gas densities in the CGM of typical quasars. 
For example \citet{Prochaska2009} used the strength of the absorption in the
collisionally excited \ion{C}{2}$^\ast$ fine-structure line to obtain an 
estimate of $n_{\rm H}\simeq 1\,{\rm cm^{-3}}$ at an impact parameter
of $R_{\perp}= 108\,{\rm kpc}$ from a foreground quasar, comparable to our lower
limit obtained from the \heii/\lya ratio.
However, photoionization modeling of a large sample of absorbers in
the quasar CGM seem to indicate that the typical gas densities are
much lower $n_{\rm H}\sim 0.01 \ll 1$ cm$^{-3}$ (\citealt{QPQ8}), although with
large uncertainties due to the unknown radiation field. 
If the typical quasar CGM has much lower values of
$n_{\rm H}\sim 0.01 \ll 1$ cm$^{-3}$ and column densities of $N_{\rm H}\sim 10^{20}$~cm$^{-2}$ (\citealt{QPQ8}), 
this would explain why quasars only rarely exhibit bright Ly$\alpha$ nebulae as in UM287.
Indeed, eqn.~(\ref{SBthin}) would then imply ${\rm SB}_{{\rm Ly}\alpha}=5.4\times10^{-20}$~\unitcgssb 
in the optically thin regime, which is far below the sensitivity of
any previous searches for extended emission
around quasars (e.g. \citealt{HuCowie1987,Heckman1991,Christensen2006}), although
these low SB levels may be reachable via stacking (\citealt{Steidel2011,FAB2015}).
In this interpretation, quasars exhibiting bright ${\rm SB} \sim 10^{-17}$~\unitcgssb giant \lya nebulae represent
the high end tail of the volume density distribution in the quasar CGM, a
conclusion supported by the analysis of another giant nebula with properties comparable to UM287 (\citealt{Jackpot})
discovered in the Quasars Probing Quasars survey (\citealt{Hennawi2013}). In this system joint
modeling of the Ly$\alpha$ nebulae and absorption lines in a background sightline piercing
the nebular gas indicate that cool gas is distributed in clouds with $R\sim40$ pc, with densities
$n_{\rm H}\simeq 2$ cm$^{-3}$, very similar to our findings for UM287.

Absorption line studies of gas around normal galaxies also provides
evidence for small-scale structure in their circumgalactic media.
Specifically, \citet{Crighton2015} conducted detailed photoionization
modeling of absorbing gas in the CGM of a Ly$\alpha$ emitter at $z
\simeq 2.5$, and deduced very small cloud sizes $<100-500$ pc,
although with much lower gas densities ($n_{H}\simeq 10^{-3}-10^{-2}\,{\rm cm}^3$)
than we find around UM287. In addition, there are multiple examples of
absorption line systems at $z\sim2-3$ in the literature
for which small sizes  $R\sim10-100$~pc have been deduced
(\citealt{Rauch1999,Simcoe2006,Schaye2007}), although the
absorbers may be larger at  $z\sim 0.2$ (\citealt{Werk2014}). Also,
compact structures
with $r\sim 50\,{\rm pc}$ have been directly resolved in 
high-velocity clouds in the CGM of the Milky Way (\citealt{BenBekhti2009}).
Given their expected sizes and masses, such
small structures are currently unresolved in simulations (see
discussion in \S~5.3 of \citealt{Crighton2015}).

\section{Model Spectra vs Current Observational Limits}
\label{sec:sens}

\begin{figure*}
\centering{\epsfig{file=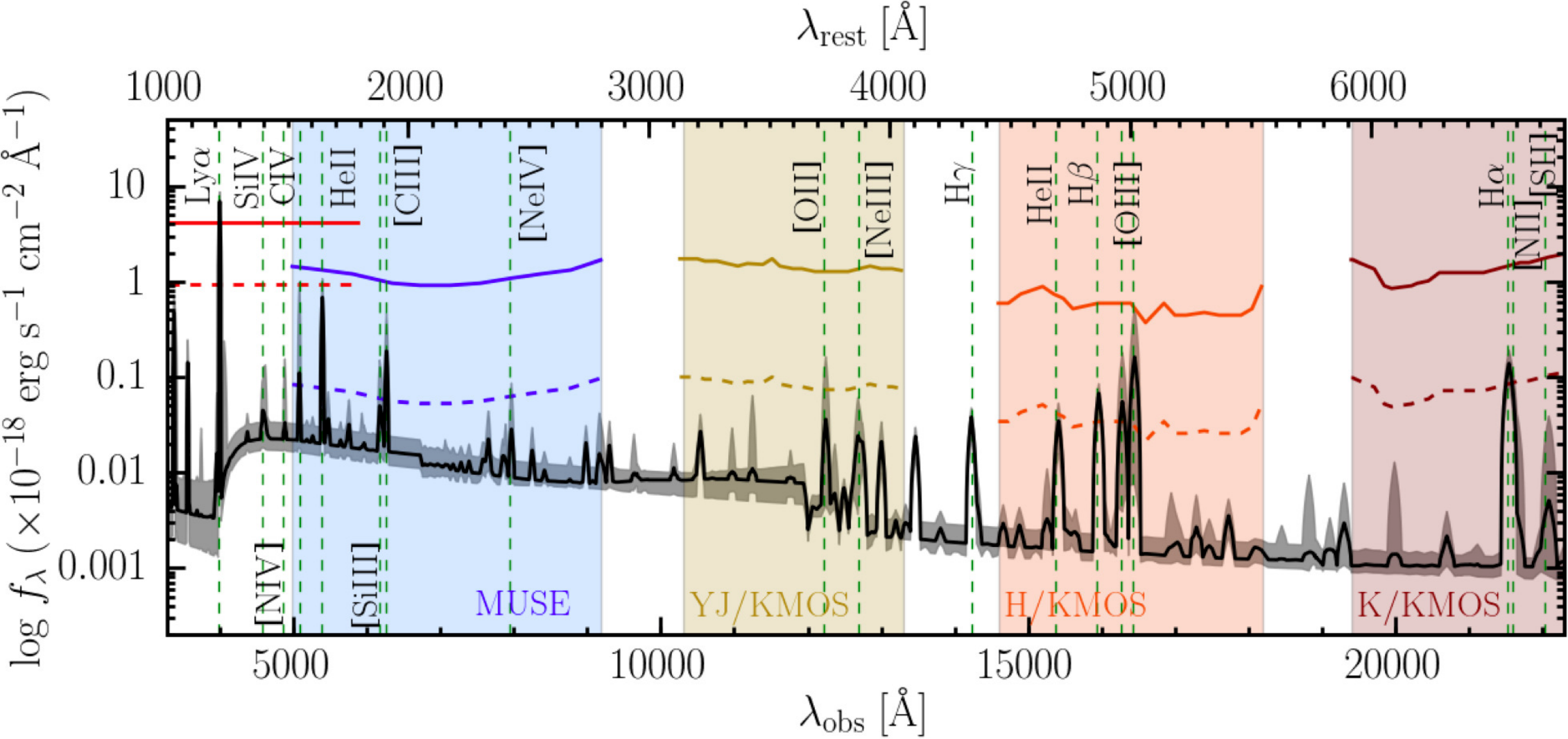, width=0.47\textwidth, clip, angle=270}}
\caption{Predicted median spectra for the models in our grid that satisfy simultaneously 
SB$_{\rm Ly\alpha}\sim7\times10^{-18}$ \unitcgssb, \heii/\lya$<0.18$, and \civ/\lya$<0.16$. 
The gray shaded area indicates the maximum and the minimum possible value for the selected models at 
each wavelength, showing the range of all the possible values, including the variation of the EUV slope, 
i.e. $\alpha_{\rm EUV}=-2.3, -1.7$, and $-1.1$ (see Appendix).
Our Keck/LRIS 3$\sigma$ sensitivity limit calculated in 1 arcsec$^2$ and over 3000~km~s$^{-1}$, is plotted as a solid red line, together with 
the 3$\sigma$ sensitivity of MUSE and KMOS (YJ, H, K gratings) for an exposure time of 10 hours (other colored solid lines). 
The red dashed line indicates our 3$\sigma$ sensitivity limit average over an aperture of 20 arcsec$^2$ (see \S\ref{sec:results}), while 
all the other dashed lines show the sensitivity averaged over an aperture of 300 arcsec$^2$, i.e. $SB_{\rm limit}=SB_{1\sigma}/\sqrt{A}$.
The principal emission lines are indicated by the green vertical dashed lines. The lines that may be detectable
in the future, given appropriate  
physical conditions (i.e. $n_{H}, Z$) in the targeted nebula are \heii, [\ciii], \civ, \siiii, [\oii], [\oiii], H$\beta$, and H$\alpha$.}
\label{FigMS}
\end{figure*}

In order to assess the feasibility of detecting other emission 
lines besides \lya from the UM287 nebula, and other similar extended \lya nebulae, 
e.g. around other quasars, HzRGS, or LABs, we construct model spectra using the output continuum and line emission data from 
Cloudy.
In Figure \ref{FigMS} we show the predicted median spectrum for the nebula at 
160 kpc from UM287, resulting from our modeling. Specifically, the solid black curve represents
the  median of all the models in our parameter grid which simultaneously satisfy the conditions
$5.5\times10^{-18}$~\unitcgssb~$<{\rm SB}_{\rm Ly\alpha}<8.5\times10^{-18}$~\unitcgssb,
such that they produce the right Ly$\alpha$ emission
level, as well as the emission line constraints \heii/\lya$<0.18$ and \civ/\lya$<0.16$ implied
by our spectroscopic limits. Following our discussion in the Appendix, this grid also includes
models with a harder (softer) $\alpha_{\rm EUV}=-1.1$ ($\alpha_{\rm EUV}=-2.3$) quasar ionizing continuum,
in addition to our fiducial value of $\alpha_{\rm EUV}=-1.7$. The gray shaded area indicates the maximum and
the minimum possible values for the selected models at each wavelength.

For comparison we show our Keck/LRIS 3$\sigma$ sensitivity limits from
\S\ref{sec:data} calculated by averaging over a 1 arcsec$^2$ aperture and
over a 3000~km~s$^{-1}$ velocity interval (solid red line), 
together with the 3$\sigma$ sensitivity limits for 10 hours of integration with
the Multi Unit Spectroscopic Explorer (MUSE) (\citealt{Bacon2010}; solid blue line), and with the
K-band Multi Object Spectrograph (KMOS) (\citealt{Sharples2006}; gold, orange, and dark-red
solid lines), on the VLT, computed for the same spatial and spectral aperture. 
Note that these sensitivity limits can be lowered by assuming a
certain amount of spatial averaging, following the relation $SB_{\rm
  limit}=SB_{1\sigma}/\sqrt{A}$, where $A$ is the area in arcsec$^2$
over which the data are averaged. Indeed, we employed this approach in
\S\ref{sec:results}, and averaged over an area of 20 arcsec$^2$ to
obtain a more sensitive constraint on the \heii/\lya and \civ/\lya
line ratios, and this lower SB level is indicated by the red dashed
line in Figure~\ref{FigMS}. In contrast with a longslit,
integral-field units like MUSE and KMOS, as well as the upcoming Keck
Cosmic Web Imager (KCWI, \citealt{Morrissey2012}), provide near
continuous spatial sampling over wide areas, and are thus the ideal
instruments for trying to detect extended line emission from the
CGM. Thus for MUSE and KMOS, we have assumed that we can average over
an area as large as $300$ arcsec$^2$, as shown by the colored dashed
lines, and indeed this approach has already been used with the Cosmic
Web Imager (\citealt{Martin2014a}) to study lower SB Ly$\alpha$ emission
(\citealt{Martin2014b}).

Given these expected sensitivities, in Figure~\ref{FigMS} 
we indicate the principal emission lines that may be detectable  
(vertical green dashed lines), whose observation would provide additional
constraints on the properties of the emitting gas. The large range of
metallicities in our grid $Z = 10^{-3}Z_{\odot}$ to $Z_{\odot}$, results in a
correspondingly large range of metal emission line strengths, whereas
the Hydrogen Balmer lines and \heii are much less sensitive to metallicity
and thus show very little variation across our model grid.

Focusing first on the primordial elements, we see that \heii is the
strongest line, and in particular it is stronger than
H$\alpha$. Indeed, if the Helium is completely doubly ionized then
\heii/H$\alpha\sim3$, and although it decreases to lower values for
lower ionization parameters (higher densities), it always remains
higher than unity.  As we have argued in \S\ref{sec:comparison}, a
detection of \heii can be used to measure the volume density $n_{\rm
  H}$ of the emitting gas.  Further, by comparing the morphology, 
size, and kinematics of the non-resonant extended \heii emission to that of \lya, one
can test whether resonant scattering of \lya plays an important role
in the structure of the nebula (\citealt{Prescott2015a}).  Naively,
one might have thought that H$\alpha$ would be ideal for this purpose
given that it is the strongest Hydrogen recombination line after
Ly$\alpha$. However, our models indicate that for photoionization by a
hard source, the \heii line is always stronger than H$\alpha$, and
given that \heii is in the optical whereas H$\alpha$ is in the
near-IR, it is also much easier to detect.

Figure~\ref{FigMS} shows that deep integrations in the near-IR with
KMOS will consistently detect the Hydrogen Balmer lines H$\alpha$ and
H$\beta$. When compared to the Ly$\alpha$ emission, these lines would
allows one to determine the extinction due to dust (\citealt{OF2006}).
Further, at the low densities we consider ($n_{\rm H}\ll 10^4$~cm$^{-3}$), any
departure of the ratios H$\alpha$/H$\beta$ and \lya/H$\beta$ from
their case B values provide information on the importance of
collisional excitation of Ly$\alpha$, which is exponentially sensitive
to the gas temperature (\citealt{OF1985}). In other words, the amount of
collisional excitation is set by the equilibrium temperature of the
gas, which is set by the balance of heating and
cooling. Photoionization by a hard source will result in a
characteristic temperature and hence ratio of \lya/H$\beta$ set by the
ionizing continuum slope, whereas an additional source of heat, as has
been postulated in gravitational cooling radiation scenarios for
Ly$\alpha$ nebulae (e.g. \citealt{Rosdahl12}), would increase the amount of
collisionally excited Ly$\alpha$ and hence the ratio of \lya/H$\beta$.

Figure \ref{FigMS} shows also that one could probably detect metal
emission lines depending on the physical conditions in the gas, which
are parameterized by $n_{\rm H}$ and $Z$. In particular, if the gas
has metallicity $Z > 0.1Z_{\odot}$, a deep integration with MUSE would
detect \civ, [\ciii], and, for metallicity close to solar, also
\siiii$\lambda1883$. In the near-IR, we see that a deep integration 
with KMOS would detect [\oiii] for $Z > 0.1Z_{\odot}$, and [\oii] for
metallicity close to solar.  Note that for similar bright nebulae
at different redshifts, it would be possible to detect other lines in
extended emission for particular $n_{\rm H}$ and $Z$ combinations,
e.g. \siiv$\lambda1394$, and \niv$\lambda1480$.

According to Figure \ref{FigMS}, a good observational strategy is thus to look for the \heii line, which appears 
to be the strongest and easiest
line to detect, and our analysis in \S\ref{sec:comparison} indicates that its detection
constrains  the gas properties to lie on a line
in the $n_{\rm H}$-$Z$ plane (see panel `a' in Figure~\ref{Fig7}). Following our
discussion of \civ (panel `b' of Figure~\ref{Fig7}), 
the detection of any metal line would define another line in the $n_{\rm H}$-$Z$ plane,
and the intersection of these curves would determine the $n_{\rm H}$ and $Z$ of the gas.
These conclusions will be somewhat sensitive to the assumed spectral slope in the UV (see Appendix),
but given the different ionization thresholds to ionized Carbon to \civ (47.9eV), and
Oxygen to \oiii\ (35.1eV) or \oii\ (13.6eV), it is clear that detections or limits on
multiple metal lines from high and low ionization states would also 
constrain the slope $\alpha_{\rm EUV}$ of the ionizing continuum.

To summarize, our photoionization modeling and analysis provide a
compelling motivation to find more bright nebulae by surveying large
samples of quasars and HzRGs, and conducting NB emission line surveys
of LABs over large areas. Armed with the brightest and largest giant
nebulae like UM287, one can conduct deep observations with IFUs, and
combined with suitable spatial averaging, this will uncover a rich
emission line spectrum from the CGM and its interface with the IGM,
which can be used to constrain the physical properties of the emitting
gas, and shed light on physical mechanism powering giant nebulae.

\section{Caveats}
\label{sec:caveats}

In section \S\ref{sec:comparison}, under the assumption
of photoionization by the central QSO, and in the context
of a simple model for the gas distribution,
we showed how our upper limits on the
\heii/\lya and \civ/\lya ratios, can set constraints on the physical
properties of the cool gas observed in emission.
However, this simple
modeling is just a zeroth-order approximation to a
more complicated problem which is beyond the scope of the
present work. In what follows we 
highlight some issues
which should be examined further.

\smallskip \noindent {\bf Radial Dependence}: for simplicity we have evaluated the ionizing flux at a
single radial location for input into Cloudy. We have tested the impact of this
assumption, by decreasing $R$ from 160 kpc to
100 kpc, and find that our lower limit on the density increases by 0.4 dex. This
results from the fact that the \heii/\lya ratio varies with ionization
parameter $U$, and our upper limit on the line ratio sets a particular
value of $U$.  By decreasing $R$, the density
$n_{\rm H}$ corresponding to this specific value of $U$ thus increases
by a factor $R^2$.  The variation of the ionizing flux with radius,
should be taken into account in a more detailed calculation.

\smallskip \noindent {\bf Slope of the Ionizing Continuum}: we have assumed $\alpha_{\rm EUV} =
- 1.7$ (\citealt{Lusso2015}). However, estimates for $\alpha_{\rm EUV}$ in
the literature vary widely (\citealt{Zheng1997,Scott2004,Shull2012}),
most likely because of uncertainties introduced when correcting for
absorption due to the IGM or because of the heterogeneity of the
samples considered. Furthermore, the shape of the ionizing continuum near the \ion{He}{2}
edge of 4 Rydberg is not well constrained. For detailed analysis
on the sensitivity of our results to the ionizing continuum slope, see the Appendix, where we consider two
different ionizing slopes, i.e. $\alpha_{\rm EUV} = - 1.1$ and $-2.3$.  We find
that a harder ionizing slope $\alpha_{\rm EUV} = -
1.1$ moves our lower limit on the density from $n_{\rm
  H}\gtrsim3$~cm$^{-3}$ to $n_{\rm H}\gtrsim1$~cm$^{-3}$.  Thus, the
uncertainty on the ionizing slope has an order unity impact on our constraints
of the volume density. As discussed at the end of \S\ref{sec:sens}, the detection of
additional metal lines with a range of ionization thresholds would further
constrain $\alpha_{\rm EUV}$.

\smallskip \noindent {\bf Covering Factor}: Based on the morphology of the emission we
argued $f_{\rm C} \gtrsim 0.5$, but assumed the value of $f_{\rm C} =
1.0$ for simplicity.  The $f_{\rm C}$ drops out of the line ratios
(see eqn.~(\ref{SBthin}) and (\ref{eqn:jthin})), however our model depends on $f_C$, since
we were selecting only models able to reproduce the observed
Ly$\alpha$ SB, which varies linearly with covering factor.  We
estimate that lowering the covering factor to $f_{\rm C}=0.4$, only
change our lower limit on the density at the $15\%$ level. As
discussed in section \S\ref{sec:comparison}, lowering $f_{\rm C}$
results in a reduction of the number of models which are able to
reproduce the observed \lya SB, because models with high $n_{\rm
  H}N_{\rm H}$ valuse become optically thick, and thus over-estimate
the \lya emission.  In particular, there are no models which reproduce
the observed \lya SB for low covering factors ($f_{\rm C}<0.3$).  Thus
our conclusions are largely insensitive to the covering factor we assumed.

\smallskip \noindent {\bf Geometry}: we have assumed the emitting clouds are spatially
uniformly distributed throughout a spherical halo. This simple
representation would need geometric corrections to take into account
more complicated gas distributions, such as variation of the covering
factor with radius or filamentary
structures. However, these corrections should be of order unity, and
are thus likely sub-dominant compared to other effects.

\smallskip \noindent {\bf Single Uniform Cloud Population}: our simple model
  assumes a single population of clouds which all have the same
  constant physical parameters $N_{\rm H}$, $n_{\rm H}$, and $Z$,
  following a uniform spatial distribution throughout the halo.  In
  reality one expects a distribution of cloud properties, and a radial
  dependence. Indeed, \citet{Binette1996} argued that a single
  population of clouds is not able to simultaneously explain both the
  high and low ionization lines in the extended emission line regions
  of HzRGs, and instead invoked a mixed population of completely
  ionized clouds and partially ionized clouds.  While for the case of
  extended emission line regions (EELRs) around quasars, which are on smaller scale $R <
  50\,{\rm kpc}$ than studied here, detailed
  photoionization modeling of spectroscopic data has demonstrated
  that at least two density phases are likely required: a diffuse
  abundant cloud population with $n_{\rm H}\sim1$~cm$^{-3}$, and much
  rarer dense clouds with $n_{\rm H}\sim500$~cm$^{-3}$
  (\citealt{Stockton2002, FuStockton2007, Hennawi2009}). Further, these clouds
  may be in pressure
  equilibrium with the ionizing radiation (\citealt{Dopita2002},
  \citealt{Stern2014}), as has been invoked in modeling the narrow-line regions of AGN.
  Future detailed modeling of multiple emission lines from giant nebulae, analogous to previous
  work on the smaller scale of EELRs (\citealt{Stockton2002, FuStockton2007}), might provide information
  on multiple density phases.

\smallskip
In order to properly address the aforementioned issues, the ideal approach would be to conduct a full radiative transfer
calculation on a three dimensional gas distribution, possibly taken from a cosmological hydrodynamical simulation. 
\citet{Cantalupo2014} carried out exactly this kind of calculation treating both ionizing and resonant radiative transfer, however this
analysis was restricted only to the Ly$\alpha$ line.  Full radiative transfer coupled to detailed photoionization modeling
as executed by Cloudy would clearly be too computationally challenging. However it would be interesting 
to introduce the solutions of 1-D Cloudy slab models into a realistic gas distribution drawn from a cosmological simulation. This would be relatively straightforward for the case of optically thin nebulae
(e.g. \citealt{vandeVoort2013}).

\section{Summary and Conclusions}
\label{sec:Conclusion}

To study the kinematics of the extended \lya line and to search for
extended \heii$\lambda1640$ and \civ$\lambda1549$ emission, we obtained
deep spectroscopy of the UM287 nebula (\citealt{Cantalupo2014}) with
the Keck/LRIS spectrograph. Our spectrum of the nebula provides
evidence for large motions suggested by the \lya line of  FWHM$_{\rm gauss}\sim500$~km~s$^{-1}$
which are spatially coherent on scales of $\sim$150 kpc.
There is no evidence  for a ``double-peaked'' line along either of the slits,
as might be expected in a scenario where resonant scattering determines
the \lya kinematic structure.  

Although our observations achieve an unprecedented sensitivity in the \heii and 
\civ line (${\rm SB}_{3\sigma}\simeq 10^{-18}$~\unitcgssb, average 
over 1\arcsec\,$\times$\,20\arcsec and $\Delta v=3000$~km~s$^{-1}$) for giant \lya nebulae, 
we do not detect extended emission in either line for both of our slit orientations.
We constrain the \heii/\lya and \civ/\lya ratios to be $<0.18$ (3$\sigma$), and 
$<0.16$ (3$\sigma$), respectively.

To interpret these non-detections, we constructed models of the emission line ratios, assuming
photoionization by the central quasar and a simple spatial distribution of cool gas in the quasar
halo. We find that:

\begin{itemize}

\item if the gas clouds emitting Ly$\alpha$ are optically thick to
  ionizing radiation, then the nebula would be $\sim$$120\times$
  brighter than observed, unless we assume an unrealistically low
  covering factor, i.e. $f_C\lesssim0.02$, which is in conflict with
  the smooth morphology of the nebula. Thus we conclude that the
  covering factor of cool gas clouds in the nebula is high $f_{\rm C}
  \gtrsim 0.5$, and that the gas in the nebula is highly ionized,
  resulting in gas clouds optically thin ($N_{\rm HI} < 17.2$) to
  ionizing radiation.

\item The \heii line is a recombination line and thus, once the
  density is fixed, its emission depends primarily on the fraction of
  Helium that is doubly ionized. On the other hand, the \civ emission
  line is an important coolant and is powered primarily by collisional
  excitation, and thus its emission depends on the amount of Carbon in
  the C$^{+3}$ ionic state.  As we know the ionizing luminosity of the
  central quasar, and the \lya SB of the nebula, constraints on the
  \heii/\lya and \civ/\lya ratios determine where the gas lives in the
  $n_{\rm H}-Z$ diagram.

\item Photoionization from the central quasar is consistent with the
  \lya emission and the \heii and
  \civ upper limits, provided that the gas distribution satisfies the following constraints:
  \begin{enumerate}
  \item[a)] $n_{\rm H} \gtrsim 3$ cm$^{-3}$, 
  \item[b)] $N_{\rm H} \lesssim 10^{20}$ cm$^{-2}$,
  \item[c)] $R \lesssim 20$ pc.   
\end{enumerate}
If these properties hold through the entire nebula, it then follows that the total 
cool gas ($T\sim10^4$~K) mass is  $M_{\rm c} \lesssim 6.4\times10^{10}$~M$_\odot$.

\end{itemize}

Because the \lya surface brightness scales as SB$_{\rm Ly\alpha}\propto n_{\rm H}N_{\rm H}$, 
whereas the total cool gas mass as $M_{\rm c}\propto N_{\rm H}$, observations of
Ly$\alpha$ emission cannot independently determine the cool gas mass and $n_{\rm H}$ (or the
gas clumping factor $C$), which limited the previous modeling by \citet{Cantalupo2014}.
Our non-detection of \heii/\lya combined with photoionization modeling allows us
to break this degeneracy, and independently constrain both $n_{\rm H}$ and $M_{\rm c}$. 

Our results point to the presence of a population of compact
($R \lesssim 20$~pc) cool gas clouds in the CGM at ISM-like densities
of $n_{\rm H}\gtrsim 3\,{\rm cm}^{-3}$ moving through the quasar
halo at velocities $\simeq 500\,{\rm km\,s^{-1}}$. It is
well known that even by $z\sim2$, the gas in the
massive $M\sim10^{12.5}$~M$_\odot$ halos hosting quasars is expected
to be dominated by a hot shock-heated plasma at the virial temperature $T\sim 10^7\,{\rm K}$. 
Cool clouds moving rapidly through a hot plasma will
be disrupted by hydrodynamic instabilities on the cloud-crushing
timescale (e.g. \citealt{Jones1994, Schaye2007, Agertz2007, Crighton2015, Scannapieco2015})
\bea
t_{\rm cc}&\approx& 1.3\,{\rm Myr}\left( \frac{R}{20\, {\rm pc}}\right)
\left( \frac{v}{500\,{\rm
    km\,s^{-1}}}\right)^{-1}\nonumber\\ &\times& \left(\frac{n_{\rm
    cl}/n_{\rm halo}}{1000}\right)^{1/2},
\eea
where we assume that the \lya line trace the kinematics of the cool clouds,
and that the cloud-halo density contrast is of the order of 1000 ($n_{\rm halo}\sim10^{-3}$~cm$^{-3}$).
If there is hot plasma present in the halo, these clouds are thus very short lived, and can only be
transported $\sim 0.7\,{\rm kpc}$ before being disrupted. These very short disruption timescales
thus require a mechanism that makes the clumps resistant to hydrodynamic instabilities, such as confinement
by magnetic fields (e.g. \citealt{McClure2010,McCourt2015}), otherwise the population
of cool dense clouds must be constantly replenished. In the latter scenario, 
the short lived clouds might be formed in situ, via cooling and fragmentation instabilities.
If the hot plasma pressure confines the
clouds, this might compresses them to high enough densities (\citealt{Fall1985,MallerBullock2004,MoMiralda1996}) to explain
our results.  Emission line nebulae from cool dense gas has also been observed at
the centers of present-day cooling flow clusters
(\citealt{Heckman1989,McDonald2010}), albeit on much smaller scales $\lesssim
50\,{\rm kpc}$. The giant Ly$\alpha$ nebula in UM287 
might be a manifestation of the same phenomenon, but with
much larger sizes and luminosities, reflecting different physical
conditions at high-redshift. Detailed study of the hydrodynamics
of cool dense gas clouds, with properties consistent with our constraints, moving 
through hot plasma are clearly required (\citealt{Scannapieco2015}).

As we showed in \S\ref{sec:sens}, deep observations ($\sim 10\,{\rm
  hr}$) of UM287 and other giant nebulae with the new integral field
units such as MUSE (\citealt{Bacon2010}), KCWI
(\citealt{Morrissey2012}), and KMOS (\citealt{Sharples2006}), combined
with spatial averaging, will be able to detect extended emission from
other lines besides \lya (see Figure \ref{FigMS}). In particular, the
strongest line will be \heii which should be routinely detectable,
and following our analysis, will enable measurements of the volume
density $n_{\rm H}$ of the gas.  Specifically, a 10 hour MUSE
integration would correspond to a sensitivity in \heii/\lya of $\sim$0.01
($3\sigma$ in 300 arcsec$^2$ 
), which would allow us
to probe gas densities as high as $n_{\rm H}=1000$~cm$^{-3}$. 
Although we have argued
that the UM287 is powered by photoionization, which is compelling 
given the presence of a hyper-luminous quasar, a non-detection of
\heii in a 10hr MUSE integration would imply such extreme gas
densities in the CGM, i.e. $n_{\rm H} > 1000$~cm$^{-3}$, that one might need
to reconsider other potential physical mechanisms for powering the
\lya nebula which do not produce \heii, such as cold-accretion
(e.g.,
\citealt{Haiman2000,Furlanetto05,Dijkstra06,Faucher2010,Rosdahl12}),
star-formation (e.g., \citealt{Cen2013}), or superwinds (e.g.,
\citealt{Taniguchi&Shioya2000, Taniguchi2001,Wilman2005}).
Furthermore, comparison of the morphology and kinematics of the nebula in \heii and \lya can be used
to test  whether resonant scattering of \lya photons is important.
Although H$\alpha$ could also be used to test the impact of resonant scattering, it is
always fainter than \heii and redshifted into the near-IR, where a detection of extended
emission is much more challenging.

In a photoionization scenario, a 10 hr observation of UM287 or a
comparable nebula with MUSE (or KCWI) and KMOS would result in a
rich emission line spectrum of the CGM, which, depending on the
properties of the gas (i.e. $n_{\rm H}$ and $Z$), could yield
detections of \lya, \niv, \siiv, \neiv, \civ, [\ciii], \siiii, [\oiii], [\oii],
H$\beta$, and H$\alpha$.  This would enable modeling of the CGM at a
comparable level of detail as models of \ion{H}{2} regions and the
narrow and broad-line regions of AGN, resulting in comparably detailed
constraints on the physical properties of the gas.

Current estimates suggest that $\sim 10-20\%$ of quasars exhibit
bright giant nebulae \citep{Jackpot} like UM287, and our results
provide a compelling motivation to expand current samples by surveying
large numbers of quasars with instruments like MUSE and KCWI. 
At the same time, this same survey data would enable one to compute
a stacked composite CGM spectrum of quasars which do not exhibit
bright nebulae, constraining the gas properties around typical quasars.

\acknowledgments 

We thank the members of the ENIGMA group\footnote{http://www.mpia-hd.mpg.de/ENIGMA/} at the
Max Planck Institute for Astronomy (MPIA) for helpful discussions.
JFH acknowledges generous support from the Alexander von Humboldt
foundation in the context of the Sofja Kovalevskaja Award. The
Humboldt foundation is funded by the German Federal Ministry for
Education and Research.  
JXP is supported by NSF grants AST-1010004 and AST-1412981.
The authors wish to recognize and acknowledge the very significant cultural role 
and reverence that the summit of Mauna Kea has always had within the indigenous 
Hawaiian community. We are most fortunate to have the opportunity to conduct 
observations from this mountain.

\bibliographystyle{apj}
\bibliography{biblio}

\begin{thebibliography}{}
\expandafter\ifx\csname natexlab\endcsname\relax\def\natexlab#1{#1}\fi

\bibitem[{{Agertz} {et~al.}(2007){Agertz}, {Moore}, {Stadel}, {Potter},
  {Miniati}, {Read}, {Mayer}, {Gawryszczak}, {Kravtsov}, {Nordlund}, {Pearce},
  {Quilis}, {Rudd}, {Springel}, {Stone}, {Tasker}, {Teyssier}, {Wadsley}, \&
  {Walder}}]{Agertz2007}
{Agertz}, O., {Moore}, B., {Stadel}, J., {et~al.} 2007, \mnras, 380, 963

\bibitem[{{Alam} \& {Miralda-Escud{\'e}}(2002)}]{Alam2002}
{Alam}, S.~M.~K., \& {Miralda-Escud{\'e}}, J. 2002, \apj, 568, 576

\bibitem[{{Arrigoni Battaia} {et~al.}(2015){Arrigoni Battaia}, {Hennawi},
  {Prochaska}, \& {Cantalupo}}]{FAB2015}
{Arrigoni Battaia}, F., {Hennawi}, J., {Prochaska}, J., \& {Cantalupo}, S.
  2015, in prep.

\bibitem[{{Arrigoni Battaia} {et~al.}(2014{\natexlab{a}}){Arrigoni Battaia},
  {Hennawi}, {Cantalupo}, \& {Prochaska}}]{FABProceeding}
{Arrigoni Battaia}, F., {Hennawi}, J.~F., {Cantalupo}, S., \& {Prochaska},
  J.~X. 2014{\natexlab{a}}, Proc. IAU Symp. \#304: Multiwavelength AGN Surveys
  and Studies, Cambridge Univ. Press, 253

\bibitem[{{Arrigoni Battaia} {et~al.}(2014{\natexlab{b}}){Arrigoni Battaia},
  {Yang}, {Hennawi}, {Prochaska}, {Matsuda}, {Yamada}, \&
  {Hayashino}}]{FAB2014}
{Arrigoni Battaia}, F., {Yang}, Y., {Hennawi}, J.~F., {et~al.}
  2014{\natexlab{b}}, ArXiv e-prints, arXiv:1407.2944

\bibitem[{{Bacon} {et~al.}(2010){Bacon}, {Accardo}, {Adjali}, {Anwand},
  {Bauer}, {Biswas}, {Blaizot}, {Boudon}, {Brau-Nogue}, {Brinchmann},
  {Caillier}, {Capoani}, {Carollo}, {Contini}, {Couderc}, {Daguis{\'e}},
  {Deiries}, {Delabre}, {Dreizler}, {Dubois}, {Dupieux}, {Dupuy}, {Emsellem},
  {Fechner}, {Fleischmann}, {Fran{\c c}ois}, {Gallou}, {Gharsa}, {Glindemann},
  {Gojak}, {Guiderdoni}, {Hansali}, {Hahn}, {Jarno}, {Kelz}, {Koehler},
  {Kosmalski}, {Laurent}, {Le Floch}, {Lilly}, {Lizon}, {Loupias}, {Manescau},
  {Monstein}, {Nicklas}, {Olaya}, {Pares}, {Pasquini}, {P{\'e}contal-Rousset},
  {Pell{\'o}}, {Petit}, {Popow}, {Reiss}, {Remillieux}, {Renault}, {Roth},
  {Rupprecht}, {Serre}, {Schaye}, {Soucail}, {Steinmetz}, {Streicher}, {Stuik},
  {Valentin}, {Vernet}, {Weilbacher}, {Wisotzki}, \& {Yerle}}]{Bacon2010}
{Bacon}, R., {Accardo}, M., {Adjali}, L., {et~al.} 2010, in Society of
  Photo-Optical Instrumentation Engineers (SPIE) Conference Series, Vol. 7735,
  Society of Photo-Optical Instrumentation Engineers (SPIE) Conference Series

\bibitem[{{Baskin} {et~al.}(2014){Baskin}, {Laor}, \& {Stern}}]{Baskin2014}
{Baskin}, A., {Laor}, A., \& {Stern}, J. 2014, \mnras, 438, 604

\bibitem[{{Ben Bekhti} {et~al.}(2009){Ben Bekhti}, {Richter}, {Winkel}, {Kenn},
  \& {Westmeier}}]{BenBekhti2009}
{Ben Bekhti}, N., {Richter}, P., {Winkel}, B., {Kenn}, F., \& {Westmeier}, T.
  2009, \aap, 503, 483

\bibitem[{{Bergeron} {et~al.}(2004){Bergeron}, {Petitjean}, {Aracil}, {Pichon},
  {Scannapieco}, {Srianand}, {Boisse}, {Carswell}, {Chand}, {Cristiani},
  {Ferrara}, {Haehnelt}, {Hughes}, {Kim}, {Ledoux}, {Richter}, \&
  {Viel}}]{Bergeron2004}
{Bergeron}, J., {Petitjean}, P., {Aracil}, B., {et~al.} 2004, The Messenger,
  118, 40

\bibitem[{{Binette} {et~al.}(1993){Binette}, {Wang}, {Zuo}, \&
  {Magris}}]{Binette1993}
{Binette}, L., {Wang}, J.~C.~L., {Zuo}, L., \& {Magris}, C.~G. 1993, \aj, 105,
  797

\bibitem[{{Binette} {et~al.}(1996){Binette}, {Wilson}, \&
  {Storchi-Bergmann}}]{Binette1996}
{Binette}, L., {Wilson}, A.~S., \& {Storchi-Bergmann}, T. 1996, \aap, 312, 365

\bibitem[{{Boesgaard} \& {Steigman}(1985)}]{Boesgaard1985}
{Boesgaard}, A.~M., \& {Steigman}, G. 1985, \araa, 23, 319

\bibitem[{{Cantalupo} {et~al.}(2014){Cantalupo}, {Arrigoni-Battaia},
  {Prochaska}, {Hennawi}, \& {Madau}}]{Cantalupo2014}
{Cantalupo}, S., {Arrigoni-Battaia}, F., {Prochaska}, J.~X., {Hennawi}, J.~F.,
  \& {Madau}, P. 2014, \nat, 506, 63

\bibitem[{{Cantalupo} {et~al.}(2012){Cantalupo}, {Lilly}, \&
  {Haehnelt}}]{Cantalupo2012}
{Cantalupo}, S., {Lilly}, S.~J., \& {Haehnelt}, M.~G. 2012, \mnras, 425, 1992

\bibitem[{{Cantalupo} {et~al.}(2007){Cantalupo}, {Lilly}, \&
  {Porciani}}]{Cantalupo2007}
{Cantalupo}, S., {Lilly}, S.~J., \& {Porciani}, C. 2007, \apj, 657, 135

\bibitem[{{Cantalupo} {et~al.}(2005){Cantalupo}, {Porciani}, {Lilly}, \&
  {Miniati}}]{Cantalupo2005}
{Cantalupo}, S., {Porciani}, C., {Lilly}, S.~J., \& {Miniati}, F. 2005, \apj,
  628, 61

\bibitem[{{Cen} \& {Zheng}(2013)}]{Cen2013}
{Cen}, R., \& {Zheng}, Z. 2013, \apj, 775, 112

\bibitem[{{Charlot} \& {Fall}(1993)}]{Charlot1993}
{Charlot}, S., \& {Fall}, S.~M. 1993, \apj, 415, 580

\bibitem[{{Christensen} {et~al.}(2006){Christensen}, {Jahnke}, {Wisotzki}, \&
  {S{\'a}nchez}}]{Christensen2006}
{Christensen}, L., {Jahnke}, K., {Wisotzki}, L., \& {S{\'a}nchez}, S.~F. 2006,
  \aap, 459, 717

\bibitem[{{Crighton} {et~al.}(2015){Crighton}, {Hennawi}, {Simcoe}, {Cooksey},
  {Murphy}, {Fumagalli}, {Prochaska}, \& {Shanks}}]{Crighton2015}
{Crighton}, N.~H.~M., {Hennawi}, J.~F., {Simcoe}, R.~A., {et~al.} 2015, \mnras,
  446, 18

\bibitem[{{Croft} {et~al.}(2002){Croft}, {Weinberg}, {Bolte}, {Burles},
  {Hernquist}, {Katz}, {Kirkman}, \& {Tytler}}]{Croft2002}
{Croft}, R.~A.~C., {Weinberg}, D.~H., {Bolte}, M., {et~al.} 2002, \apj, 581, 20

\bibitem[{{Dere} {et~al.}(1997){Dere}, {Landi}, {Mason}, {Monsignori Fossi}, \&
  {Young}}]{Dere1997}
{Dere}, K.~P., {Landi}, E., {Mason}, H.~E., {Monsignori Fossi}, B.~C., \&
  {Young}, P.~R. 1997, \aaps, 125, 149

\bibitem[{{Dey} {et~al.}(2005){Dey}, {Bian}, {Soifer}, {Brand}, {Brown},
  {Chaffee}, {Le Floc'h}, {Hill}, {Houck}, {Jannuzi}, {Rieke}, {Weedman},
  {Brodwin}, \& {Eisenhardt}}]{Dey2005}
{Dey}, A., {Bian}, C., {Soifer}, B.~T., {et~al.} 2005, \apj, 629, 654

\bibitem[{{Dijkstra} {et~al.}(2006{\natexlab{a}}){Dijkstra}, {Haiman}, \&
  {Spaans}}]{Dijkstra06}
{Dijkstra}, M., {Haiman}, Z., \& {Spaans}, M. 2006{\natexlab{a}}, \apj, 649, 14

\bibitem[{{Dijkstra} {et~al.}(2006{\natexlab{b}}){Dijkstra}, {Haiman}, \&
  {Spaans}}]{Dijkstra2006}
---. 2006{\natexlab{b}}, \apj, 649, 37

\bibitem[{{Dijkstra} \& {Loeb}(2008)}]{Dijkstra2008}
{Dijkstra}, M., \& {Loeb}, A. 2008, \mnras, 386, 492

\bibitem[{{Dopita} {et~al.}(2002){Dopita}, {Groves}, {Sutherland}, {Binette},
  \& {Cecil}}]{Dopita2002}
{Dopita}, M.~A., {Groves}, B.~A., {Sutherland}, R.~S., {Binette}, L., \&
  {Cecil}, G. 2002, \apj, 572, 753

\bibitem[{{Fall} \& {Rees}(1985)}]{Fall1985}
{Fall}, S.~M., \& {Rees}, M.~J. 1985, \apj, 298, 18

\bibitem[{{Fardal} {et~al.}(2001){Fardal}, {Katz}, {Gardner}, {Hernquist},
  {Weinberg}, \& {Dav{\'e}}}]{Fardal2001}
{Fardal}, M.~A., {Katz}, N., {Gardner}, J.~P., {et~al.} 2001, \apj, 562, 605

\bibitem[{{Farina} {et~al.}(2013){Farina}, {Falomo}, {Decarli}, {Treves}, \&
  {Kotilainen}}]{Farina2013}
{Farina}, E.~P., {Falomo}, R., {Decarli}, R., {Treves}, A., \& {Kotilainen},
  J.~K. 2013, \mnras, 429, 1267

\bibitem[{{Faucher-Giguere} {et~al.}(2014){Faucher-Giguere}, {Hopkins},
  {Keres}, {Muratov}, {Quataert}, \& {Murray}}]{Faucher-Giguere2014}
{Faucher-Giguere}, C.-A., {Hopkins}, P.~F., {Keres}, D., {et~al.} 2014, ArXiv
  e-prints, arXiv:1409.1919

\bibitem[{{Faucher-Gigu{\`e}re} {et~al.}(2010){Faucher-Gigu{\`e}re}, {Kere{\v
  s}}, {Dijkstra}, {Hernquist}, \& {Zaldarriaga}}]{Faucher2010}
{Faucher-Gigu{\`e}re}, C.-A., {Kere{\v s}}, D., {Dijkstra}, M., {Hernquist},
  L., \& {Zaldarriaga}, M. 2010, \apj, 725, 633

\bibitem[{{Ferland} \& {Osterbrock}(1985)}]{OF1985}
{Ferland}, G.~J., \& {Osterbrock}, D.~E. 1985, \apj, 289, 105

\bibitem[{{Ferland} {et~al.}(2013){Ferland}, {Porter}, {van Hoof}, {Williams},
  {Abel}, {Lykins}, {Shaw}, {Henney}, \& {Stancil}}]{Ferland2013}
{Ferland}, G.~J., {Porter}, R.~L., {van Hoof}, P.~A.~M., {et~al.} 2013, \rmxaa,
  49, 137

\bibitem[{{Francis} \& {Bland-Hawthorn}(2004)}]{Francis2004}
{Francis}, P.~J., \& {Bland-Hawthorn}, J. 2004, \mnras, 353, 301

\bibitem[{{Fu} \& {Stockton}(2007)}]{FuStockton2007}
{Fu}, H., \& {Stockton}, A. 2007, \apj, 666, 794

\bibitem[{{Fumagalli} {et~al.}(2014){Fumagalli}, {Hennawi}, {Prochaska},
  {Kasen}, {Dekel}, {Ceverino}, \& {Primack}}]{Fumagalli2014}
{Fumagalli}, M., {Hennawi}, J.~F., {Prochaska}, J.~X., {et~al.} 2014, \apj,
  780, 74

\bibitem[{{Furlanetto} {et~al.}(2005){Furlanetto}, {Schaye}, {Springel}, \&
  {Hernquist}}]{Furlanetto05}
{Furlanetto}, S.~R., {Schaye}, J., {Springel}, V., \& {Hernquist}, L. 2005,
  \apj, 622, 7

\bibitem[{{Fynbo} {et~al.}(1999){Fynbo}, {M{\o}ller}, \& {Warren}}]{Fynbo1999}
{Fynbo}, J.~U., {M{\o}ller}, P., \& {Warren}, S.~J. 1999, \mnras, 305, 849

\bibitem[{{Geach} {et~al.}(2007){Geach}, {Smail}, {Chapman}, {Alexander},
  {Blain}, {Stott}, \& {Ivison}}]{Geach2007}
{Geach}, J.~E., {Smail}, I., {Chapman}, S.~C., {et~al.} 2007, \apjl, 655, L9

\bibitem[{{Geach} {et~al.}(2009){Geach}, {Alexander}, {Lehmer}, {Smail},
  {Matsuda}, {Chapman}, {Scharf}, {Ivison}, {Volonteri}, {Yamada}, {Blain},
  {Bower}, {Bauer}, \& {Basu-Zych}}]{Geach2009}
{Geach}, J.~E., {Alexander}, D.~M., {Lehmer}, B.~D., {et~al.} 2009, \apj, 700,
  1

\bibitem[{{Gould} \& {Weinberg}(1996)}]{Gould1996}
{Gould}, A., \& {Weinberg}, D.~H. 1996, \apj, 468, 462

\bibitem[{{Greene} {et~al.}(2011){Greene}, {Zakamska}, {Ho}, \&
  {Barth}}]{Greene2011}
{Greene}, J.~E., {Zakamska}, N.~L., {Ho}, L.~C., \& {Barth}, A.~J. 2011, \apj,
  732, 9

\bibitem[{{Groves} {et~al.}(2004){Groves}, {Dopita}, \&
  {Sutherland}}]{Groves2004}
{Groves}, B.~A., {Dopita}, M.~A., \& {Sutherland}, R.~S. 2004, \apjs, 153, 75

\bibitem[{{Haiman} \& {Rees}(2001)}]{HaimanRees2001}
{Haiman}, Z., \& {Rees}, M.~J. 2001, \apj, 556, 87

\bibitem[{{Haiman} {et~al.}(2000){Haiman}, {Spaans}, \&
  {Quataert}}]{Haiman2000}
{Haiman}, Z., {Spaans}, M., \& {Quataert}, E. 2000, \apjl, 537, L5

\bibitem[{{Hayes} {et~al.}(2011){Hayes}, {Scarlata}, \& {Siana}}]{Hayes2011}
{Hayes}, M., {Scarlata}, C., \& {Siana}, B. 2011, \nat, 476, 304

\bibitem[{{Heckman} {et~al.}(1989){Heckman}, {Baum}, {van Breugel}, \&
  {McCarthy}}]{Heckman1989}
{Heckman}, T.~M., {Baum}, S.~A., {van Breugel}, W.~J.~M., \& {McCarthy}, P.
  1989, \apj, 338, 48

\bibitem[{{Heckman} {et~al.}(1991{\natexlab{a}}){Heckman}, {Lehnert}, {Miley},
  \& {van Breugel}}]{Heckman1991spec}
{Heckman}, T.~M., {Lehnert}, M.~D., {Miley}, G.~K., \& {van Breugel}, W.
  1991{\natexlab{a}}, \apj, 381, 373

\bibitem[{{Heckman} {et~al.}(1991{\natexlab{b}}){Heckman}, {Miley}, {Lehnert},
  \& {van Breugel}}]{Heckman1991}
{Heckman}, T.~M., {Miley}, G.~K., {Lehnert}, M.~D., \& {van Breugel}, W.
  1991{\natexlab{b}}, \apj, 370, 78

\bibitem[{{Hennawi} {et~al.}(2015){Hennawi}, {Prochaska}, {Cantalupo}, \&
  {Arrigoni Battaia}}]{Jackpot}
{Hennawi}, J., {Prochaska}, J., {Cantalupo}, S., \& {Arrigoni Battaia}, F.
  2015, submitted to Science

\bibitem[{{Hennawi} \& {Prochaska}(2007)}]{Hennawi2007}
{Hennawi}, J.~F., \& {Prochaska}, J.~X. 2007, \apj, 655, 735

\bibitem[{{Hennawi} \& {Prochaska}(2013)}]{Hennawi2013}
---. 2013, \apj, 766, 58

\bibitem[{{Hennawi} {et~al.}(2009){Hennawi}, {Prochaska}, {Kollmeier}, \&
  {Zheng}}]{Hennawi2009}
{Hennawi}, J.~F., {Prochaska}, J.~X., {Kollmeier}, J., \& {Zheng}, Z. 2009,
  \apjl, 693, L49

\bibitem[{{Hennawi} {et~al.}(2006){Hennawi}, {Prochaska}, {Burles}, {Strauss},
  {Richards}, {Schlegel}, {Fan}, {Schneider}, {Zakamska}, {Oguri}, {Gunn},
  {Lupton}, \& {Brinkmann}}]{Hennawi2006}
{Hennawi}, J.~F., {Prochaska}, J.~X., {Burles}, S., {et~al.} 2006, \apj, 651,
  61

\bibitem[{{Hogan} \& {Weymann}(1987)}]{Hogan1987}
{Hogan}, C.~J., \& {Weymann}, R.~J. 1987, \mnras, 225, 1P

\bibitem[{{Hu} \& {Cowie}(1987)}]{HuCowie1987}
{Hu}, E.~M., \& {Cowie}, L.~L. 1987, \apjl, 317, L7

\bibitem[{{Humphrey} {et~al.}(2006){Humphrey}, {Villar-Mart{\'{\i}}n},
  {Fosbury}, {Vernet}, \& {di Serego Alighieri}}]{Humphrey2006}
{Humphrey}, A., {Villar-Mart{\'{\i}}n}, M., {Fosbury}, R., {Vernet}, J., \& {di
  Serego Alighieri}, S. 2006, \mnras, 369, 1103

\bibitem[{{Humphrey} {et~al.}(2008){Humphrey}, {Villar-Mart{\'{\i}}n},
  {Vernet}, {Fosbury}, {di Serego Alighieri}, \& {Binette}}]{Humphrey2008}
{Humphrey}, A., {Villar-Mart{\'{\i}}n}, M., {Vernet}, J., {et~al.} 2008,
  \mnras, 383, 11

\bibitem[{{Husemann} {et~al.}(2013){Husemann}, {Wisotzki}, {S{\'a}nchez}, \&
  {Jahnke}}]{Husemann2013}
{Husemann}, B., {Wisotzki}, L., {S{\'a}nchez}, S.~F., \& {Jahnke}, K. 2013,
  \aap, 549, A43

\bibitem[{{Iocco} {et~al.}(2009){Iocco}, {Mangano}, {Miele}, {Pisanti}, \&
  {Serpico}}]{Iocco2009}
{Iocco}, F., {Mangano}, G., {Miele}, G., {Pisanti}, O., \& {Serpico}, P.~D.
  2009, \physrep, 472, 1

\bibitem[{{Izotov} {et~al.}(1999){Izotov}, {Chaffee}, {Foltz}, {Green},
  {Guseva}, \& {Thuan}}]{Izotov1999}
{Izotov}, Y.~I., {Chaffee}, F.~H., {Foltz}, C.~B., {et~al.} 1999, \apj, 527,
  757

\bibitem[{{Jones} {et~al.}(1994){Jones}, {Kang}, \& {Tregillis}}]{Jones1994}
{Jones}, T.~W., {Kang}, H., \& {Tregillis}, I.~L. 1994, \apj, 432, 194

\bibitem[{{Landi} {et~al.}(2013){Landi}, {Young}, {Dere}, {Del Zanna}, \&
  {Mason}}]{Landi2013}
{Landi}, E., {Young}, P.~R., {Dere}, K.~P., {Del Zanna}, G., \& {Mason}, H.~E.
  2013, \apj, 763, 86

\bibitem[{{Lau} {et~al.}(2015){Lau}, {Prochaska}, \& {Hennawi}}]{QPQ8}
{Lau}, M., {Prochaska}, J., \& {Hennawi}, J. 2015, submitted to ApJ

\bibitem[{{Lee} {et~al.}(2014){Lee}, {Hennawi}, {White}, {Croft}, {Prochaska},
  {Schlegel}, {Suzuki}, {Kneib}, {Bailey}, {Spergel}, {Rix}, \&
  {Strauss}}]{Lee2014}
{Lee}, K.-G., {Hennawi}, J.~F., {White}, M., {et~al.} 2014, in American
  Astronomical Society Meeting Abstracts, Vol. 223, American Astronomical
  Society Meeting Abstracts, 457.11

\bibitem[{{Lusso} {et~al.}(2015){Lusso}, {Worseck}, {Hennawi}, {Prochaska},
  {Vignali}, {Stern}, \& {O'Meara}}]{Lusso2015}
{Lusso}, E., {Worseck}, G., {Hennawi}, J.~F., {et~al.} 2015, ArXiv e-prints,
  arXiv:1503.02075

\bibitem[{{Maller} \& {Bullock}(2004)}]{MallerBullock2004}
{Maller}, A.~H., \& {Bullock}, J.~S. 2004, \mnras, 355, 694

\bibitem[{{Martin} {et~al.}(2014{\natexlab{a}}){Martin}, {Chang},
  {Matuszewski}, {Morrissey}, {Rahman}, {Moore}, \& {Steidel}}]{Martin2014a}
{Martin}, D.~C., {Chang}, D., {Matuszewski}, M., {et~al.} 2014{\natexlab{a}},
  \apj, 786, 106

\bibitem[{{Martin} {et~al.}(2014{\natexlab{b}}){Martin}, {Chang},
  {Matuszewski}, {Morrissey}, {Rahman}, {Moore}, {Steidel}, \&
  {Matsuda}}]{Martin2014b}
---. 2014{\natexlab{b}}, \apj, 786, 107

\bibitem[{{Matsuda} {et~al.}(2004){Matsuda}, {Yamada}, {Hayashino}, {Tamura},
  {Yamauchi}, {Ajiki}, {Fujita}, {Murayama}, {Nagao}, {Ohta}, {Okamura},
  {Ouchi}, {Shimasaku}, {Shioya}, \& {Taniguchi}}]{Matsuda2004}
{Matsuda}, Y., {Yamada}, T., {Hayashino}, T., {et~al.} 2004, \aj, 128, 569

\bibitem[{{McCarthy}(1993)}]{McCarthy1993}
{McCarthy}, P.~J. 1993, \araa, 31, 639

\bibitem[{{McClure-Griffiths} {et~al.}(2010){McClure-Griffiths}, {Madsen},
  {Gaensler}, {McConnell}, \& {Schnitzeler}}]{McClure2010}
{McClure-Griffiths}, N.~M., {Madsen}, G.~J., {Gaensler}, B.~M., {McConnell},
  D., \& {Schnitzeler}, D.~H.~F.~M. 2010, \apj, 725, 275

\bibitem[{{McCourt} {et~al.}(2015){McCourt}, {O'Leary}, {Madigan}, \&
  {Quataert}}]{McCourt2015}
{McCourt}, M., {O'Leary}, R.~M., {Madigan}, A.-M., \& {Quataert}, E. 2015,
  \mnras, 449, 2

\bibitem[{{McDonald} {et~al.}(2010){McDonald}, {Veilleux}, {Rupke}, \&
  {Mushotzky}}]{McDonald2010}
{McDonald}, M., {Veilleux}, S., {Rupke}, D.~S.~N., \& {Mushotzky}, R. 2010,
  \apj, 721, 1262

\bibitem[{{McIntosh} {et~al.}(1999){McIntosh}, {Rieke}, {Rix}, {Foltz}, \&
  {Weymann}}]{McIntosh1999}
{McIntosh}, D.~H., {Rieke}, M.~J., {Rix}, H.-W., {Foltz}, C.~B., \& {Weymann},
  R.~J. 1999, \apj, 514, 40

\bibitem[{{Miley} \& {De Breuck}(2008)}]{Miley2008}
{Miley}, G., \& {De Breuck}, C. 2008, \aapr, 15, 67

\bibitem[{{Mo} \& {Miralda-Escude}(1996)}]{MoMiralda1996}
{Mo}, H.~J., \& {Miralda-Escude}, J. 1996, \apj, 469, 589

\bibitem[{{Morrissey} {et~al.}(2012){Morrissey}, {Matuszewski}, {Martin},
  {Moore}, {Adkins}, {Epps}, {Bartos}, {Cabak}, {Cowley}, {Davis}, {Delacroix},
  {Fucik}, {Hilliard}, {James}, {Kaye}, {Lingner}, {Neill}, {Pistor},
  {Phillips}, {Rockosi}, \& {Weber}}]{Morrissey2012}
{Morrissey}, P., {Matuszewski}, M., {Martin}, C., {et~al.} 2012, in Society of
  Photo-Optical Instrumentation Engineers (SPIE) Conference Series, Vol. 8446,
  Society of Photo-Optical Instrumentation Engineers (SPIE) Conference Series,
  13

\bibitem[{{Nelson} {et~al.}(2015){Nelson}, {Genel}, {Pillepich},
  {Vogelsberger}, {Springel}, \& {Hernquist}}]{Nelson2015}
{Nelson}, D., {Genel}, S., {Pillepich}, A., {et~al.} 2015, ArXiv e-prints,
  arXiv:1503.02665

\bibitem[{{Neufeld}(1990)}]{Neufeld1990}
{Neufeld}, D.~A. 1990, \apj, 350, 216

\bibitem[{{North} {et~al.}(2012){North}, {Courbin}, {Eigenbrod}, \&
  {Chelouche}}]{North2012}
{North}, P.~L., {Courbin}, F., {Eigenbrod}, A., \& {Chelouche}, D. 2012, \aap,
  542, A91

\bibitem[{{Oke}(1974)}]{Oke1974}
{Oke}, J.~B. 1974, \apjs, 27, 21

\bibitem[{{Oke} {et~al.}(1995){Oke}, {Cohen}, {Carr}, {Cromer}, {Dingizian},
  {Harris}, {Labrecque}, {Lucinio}, {Schaal}, {Epps}, \& {Miller}}]{Oke1995}
{Oke}, J.~B., {Cohen}, J.~G., {Carr}, M., {et~al.} 1995, \pasp, 107, 375

\bibitem[{{Osterbrock} \& {Ferland}(2006)}]{OF2006}
{Osterbrock}, D.~E., \& {Ferland}, G.~J. 2006, {Astrophysics of Gaseous Nebulae
  and Active Galactic Nuclei (2nd ed.: Sausalito, CA: Univ. Science Books)}

\bibitem[{{Overzier} {et~al.}(2013){Overzier}, {Nesvadba}, {Dijkstra}, {Hatch},
  {Lehnert}, {Villar-Mart{\'{\i}}n}, {Wilman}, \& {Zirm}}]{Overzier2013}
{Overzier}, R.~A., {Nesvadba}, N.~P.~H., {Dijkstra}, M., {et~al.} 2013, \apj,
  771, 89

\bibitem[{{Planck Collaboration} {et~al.}(2014){Planck Collaboration}, {Ade},
  {Aghanim}, {Armitage-Caplan}, {Arnaud}, {Ashdown}, {Atrio-Barandela},
  {Aumont}, {Baccigalupi}, {Banday}, \& et~al.}]{PlanckColl2014}
{Planck Collaboration}, {Ade}, P.~A.~R., {Aghanim}, N., {et~al.} 2014, \aap,
  571, A16

\bibitem[{{Prescott} {et~al.}(2009){Prescott}, {Dey}, \&
  {Jannuzi}}]{Prescott2009}
{Prescott}, M.~K.~M., {Dey}, A., \& {Jannuzi}, B.~T. 2009, \apj, 702, 554

\bibitem[{{Prescott} {et~al.}(2013){Prescott}, {Dey}, \&
  {Jannuzi}}]{Prescott2013}
---. 2013, \apj, 762, 38

\bibitem[{{Prescott} {et~al.}(2015{\natexlab{a}}){Prescott}, {Martin}, \&
  {Dey}}]{Prescott2015a}
{Prescott}, M.~K.~M., {Martin}, C.~L., \& {Dey}, A. 2015{\natexlab{a}}, \apj,
  799, 62

\bibitem[{{Prescott} {et~al.}(2015{\natexlab{b}}){Prescott}, {Momcheva},
  {Brammer}, {Fynbo}, \& {M{\o}ller}}]{Prescott2015}
{Prescott}, M.~K.~M., {Momcheva}, I., {Brammer}, G.~B., {Fynbo}, J.~P.~U., \&
  {M{\o}ller}, P. 2015{\natexlab{b}}, ArXiv e-prints, arXiv:1501.05312

\bibitem[{{Prochaska} \& {Hennawi}(2009)}]{Prochaska2009}
{Prochaska}, J.~X., \& {Hennawi}, J.~F. 2009, \apj, 690, 1558

\bibitem[{{Prochaska} {et~al.}(2013{\natexlab{a}}){Prochaska}, {Hennawi}, \&
  {Simcoe}}]{Prochaska2013}
{Prochaska}, J.~X., {Hennawi}, J.~F., \& {Simcoe}, R.~A. 2013{\natexlab{a}},
  \apjl, 762, L19

\bibitem[{{Prochaska} {et~al.}(2014){Prochaska}, {Wingyee Lau}, \&
  {Hennawi}}]{Prochaska2014}
{Prochaska}, J.~X., {Wingyee Lau}, M., \& {Hennawi}, J.~F. 2014, \apj, 796, 140

\bibitem[{{Prochaska} {et~al.}(2013{\natexlab{b}}){Prochaska}, {Hennawi},
  {Lee}, {Cantalupo}, {Bovy}, {Djorgovski}, {Ellison}, {Lau}, {Martin},
  {Myers}, {Rubin}, \& {Simcoe}}]{Prochaska2013b}
{Prochaska}, J.~X., {Hennawi}, J.~F., {Lee}, K.-G., {et~al.}
  2013{\natexlab{b}}, \apj, 776, 136

\bibitem[{{Rauch} {et~al.}(1999){Rauch}, {Sargent}, \& {Barlow}}]{Rauch1999}
{Rauch}, M., {Sargent}, W.~L.~W., \& {Barlow}, T.~A. 1999, \apj, 515, 500

\bibitem[{{Rauch} {et~al.}(2008){Rauch}, {Haehnelt}, {Bunker}, {Becker},
  {Marleau}, {Graham}, {Cristiani}, {Jarvis}, {Lacey}, {Morris}, {Peroux},
  {R{\"o}ttgering}, \& {Theuns}}]{Rauch2008}
{Rauch}, M., {Haehnelt}, M., {Bunker}, A., {et~al.} 2008, \apj, 681, 856

\bibitem[{{Rees}(1988)}]{Rees1988}
{Rees}, M.~J. 1988, \mnras, 231, 91P

\bibitem[{{Reuland} {et~al.}(2003){Reuland}, {van Breugel}, {R{\"o}ttgering},
  {de Vries}, {Stanford}, {Dey}, {Lacy}, {Bland-Hawthorn}, {Dopita}, \&
  {Miley}}]{Reuland2003}
{Reuland}, M., {van Breugel}, W., {R{\"o}ttgering}, H., {et~al.} 2003, \apj,
  592, 755

\bibitem[{{Reuland} {et~al.}(2007){Reuland}, {van Breugel}, {de Vries},
  {Dopita}, {Dey}, {Miley}, {R{\"o}ttgering}, {Venemans}, {Stanford}, {Lacy},
  {Spinrad}, {Dawson}, {Stern}, \& {Bunker}}]{Reuland2007}
{Reuland}, M., {van Breugel}, W., {de Vries}, W., {et~al.} 2007, \aj, 133, 2607

\bibitem[{{Richards} {et~al.}(2006){Richards}, {Lacy}, {Storrie-Lombardi},
  {Hall}, {Gallagher}, {Hines}, {Fan}, {Papovich}, {Vanden Berk}, {Trammell},
  {Schneider}, {Vestergaard}, {York}, {Jester}, {Anderson}, {Budav{\'a}ri}, \&
  {Szalay}}]{Richards2006}
{Richards}, G.~T., {Lacy}, M., {Storrie-Lombardi}, L.~J., {et~al.} 2006, \apjs,
  166, 470

\bibitem[{{Rosdahl} \& {Blaizot}(2012)}]{Rosdahl12}
{Rosdahl}, J., \& {Blaizot}, J. 2012, \mnras, 423, 344

\bibitem[{{Rudie} {et~al.}(2012){Rudie}, {Steidel}, {Trainor}, {Rakic},
  {Bogosavljevi{\'c}}, {Pettini}, {Reddy}, {Shapley}, {Erb}, \&
  {Law}}]{Rudie2012}
{Rudie}, G.~C., {Steidel}, C.~C., {Trainor}, R.~F., {et~al.} 2012, \apj, 750,
  67

\bibitem[{{Scannapieco} \& {Br{\"u}ggen}(2015)}]{Scannapieco2015}
{Scannapieco}, E., \& {Br{\"u}ggen}, M. 2015, ArXiv e-prints, arXiv:1503.06800

\bibitem[{{Schaye} {et~al.}(2003){Schaye}, {Aguirre}, {Kim}, {Theuns}, {Rauch},
  \& {Sargent}}]{Schaye2003}
{Schaye}, J., {Aguirre}, A., {Kim}, T.-S., {et~al.} 2003, \apj, 596, 768

\bibitem[{{Schaye} {et~al.}(2007){Schaye}, {Carswell}, \& {Kim}}]{Schaye2007}
{Schaye}, J., {Carswell}, R.~F., \& {Kim}, T.-S. 2007, \mnras, 379, 1169

\bibitem[{{Scott} {et~al.}(2004){Scott}, {Kriss}, {Brotherton}, {Green},
  {Hutchings}, {Shull}, \& {Zheng}}]{Scott2004}
{Scott}, J.~E., {Kriss}, G.~A., {Brotherton}, M., {et~al.} 2004, \apj, 615, 135

\bibitem[{{Sharples} {et~al.}(2006){Sharples}, {Bender}, {Bennett}, {Burch},
  {Carter}, {Clark}, {Content}, {Davies}, {Davies}, {Dubbeldam}, {Genzel},
  {Hess}, {Laidlaw}, {Lehnert}, {Lewis}, {Muschielok}, {Ramsey-Howat}, {Rees},
  {Robertson}, {Robson}, {Saglia}, {Tecza}, {Thatte}, {Todd}, {Wall}, \&
  {Wegner}}]{Sharples2006}
{Sharples}, R., {Bender}, R., {Bennett}, R., {et~al.} 2006, \nar, 50, 370

\bibitem[{{Shull} {et~al.}(2012){Shull}, {Stevans}, \& {Danforth}}]{Shull2012}
{Shull}, J.~M., {Stevans}, M., \& {Danforth}, C.~W. 2012, \apj, 752, 162

\bibitem[{{Simcoe} {et~al.}(2006){Simcoe}, {Sargent}, {Rauch}, \&
  {Becker}}]{Simcoe2006}
{Simcoe}, R.~A., {Sargent}, W.~L.~W., {Rauch}, M., \& {Becker}, G. 2006, \apj,
  637, 648

\bibitem[{{Smith} \& {Jarvis}(2007)}]{Smith2007}
{Smith}, D.~J.~B., \& {Jarvis}, M.~J. 2007, \mnras, 378, L49

\bibitem[{{Steidel} {et~al.}(2000){Steidel}, {Adelberger}, {Shapley},
  {Pettini}, {Dickinson}, \& {Giavalisco}}]{Steidel2000}
{Steidel}, C.~C., {Adelberger}, K.~L., {Shapley}, A.~E., {et~al.} 2000, \apj,
  532, 170

\bibitem[{{Steidel} {et~al.}(2011){Steidel}, {Bogosavljevi{\'c}}, {Shapley},
  {Kollmeier}, {Reddy}, {Erb}, \& {Pettini}}]{Steidel2011}
{Steidel}, C.~C., {Bogosavljevi{\'c}}, M., {Shapley}, A.~E., {et~al.} 2011,
  \apj, 736, 160

\bibitem[{{Stern} {et~al.}(2014){Stern}, {Laor}, \& {Baskin}}]{Stern2014}
{Stern}, J., {Laor}, A., \& {Baskin}, A. 2014, \mnras, 438, 901

\bibitem[{{Stockton} {et~al.}(2006){Stockton}, {Fu}, \&
  {Canalizo}}]{Stockton2006}
{Stockton}, A., {Fu}, H., \& {Canalizo}, G. 2006, \nar, 50, 694

\bibitem[{{Stockton} {et~al.}(2002){Stockton}, {MacKenty}, {Hu}, \&
  {Kim}}]{Stockton2002}
{Stockton}, A., {MacKenty}, J.~W., {Hu}, E.~M., \& {Kim}, T.-S. 2002, \apj,
  572, 735

\bibitem[{{Strateva} {et~al.}(2005){Strateva}, {Brandt}, {Schneider}, {Vanden
  Berk}, \& {Vignali}}]{Strateva2005}
{Strateva}, I.~V., {Brandt}, W.~N., {Schneider}, D.~P., {Vanden Berk}, D.~G.,
  \& {Vignali}, C. 2005, \aj, 130, 387

\bibitem[{{Taniguchi} \& {Shioya}(2000)}]{Taniguchi&Shioya2000}
{Taniguchi}, Y., \& {Shioya}, Y. 2000, \apjl, 532, L13

\bibitem[{{Taniguchi} {et~al.}(2001){Taniguchi}, {Shioya}, \&
  {Kakazu}}]{Taniguchi2001}
{Taniguchi}, Y., {Shioya}, Y., \& {Kakazu}, Y. 2001, \apjl, 562, L15

\bibitem[{{van de Voort} \& {Schaye}(2013)}]{vandeVoort2013}
{van de Voort}, F., \& {Schaye}, J. 2013, \mnras, 430, 2688

\bibitem[{{van Ojik} {et~al.}(1997){van Ojik}, {Roettgering}, {Miley}, \&
  {Hunstead}}]{vanOjik1997}
{van Ojik}, R., {Roettgering}, H.~J.~A., {Miley}, G.~K., \& {Hunstead}, R.~W.
  1997, \aap, 317, 358

\bibitem[{{Vanden Berk} {et~al.}(2001){Vanden Berk}, {Richards}, {Bauer},
  {Strauss}, {Schneider}, {Heckman}, {York}, {Hall}, {Fan}, {Knapp},
  {Anderson}, {Annis}, {Bahcall}, {Bernardi}, {Briggs}, {Brinkmann}, {Brunner},
  {Burles}, {Carey}, {Castander}, {Connolly}, {Crocker}, {Csabai}, {Doi},
  {Finkbeiner}, {Friedman}, {Frieman}, {Fukugita}, {Gunn}, {Hennessy},
  {Ivezi{\'c}}, {Kent}, {Kunszt}, {Lamb}, {Leger}, {Long}, {Loveday}, {Lupton},
  {Meiksin}, {Merelli}, {Munn}, {Newberg}, {Newcomb}, {Nichol}, {Owen}, {Pier},
  {Pope}, {Rockosi}, {Schlegel}, {Siegmund}, {Smee}, {Snir}, {Stoughton},
  {Stubbs}, {SubbaRao}, {Szalay}, {Szokoly}, {Tremonti}, {Uomoto}, {Waddell},
  {Yanny}, \& {Zheng}}]{VandenBerk2001}
{Vanden Berk}, D.~E., {Richards}, G.~T., {Bauer}, A., {et~al.} 2001, \aj, 122,
  549

\bibitem[{{Verhamme} {et~al.}(2006){Verhamme}, {Schaerer}, \&
  {Maselli}}]{Verhamme2006}
{Verhamme}, A., {Schaerer}, D., \& {Maselli}, A. 2006, \aap, 460, 397

\bibitem[{{Villar-Mart{\'{\i}}n} {et~al.}(2007){Villar-Mart{\'{\i}}n},
  {Humphrey}, {De Breuck}, {Fosbury}, {Binette}, \& {Vernet}}]{VillarM2007}
{Villar-Mart{\'{\i}}n}, M., {Humphrey}, A., {De Breuck}, C., {et~al.} 2007,
  \mnras, 375, 1299

\bibitem[{{Villar-Mart{\'{\i}}n}
  {et~al.}(2003{\natexlab{a}}){Villar-Mart{\'{\i}}n}, {Vernet}, {di Serego
  Alighieri}, {Fosbury}, {Humphrey}, \& {Pentericci}}]{VillarM2003b}
{Villar-Mart{\'{\i}}n}, M., {Vernet}, J., {di Serego Alighieri}, S., {et~al.}
  2003{\natexlab{a}}, \mnras, 346, 273

\bibitem[{{Villar-Mart{\'{\i}}n}
  {et~al.}(2003{\natexlab{b}}){Villar-Mart{\'{\i}}n}, {Vernet}, {di Serego
  Alighieri}, {Fosbury}, {Humphrey}, {Pentericci}, \& {Cohen}}]{VillarM2003}
---. 2003{\natexlab{b}}, \nar, 47, 291

\bibitem[{{Werk} {et~al.}(2014){Werk}, {Prochaska}, {Tumlinson}, {Peeples},
  {Tripp}, {Fox}, {Lehner}, {Thom}, {O'Meara}, {Ford}, {Bordoloi}, {Katz},
  {Tejos}, {Oppenheimer}, {Dav{\'e}}, \& {Weinberg}}]{Werk2014}
{Werk}, J.~K., {Prochaska}, J.~X., {Tumlinson}, J., {et~al.} 2014, \apj, 792, 8

\bibitem[{{White} {et~al.}(2012){White}, {Myers}, {Ross}, {Schlegel},
  {Hennawi}, {Shen}, {McGreer}, {Strauss}, {Bolton}, {Bovy}, {Fan},
  {Miralda-Escude}, {Palanque-Delabrouille}, {Paris}, {Petitjean}, {Schneider},
  {Viel}, {Weinberg}, {Yeche}, {Zehavi}, {Pan}, {Snedden}, {Bizyaev},
  {Brewington}, {Brinkmann}, {Malanushenko}, {Malanushenko}, {Oravetz},
  {Simmons}, {Sheldon}, \& {Weaver}}]{White2012}
{White}, M., {Myers}, A.~D., {Ross}, N.~P., {et~al.} 2012, \mnras, 424, 933

\bibitem[{{Wilman} {et~al.}(2005){Wilman}, {Gerssen}, {Bower}, {Morris},
  {Bacon}, {de Zeeuw}, \& {Davies}}]{Wilman2005}
{Wilman}, R.~J., {Gerssen}, J., {Bower}, R.~G., {et~al.} 2005, \nat, 436, 227

\bibitem[{{Yang} {et~al.}(2014){Yang}, {Zabludoff}, {Jahnke}, \&
  {Dav{\'e}}}]{Yang2014}
{Yang}, Y., {Zabludoff}, A., {Jahnke}, K., \& {Dav{\'e}}, R. 2014, ArXiv
  e-prints, arXiv:1407.6801

\bibitem[{{Yang} {et~al.}(2011){Yang}, {Zabludoff}, {Jahnke}, {Eisenstein},
  {Dav{\'e}}, {Shectman}, \& {Kelson}}]{Yang2011}
{Yang}, Y., {Zabludoff}, A., {Jahnke}, K., {et~al.} 2011, \apj, 735, 87

\bibitem[{{Yang} {et~al.}(2006){Yang}, {Zabludoff}, {Dav{\'e}}, {Eisenstein},
  {Pinto}, {Katz}, {Weinberg}, \& {Barton}}]{Yang2006}
{Yang}, Y., {Zabludoff}, A.~I., {Dav{\'e}}, R., {et~al.} 2006, \apj, 640, 539

\bibitem[{{Yang} {et~al.}(2012){Yang}, {Decarli}, {Dannerbauer}, {Walter},
  {Weiss}, {Leipski}, {Dey}, {Chapman}, {Le Floc'h}, {Prescott}, {Neri},
  {Borys}, {Matsuda}, {Yamada}, {Hayashino}, {Tapken}, \& {Menten}}]{Yang2012}
{Yang}, Y., {Decarli}, R., {Dannerbauer}, H., {et~al.} 2012, \apj, 744, 178

\bibitem[{{Zheng} {et~al.}(1997){Zheng}, {Kriss}, {Telfer}, {Grimes}, \&
  {Davidsen}}]{Zheng1997}
{Zheng}, W., {Kriss}, G.~A., {Telfer}, R.~C., {Grimes}, J.~P., \& {Davidsen},
  A.~F. 1997, \apj, 475, 469

\end{thebibliography}

\clearpage

\appendix

\section{Effects of a Variation of the EUV Slope of the Input Spectrum}

We test the robustness of our results to the change of the slope of the EUV as mentioned in 
Section \S\ref{SED}. 
In particular, we model the extremes of the range allowed by the recent estimates of \citet{Lusso2015}, i.e. $\alpha_{\rm EUV}=-1.7\pm0.6$. 
To fulfill the $\alpha_{\rm OX}$ requirement of \citet{Strateva2005} as explained in Section \S\ref{SED}, 
the value $\alpha_{\rm EUV}=-2.3$ and -1.1 imply at higher energies ($30\, {\rm Ryd} < h\nu < 2 {\rm keV}$) a slope $\alpha=-0.36$ and -2.93, respectively. 
In our fiducial input spectrum ($\alpha_{\rm EUV}=-1.7$), the photoionization rate at the Lyman limit is
\begin{equation}
\Gamma = \frac{1}{4\pi r^2}\int_{\nu_{\rm LL}}^{\infty} 
\frac{L_{\nu}}{h\nu}\sigma_\nu d\nu = 6.7\times10^{-9} \, {\rm s^{-1}},
\end{equation}
while at 4 Ryd, i.e. at the ionization energy of \heii, the photoionization rate is
$\Gamma_{\rm 4 Ryd}\sim1.0\times10^{-11}$ s$^{-1}$. 
By changing the slope in the extreme ultraviolet from $\alpha_{\rm EUV}=-1.7$, to -1.1 and to -2.3, we increase the photoionization
rate by $\sim$15\% and decrease it by $\sim$13\%, respectively.
Instead, for the same change, the $\Gamma_{\rm 4 Ryd}$ is increased/decreased by a factor of 2.6, respectively.
As it is clear from the small changes in $\Gamma$, the Hydrogen ionization state is not affected by the change in slope, and the 
models are always optically thin. Conversely, as expected, the changes in $\Gamma_{\rm 4 Ryd}$ affect \heii and \civ.
The general trend is that a softer slope, e.g. $\alpha_{\rm EUV}=-2.3$, produces fewer \heii ionizing photons, and thus 
at fixed density the \heiii fraction will be lower, resulting in lower \heii recombination emission. This thus leads to a lower 
\heii/\lya ratio.
Similarly, a softer slope is less effective in ionizing Carbon. In particular, at fixed ionization parameter $U$, 
the amount of Carbon in the C$^{+3}$ phase is lower for a softer slope.

\begin{figure}
\centering
\epsfig{file=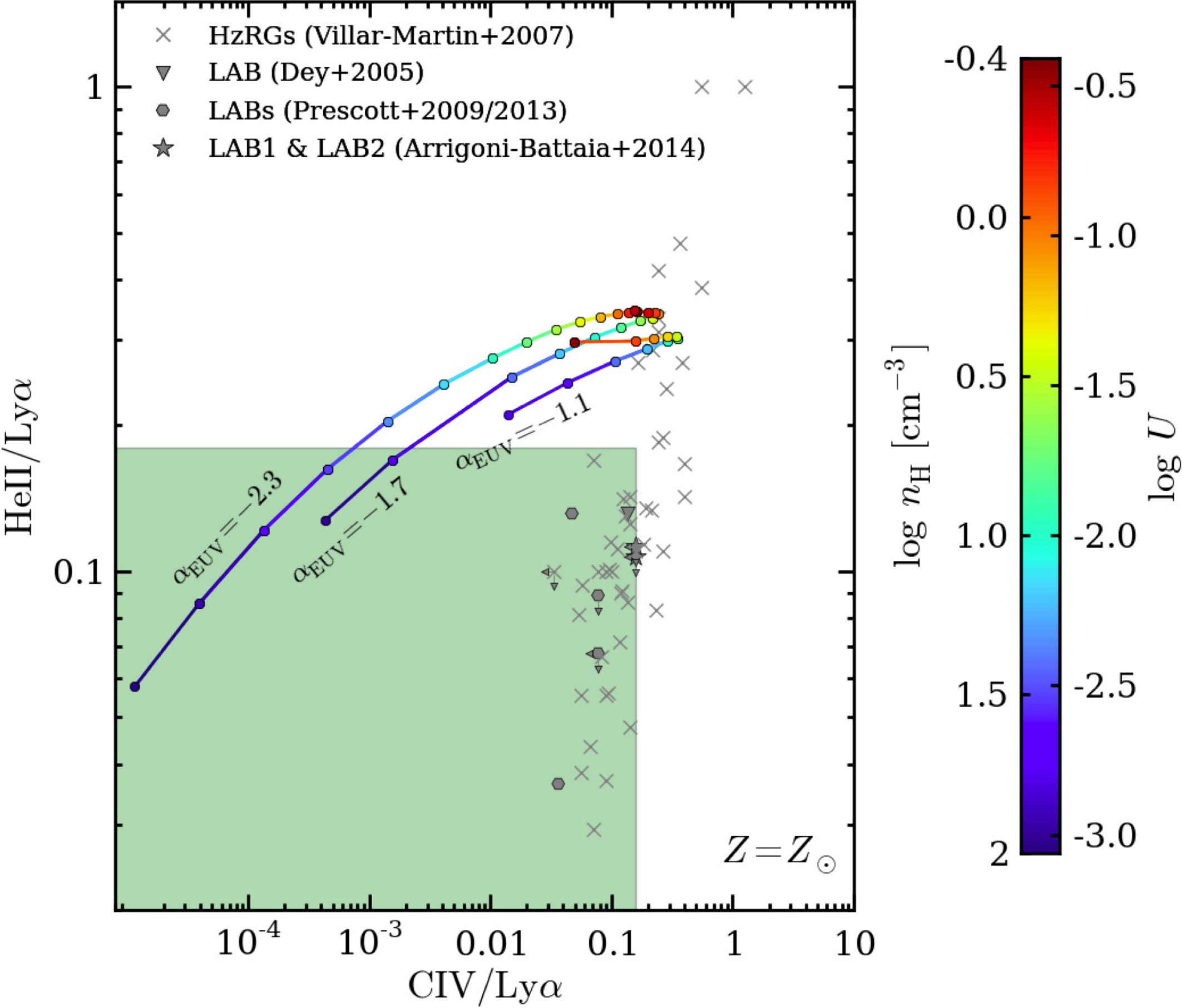, width=0.5\columnwidth, angle=270, clip} \\
\epsfig{file=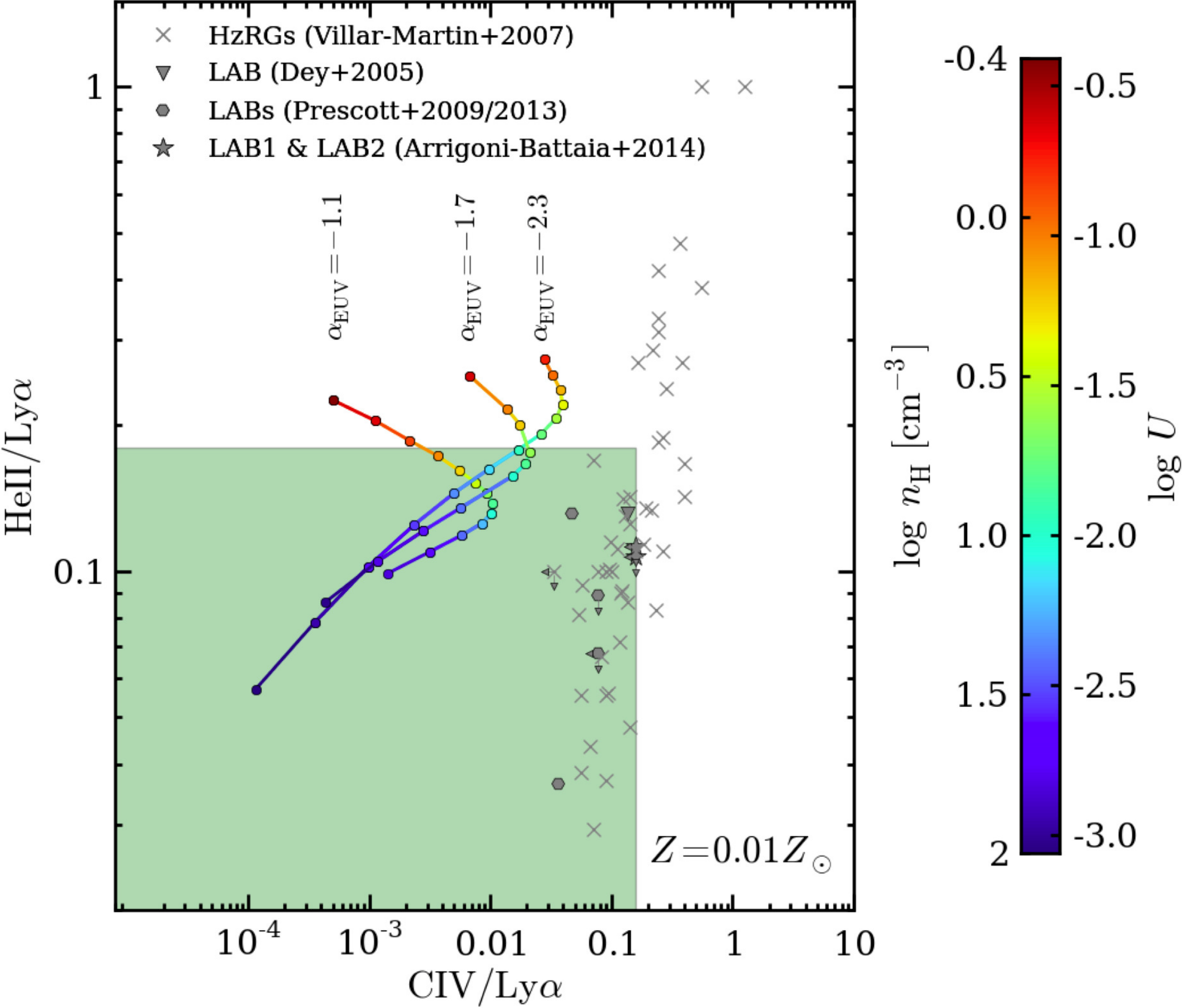, width=0.5\columnwidth, angle=270, clip} 
\caption{HeII/Ly$\alpha$ versus CIV/Ly$\alpha$ log-log plot. Our upper limits on the HeII/Ly$\alpha$ and CIV/Ly$\alpha$ ratios are compared with 
the Cloudy photoionization models that reproduce the observed SB$_{\rm Ly\alpha}\sim7\times10^{-18}$ \unitcgssb. 
{\bf Upper panel:} comparison of the model grids for different EUV slopes ($\alpha_{\rm EUV}=-1.1,-1.7,-2.3$) at $Z=Z_{\odot}$. 
A harder $\alpha_{\rm EUV}$ completely doubly ionize Helium at higher density.
{\bf Bottom panel:} same as the upper panel, but at $Z=0.01Z_{\odot}$. 
In this case, the \lya line is also powered by collisions, reshaping the trajectories (see text for explanation on the trends in this Figure).
In both panels, the models are color coded following the ionization parameter $U$, 
or equivalently the volume density $n_{\rm H}$ (see color bar on the right).
The green shaded area represents the region defined by the upper limits of the UM287 nebula. 
See Figure \ref{Fig2App} for a better visualization of the constraints on the physical parameters.}
\label{Fig1App}
\end{figure}

In Figure \ref{Fig1App} we compare our grids of models with different EUV
slopes at two different metallicities, i.e.  $Z=Z_{\odot}$, and $0.01
Z_{\odot}$, in the \heii/\lya versus \civ/\lya plot.  The dependencies
outlined above, are better visible in the plot for solar metallicity
(upper panel) because the \lya line is mainly produced by
recombinations and its behavior is not influencing the general trends.
From the figure,  it is clear that a grid with a softer slope (see grid
with $\alpha_{\rm EUV}=-2.3$) can reach lower \heii/\lya ratios
because the fraction of doubly ionized Helium is lower at high
densities.  In the same upper panel of Figure \ref{Fig1App} it is also
evident that the simulation grids for different UV slopes all
asymptote to a fixed \heii/\lya ratio when  Helium is completely
doubly ionized, which occurs at slightly different $n_{\rm H}$ (or equivalently $U$)
for each slope. Note that the value of the asymptotic \heii/\lya ratio
varies slightly with slope. Indeed, as mentioned in section \S\ref{sec:comparison}, since this
asymptotic value is proportional to the ratio of the recombination coefficients
of \heii and \lya, the value depends on temperature (eqn.~(\ref{ratioHeIILyalpha})). Higher
temperatures, which arise for a harder slope, lead to a lower
asymptotic \heii/\lya ratio.

In the bottom panel of
Figure \ref{Fig1App}, we show the same comparison at $Z=0.01Z_{\odot}$.
In this case the trends are masked by the \lya line, which is powered
also by collisions.  Indeed, the saturation in the \heii/\lya ratio is
not appreciable because, given the dependence on density of the
collisional contribution to the \lya line, the ratio is progressively
lowered at higher density.
However, it is still appreciable that the
\civ/\lya ratio is moved to lower ratios for higher slopes above log$U\sim-1.5$.
This is mainly due
to the fact that Carbon goes to higher ionization state, lowering the
fraction of Carbon in the C$^{+3}$ species. 
Thus, in our case study, where the input spectrum is not well known, 
the dependence of the amount of C$^{+3}$ on the slope of the EUV makes the \civ line
a weak metallicity indicator.

\begin{figure}
\centering
\epsfig{file=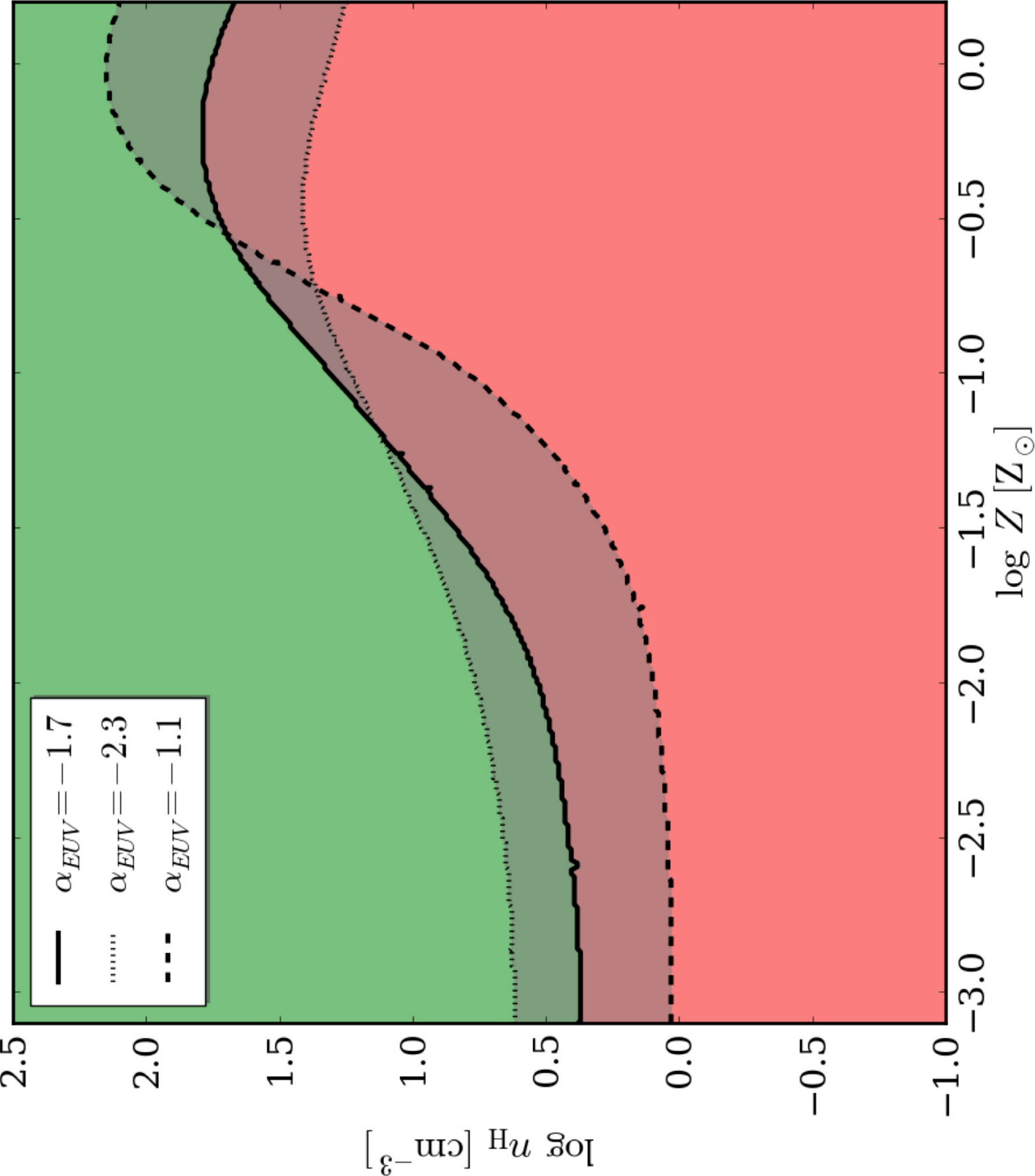, width=0.5\columnwidth, angle=270, clip} 
\caption{Schematic representation on how a variation in $\alpha_{\rm EUV}$ affects the constraints in $n_{\rm H}$ 
and $Z$. The green area highlights the region of the parameter space selected by the upper limit \heii/\lya$<0.18$ (see 
panel `a' of Figure \ref{Fig7}). The solid, dashed, and dotted lines show the location of this upper limit for $\alpha_{\rm EUV}=-1.7,-1.1$
, and $-2.3$, respectively. It is evident that a change in the ionizing slope do not affect our main conclusions. 
Namely, if the nebula is photoionized by the UM287 quasar, there should be a population of dense cool gas clumps
with very small sizes ($\lesssim$tens of pc).}
\label{Fig2App}
\end{figure}

Changes in the slope $\alpha_{\rm EUV}$ only slightly
modifies the constraints on $n_{\rm H}$ that we previously
obtained. In particular, since the \heii/\lya ratio gives the stronger
constraints, in Figure \ref{Fig2App} we show how a variation in the EUV
slope affects the selection of $n_{\rm H}$ (compare Figure
\ref{Fig1App} and \ref{Fig2App}).  This Figure highlights in green the
parameter space favored by our upper limits (the lines show the
location of the upper limit \heii/\lya$=0.18$).  
The mild change in the location of the line is explained by the dependencies outlined above.
At a fixed low metallicity, where the \lya line is 
an important coolant, i.e. log$Z<-1.5Z_{\odot}$, 
a harder slope moves the lower limit boundary implied by our measurement on the \heii/\lya ratio 
to lower densities. Indeed, the expected increase of the \heii line due to a harder slope is 
washed out by the increase in the emission in the \lya line due to collisions. Thus,
our constraint on the density that we quote in the main text is weakened from $n_{\rm H}\gtrsim 3$ cm$^{-3}$ to $n_{\rm H}\gtrsim 1$ cm$^{-3}$.
On the other hand, at higher metallicities, a harder UV slope will doubly ionize Helium
at higher density, moving the lower limit boundary implied by our measurement to higher densities. For example, 
at solar metallicity, the limit is moved to $\gtrsim100$ cm$^{-3}$ from $\gtrsim40$ cm$^{-3}$.

Thus, in conclusion, our ignorance on the slope of the EUV has a small effect on our density constraints and 
makes the \civ line a weak metallicity indicator. However, as discussed at the end of \S\ref{sec:sens}, the detection of
multiple metal lines with a range of ionization energies would indirectly constrain $\alpha_{\rm EUV}$, and simultaneously 
constrain the metallicity of the gas.

\clearpage

\end{document}